\documentclass[11pt]{article}
\pdfoutput=1

\usepackage{graphicx,multirow,subfig}
\usepackage{float}
\usepackage{bm}
\usepackage{amsmath}
\usepackage{amssymb}
\usepackage{amscd}
\usepackage{slashed}
\usepackage{graphicx}
\usepackage{color}
\usepackage{mathrsfs}
\usepackage{hyperref} 
\hypersetup{
    colorlinks=true,       
    linkcolor=black,          
    citecolor=blue,        
    filecolor=magenta,      
    urlcolor=blue           
}
\newcommand{\be}{\begin{equation}}
\newcommand{\ee}{\end{equation}}

\newcommand{\br}{\mathrm{BR}}
\newcommand{\Hg}{\mathcal{H}}

\begin{document}

\title{\bf A Concrete Composite 2-Higgs Doublet Model}

\author{\large {\large{\hspace{-0.5cm}} Stefania De Curtis$^a$, Luigi Delle Rose$^{a,b}$, Stefano Moretti$^{b,c}$, Kei Yagyu$^{d}$,}\\[0.35cm]
\small {\it $^a$INFN, Sezione di Firenze, and Department of Physics and Astronomy, University of Florence,} \\
\small{\it Via G. Sansone 1, 50019 Sesto Fiorentino, Italy}\\
\small{\it $^b$School of Physics and Astronomy, University of Southampton,} \\
\small{\it Highfield, Southampton SO17 1BJ, United Kingdom}\\
\small{\it $^c$Particle Physics Department, Rutherford Appleton Laboratory,} \\
\small{\it Chilton, Didcot, Oxon OX11 0QX, United Kingdom}\\
\small{\it $^d$  Seikei University, Musashino, Tokyo 180-8633, Japan  }}
\date{\today}

\maketitle

\begin{abstract}
\noindent
We  consider a  Composite Higgs Model  (CHM) with two isospin doublet Higgs fields arising as pseudo Nambu-Goldstone bosons from a ${\rm SO}(6)\to {\rm SO}(4)\times {\rm SO}(2)$ breaking.
The main focus of this work is to explicitly compute the properties of these Higgses    in terms of the fundamental parameters of the composite sector  such as masses, Yukawa and gauge couplings of the new spin-1 and spin-1/2 resonances. Concretely,  we calculate the Higgs potential at one-loop level through the Coleman-Weinberg mechanism from the explicit breaking of the ${\rm SO(6)}$ global symmetry by the partial compositeness of fermions and gauge bosons. 
We derive then the phenomenological properties of the Higgs states and highlight the main  signatures of this Composite 2-Higgs Doublet Model  at the Large Hadron Collider, including modifications to the SM-like Higgs couplings as well as production and decay channels of  heavier Higgs bosons. We also consider flavour bounds that are typical of CHMs  with more than one Higgs doublet.
\end{abstract}

\newpage

\tableofcontents
\section{Introduction}
\label{sec:introduction}
While the discovered Higgs boson is consistent with the Standard Model (SM) one, this could just be the first manifestation of an Electro-Weak Symmetry Breaking (EWSB) dynamics that is far richer than the minimal one existing in the current prevalent description of Nature. On the one hand,  the latter is non-minimal in both its matter (as there are three generations of  quarks and leptons) and interaction (as multiple gauge bosons states of different multiplicities exist) content, so that one is well motivated to postulate a non-minimal Higgs sector too.
On the other hand,  bearing in mind that the discovered Higgs state has a doublet construction, one is well justified in pursuing first, in the quest for some Beyond the SM (BSM) physics, the study of 
 2-Higgs Doublet Models (2HDMs). 
In fact, these scenarios always include a neutral scalar Higgs state that can play the
role of the detected one. Furthermore, these constructs offer additional (pseudo)scalar states, both neutral and charged, amenable to discovery by the ATLAS and CMS collaborations, which are now substantially engaged in direct searches for new Higgs bosons, in parallel with  extracting a possible BSM dynamics indirectly from the  precision measurements of the detected one.

However, 2HDMs  do not have the ability to solve the so-called hierarchy problem of the SM.  An elegant way to overcome it  is  
to presume that the revealed Higgs state  and its possible 2HDM companions are not fundamental particles, just like any spin-0 object discovered so far in Nature. In this sense, one would be interpreting these (pseudo)scalar states belonging to a Composite 2HDM (C2HDM) as (fermion) composites, i.e.,  mesonic states of a new
 theory of strong interactions not dissimilar from QCD.   A
phenomenologically viable possibility,   wherein the mass of the lightest Higgs state is kept naturally lighter than a new strong scale (of compositeness, $f$, in the $\sim $ TeV region)  is, in particular, the one of assigning to these QCD-like states a pseudo-Nambu-Goldstone Boson
(pNGB) nature, like   in  Composite Higgs Models (CHMs)  arising from the spontaneous symmetry breaking around the TeV scale of the global symmetry of such a new strong sector \cite{Dugan:1984hq}. The residual symmetry is then explicitly broken by the SM  interactions through the \textit{partial compositeness} paradigm  \cite{Kaplan:1991dc,Contino:2006nn}.
In the minimal CHM \cite{Agashe:2004rs,Contino:2006qr,DeCurtis:2011yx,Matsedonskyi:2012ym,Pomarol:2012qf,Marzocca:2012zn,Matsedonskyi:2015dns,Matsedonskyi:2014mna} 
the lone Higgs state is a pNGB (surrounded by various composite resonances, both spin-1/2 and
spin-1, generally heavier). Hence, it is natural to assume  that the new (pseudo)scalar Higgs states of a C2HDM can also be created as  pNGBs. 

Such C2HDMs embedding pNGBs, which arise from a new confining strong dynamics, can be constructed  by explicitly imposing a specific symmetry breaking structure. Herein, we will analyse 2HDMs based on  the spontaneous global symmetry breaking of a ${\rm SO}(6)\to {\rm SO}(4)\times {\rm SO}(2)$ symmetry \cite{Mrazek:2011iu}. Within this construct, one can then study both the deviations of C2HDM couplings from those of a generic renormalisable Elementary 2HDM  (E2HDM) \cite{Branco:2011iw} as well as pursue searches for new non-SM-like Higgs signals. 

We explicitly construct here a C2HDM making only a few specific assumptions about the strong sector, namely, the global symmetries, their pattern of spontaneous breaking and the sources of explicit breaking (as intimated, in our approach, they come from the couplings of the new strong sector with the SM fields) and by generalising to the coset ${\rm SO}(6)/{\rm SO}(4)\times {\rm SO}(2)$ the 2-site minimal construction developed in \cite{DeCurtis:2011yx}. (We will also show in Appendix \ref{sec:ccwz} the equivalence with the standard prescription by Callan, Coleman, Wess and Zumino (CCWZ) \cite{Coleman:1969sm,Callan:1969sn}.)
The scalar potential is in the end generated by loop effects a la Coleman-Weinberg (CW) \cite{Coleman:1973jx} and, at the lowest order,  is mainly determined by the free parameters associated to the sole top-quark and the complete gauge sector \cite{Mrazek:2011iu}. 

Calculations of the Higgs potential in the composite realisation of 2HDMs have been performed in the pioneering work~\cite{Mrazek:2011iu} based on the CCWZ technique, in which 
each coefficient of the potential is expressed in terms of invariants of the global symmetry computed as an expansion in the parameters responsible for the partial compositeness. 
This approach is quite general, but the undetermined ${\cal O}(1)$ coefficients associated to each invariant prevent one to exploit the dependence on the masses and couplings of the resonances generated by the new strong sector.
Furthermore, the computation of such coefficients is crucial to make a clear connection with the parameters of the strong dynamics which depends on the choice of the model setup.

As mentioned before, we adopt an explicit 2-site model based on \cite{DeCurtis:2011yx} originally developed in the context of minimal CHMs governed by the $\rm SO(5)$ symmetry and here extended to $\rm SO(6)$. 
These models are composed of two sectors, i.e., an elementary one including 
particles whose quantum numbers under the $\rm SU(2)_L\times U(1)_Y$ gauge symmetry are the same as those of the SM fermions and gauge bosons plus  
a composite sector having new spin-1 and spin-1/2 resonances introduced as multiplets of the global group.
The mixing between states in these two sectors realise the partial compositeness.  
With this construction, we can evaluate observables in the Higgs sector such as masses and couplings. 
This analysis also allows to clarify the differences between these observables and those from renormalisable E2HDMs \cite{DeCurtis:2018iqd}.
The aim of the paper is to show that a composite scenario could give rise to a concrete realisation of a 2HDM and also to highlight the phenomenological aspects which could reveal at the Large Hadron Collider (LHC) the composite nature of the Higgs states described by our construction.

The plan is as follows. In Sec.~\ref{sec:gauge} we describe the general features of a C2HDM based on ${\rm SO}(6)\to {\rm SO}(4)\times {\rm SO}(2)$ and we discuss the corresponding symmetries. In Sec.~\ref{sec:model} 
we present the explicit model on which our analysis is based. In Sec.~\ref{sec:CW} we compute explicitly the Higgs potential and we discuss in Sec.~\ref{sec:higgscouplings} the Higgs boson couplings to fermions and bosons as well as amongst themselves. 
In Sec.~\ref{sec:pheno} we present the Higgs spectrum of the model and discuss some phenomenological results which may act as smoking gun signals of the C2HDM. In addition, we comment on the implications from flavour constraints.
Finally, we conclude in Sec.~\ref{sec:conclusions}. We leave to the Appendices the connection with the CCWZ construction and other technical details.

\section{The {\rm SO}(6)$\to${\rm SO}(4)$\times${\rm SO}(2) symmetry breaking}
\label{sec:gauge}
In this section we discuss the main aspects of C2HDMs highlighting the general properties that follow by their constructions as effective field theories.
The scenarios we consider are schematically characterised by the following structure:
\be\label{reference}
\mathscr{L_\textrm{Composite}}=\mathscr{L}_{\rm 2HDM} + \mathscr{L}_{d \ge 6},
\ee
where $\mathscr{L}_{\rm 2HDM}$ has the same form as the Lagrangian of the E2HDM and contains the kinetic terms, the scalar potential (up to quartic terms) and Yukawa interactions,
\be
\mathscr{L}_{\rm 2HDM} = {\rm{kinetic~terms}} + V(H_1,H_2) +  \mathscr{L}_{\rm Yukawa}, 
\ee
with $H_1$ and $H_2$ being the isospin scalar doublets and
\begin{align}\label{2HDM-potential}
V(H_1, H_2) & = m_1^2 H_1^\dagger H_1 + m_2^2 H_2^\dagger H_2 
- \left[m_3^2 H_1^\dagger H_2 + \text{h.c.} \right] \notag\\
& +\frac{\lambda_1}{2} (H_1^\dagger H_1)^2 + \frac{\lambda_2}{2} (H_2^\dagger H_2)^2 
+ \lambda_3 (H_1^\dagger H_1)(H_2^\dagger H_2) 
+ \lambda_4 (H_1^\dagger H_2)(H_2^\dagger H_1) \notag\\
& + \frac{\lambda_5}{2} (H_1^\dagger H_2)^2 
+ \lambda_6 (H_1^\dagger H_1)(H_1^\dagger H_2) 
+ \lambda_7(H_2^\dagger H_2)(H_1^\dagger H_2) + \text{h.c.} 
\end{align}
The $ \mathscr{L}_{d \ge 6}$ element  includes effective operators (starting from dimension 6) that can generate modifications to the Higgs couplings to bosons and fermions, hence effects in specific experimental observables in Higgs and  
flavour physics as well as global Electro-Weak (EW) precision tests. In general, these effective operators generate effects that are suppressed by $v^2/f^2$ (with $v$ being an EW scale parameter connected to the Higgs doublet Vacuum Expectation Values (VEVs) and $f$ the scale of compositeness), however, larger suppressions can be achieved by virtue of some approximate symmetries of the underlying composite dynamics. 
We will compute here, through an expansion in $v^2/f^2$, the leading contributions to the 2HDM parameters $m^2_i$ ($i=1, ... 3)$ and $ \lambda_j$ ($j=1, ... 7$) originating from the explicit breaking of the global symmetry. We then obtain the phenomenological observables, such as masses and couplings, that were only estimated in \cite{Mrazek:2011iu} on the basis of symmetry arguments and produce {\it explicitly} the low energy particle spectrum 
of the C2HDM.

In order to be concrete we need to choose a coset space and describe how the global symmetries are explicitly broken by the elementary sector. In the remainder of this work we have as a main focus the model
\be
\frac{\mathcal{G}}{\mathcal{H}} = \frac{\mathrm{{\rm SO}(6)}}{\mathrm{{\rm SO}(4)}\times\mathrm{{\rm SO}(2)}},
\ee
expanding upon the work presented in \cite{Mrazek:2011iu} and recently discussed in \cite{DeCurtis:2018iqd}.
The NGB fluctuations are described by the matrix $U$ in the vector representation of {\rm SO}(6)
\be
U \equiv \exp(i \frac{\Pi}{f}), \quad \Pi=\bigg(\begin{array}{ccc} 0_{4 \times 4} & \phi_1 & \phi_2 \\ -\phi_1^T  & 0 &0 \\ -\phi_2^T  & 0 & 0 \end{array} \bigg),
\label{Pi}
\ee
where $\phi_{1,2}$ are two real fourplets (the two Higgs doublets)
\be\label{definition-h1-h2}
\phi_1^T=(\vec{\pi}_1,h_1), \quad \phi_2^T=(\vec{\pi}_2,h_2),  
\ee 
that can be rearranged into two $SU(2)$ doublets as
\be
H_\alpha = \frac{1}{\sqrt{2}} \left( \begin{array}{c} \phi_\alpha^2 + i \phi_\alpha^1 \\ \phi_\alpha^4 - i \phi_\alpha^3 \end{array} \right),  \quad  \alpha=1,2.
\ee

Besides the NGBs and the elementary SM fields, the model describes the extra spin-1 and spin-1/2 resonances. 
While the representations of the spin-1 states are fixed by the gauge symmetry, the model allows for some freedom in the choice of the fermion ones and in the embedding of the elementary fermions in representations of $\mathcal{G}$. We choose:
\begin{itemize}
\item chiral elementary fermions in the \textbf{6} of {\rm SO}(6),
\item vector-like composite fermions in the fourplet $\psi^I$ and doublet $\psi^\alpha$ of {\rm SO}(4)$\times${\rm SO}(2). 
\end{itemize}

The resonances and their interactions with the elementary sector are then fully described by the following Lagrangian
\be\label{C2HDM}
\mathscr{L_{\rm C2HDM}} = \mathscr{L}_{\rm elementary} + \mathscr{L}_{\rm mixing} + \mathscr{L}_{\rm resonances}.
\ee
The last part  describes also the derivative interactions of the NGBs with the composite matter fields and can be parameterised by means of the CCWZ formalism as shown in Appendix \ref{sec:ccwz}. The interactions of the NGBs are suppressed by $1/f$ and are $\mathcal{H}$-symmetric while the resonances have an overall mass scale of size $m_*$. The elementary sector contains the gauge and fermionic kinetic terms for the SM-like fields while the mixing term is the crucial ingredient since all the phenomenology strongly depends upon the interactions between the elementary and composite sectors. 

Concretely, the mixing Lagrangian contains the partial compositeness terms that generate masses for the SM fermions,
\be
\mathscr{L}_{\rm mixing} = y_L^{ij} f \bar{q}^i_L U \cdot (\psi^I)^j +  \tilde y_L^{ij} f \bar{q}^i_L\, U \cdot (\psi^\alpha)^j + (q_L \to u_R,d_R,l_L,e_R),
\label{Lmixing}
\ee
where $i,j$ are flavour indices that run over the three families and we wrote schematically with a dot all the possible invariants that can be formed (see the next section for the actual implementation).
Then, 
upon integrating out the resonances of the composite sector, we generate several effects that would allow us to match $\mathscr{L}_{\rm C2HDM}$ in Eq.~\eqref{C2HDM} to the reference Lagrangian $\mathscr{L_\textrm{Composite}}$ of Eq.~\eqref{reference}. 

In the spirit of CHMs with partial compositeness, the parameters that enter the two sectors in $\mathscr{L_\textrm{Composite}}$, i.e., the usual part $\mathscr{L}_{\rm 2HDM}$ and the new one $\mathscr{L}_{d \ge 6}$, are related to each other, since all the Higgs interactions and effective operators with SM fields are mainly generated by the explicit breaking of the global symmetry under which the Higgs doublets then behave as pNGBs. In order to set the stage for the discussion of our C2HDM we now quickly recall the main aspects of CHMs with partial compositeness.

\subsection{Custodial and discrete symmetries}
A renormalisable 2HDM never faces custodial breaking effects at tree level. This can be traced back to the presence, when the hypercharge coupling is neglected, of a large {\rm SU}(2)$_L\times$Sp(4) symmetry in the kinetic terms of the two Higgs doublets. Since in the renormalisable E2HDM there are no terms in the Lagrangian that contribute to the $\hat T$ parameter other than the kinetic terms, no custodial violation is present for any number of Higgs doublets.

In CHMs, the non-linearities of the effective Lagrangian for NGBs contribute with operators of dimension 6 of the following form
\be\label{cust}
 \mathscr{L}_{d \ge 6}\supset \frac{ c_{ij} \tilde c_{kl} }{f^2} (H^\dag_i \overleftrightarrow{D}_\mu H_j) (H^\dag_k \overleftrightarrow{D}_\mu H_l) + {\rm h.c.}, 
\ee
which do not respect the Sp(4) symmetry and contribute to the $\hat T$ parameter for generic VEVs of the two Higgs doublets. However, the value of the coefficients $c,\tilde c$'s is constrained by the symmetry of the subgroup $\mathcal{H}$. This in turn suggests that only models where the unbroken group contains $\Hg\supset$ {\rm SU}(2)$_L\times$Sp(4) are free from tree level violation of custodial symmetry for any form of the Higgs VEVs.
This is not the case for {\rm SO}(4)$\times${\rm SO}(2), which does not contain the full symmetry of the renormalisable kinetic terms, therefore, in our case the coefficients in Eq.~(\ref{cust}) are non-vanishing and fixed by the symmetries,  which then predict a $\hat T$ parameter \cite{Mrazek:2011iu} such that
\be\label{T}
\hat T \propto 16\times \frac{v^2}{f^2} \times \frac{\mathrm{Im}[\langle H_1\rangle ^\dag \langle H_2\rangle]^2}{(|\langle H_1\rangle|^2+|\langle H_2\rangle|^2)^2}.
\ee
Since custodial breaking is sensitive to the combination $\mathrm{Im}[\langle H_1\rangle ^\dag \langle H_2\rangle]$ there are two approximate symmetries, discussed in the following, that can be used to reduce these effects: $i)$ CP, which is well approximated in the SM; $ii)$ a new symmetry, $C_2$, that forbids a VEV for one of the two Higgs doublets. 

\subsection*{CP invariance}
In this case we realise a scenario where the two Higgs doublets have VEVs aligned in phase as required by the vanishing of the contribution in Eq.~\eqref{T}.
Without a very accurate alignment, the bound coming from precision tests can be roughly estimated as $\delta \hat T <10^{-3}$, which then constrains the phase misalignment $\Delta \phi =\phi_1-\phi_2$, defined through $\langle H_{1,2}\rangle^T=1/\sqrt{2}(0,v_{1,2})\exp(i\phi_{1,2})$, to be
\be
\Delta \phi \lesssim 0.03\, \big(\frac{f}{600\ \mathrm{GeV}}\big),
\ee
assuming $\tan\beta=v_2/v_1 \sim O(1)$. Such a value can be achieved by assuming an approximate CP symmetry in the scalar potential. Interestingly, the interactions of the NGBs among themselves and with other composite fields respect automatically charge conjugation C since $H_{i}\to H_i^*$ is realised on the real degrees of freedom $\phi_{1,2}$ encoded in the matrix $U$ as
\be
C=\mathrm{diag}[1,-1,1,-1,1,1],
\ee
which is  an element of {\rm SO}(4). Because of this argument, we find it rather natural to consider the scenario where CP is a good symmetry of the composite sector and very well approximated in the elementary couplings (needed to comply with flavour constraints). 
\subsection*{$C_2$ invariance}
Another possibility to control the deviations in Eq. \eqref{T}, as extensively discussed in \cite{Mrazek:2011iu}, is to make the stronger assumption that one of the two Higgs doublets has a discrete symmetry that forbids any VEV, e.g., 
\be\label{C2}
C_2: \quad H_1 \to H_1,\quad H_2 \to - H_2,
\ee
which, contrary to C, is not a symmetry of the composite sector in the sense that it is not an element of $\mathcal H$\footnote{Notice that, limitedly to the Higgs sector, the $C_2$ symmetry of the C2HDM coincides with the $Z_2$ one of the E2HDM.}.
Although this condition is not what is strictly required from Eq.~\eqref{T}, it has an interesting byproduct since it selects a specific pattern of Higgs couplings to fermions, because only one Higgs doublet is coupled to them. As well known, in this case,  any tree level mediations of Flavour Changing Neutral Currents (FCNCs) from the scalar sector are absent for a generic flavour structure of the Yukawa couplings.

\subsection{Flavour structure}
When CP is the only discrete symmetry acting on the Higgs doublets, the Yukawa couplings of the renormalisable 2HDM are of the following form:
\be\label{yukawa-aligned}
\mathscr{L}_{\rm 2HDM} \supset  Y_{u}^{ij} \bar q_L^i  \big( a_{1u} \tilde H_1 + a_{2u} \tilde H_2) u_R^j+ Y_{d}^{ij} \bar q_L^i  \big( a_{1d} H_1 + a_{2d} H_2) d_R^j + Y_{e}^{ij} \bar l_L^i  \big( a_{1e} H_1 + a_{2e} H_2) e_R^j+ {\rm h.c.}
\ee
Therefore, only under the assumption that the coefficients $a$'s are the identity in flavour space, the above interactions do not generate Higgs-mediated FCNCs at tree level. Under this assumption, FCNCs are therefore confined to loop effects as in the SM.

In a C2HDM the above description is modified by the presence of higher dimension operators that contribute to the Yukawas of Eq. \eqref{yukawa-aligned} and in general one would expect any kind of operator of the form $\kappa_{ijk}\psi \psi H_i H^\dag_j H_k$ + h.c. However, thanks to the pNGB nature of the Higgs doublets, the structure of the higher-dimensional operators is highly constrained by the symmetry of the theory and in general \cite{Agashe:2009di} the Yukawa terms including all the non-linearities are simply
\be\label{full-yuk}
  Y_{u}^{ij} \bar q_L^i  \big( a_{1u} F^u_{1}[H_i] + a_{2u} F^u_2[H_i]) u_R^j + ...,
\ee
where the functions $F^u_{1,2}$ are trigonometric invariants of $H_i$ and start with a linear term in $H_{1,2}$, respectively. Therefore, the elementary case of Eq.~\eqref{yukawa-aligned} is automatically included as a specific case in Eq.~\eqref{full-yuk} and this shows that the assumption of aligned Yukawa couplings is not a stronger requirement in the composite scenario than in the elementary one.

A difference between the elementary and composite cases arises though when one considers the additional constraints on the theory induced by the alignment of Eq.~\eqref{full-yuk}. In other words, while in the elementary case 
Eq.~\eqref{yukawa-aligned} is the only possible source of flavour violation, in the composite one $\mathscr{L}_{\rm Composite}$ contains four-fermion operators generated by  integrating out the composite fermions and vectors of the form
\be\label{4fermi}
\frac{x_{ijkl}}{f^2} \psi_i \psi_j \psi_k \psi_l,
\ee
with $\psi$ being a SM fermion, which can mediate FCNCs at tree level if the flavour coefficients $x_{ijkl}$ are generic, and where we neglected the precise chirality structure of fermions. This shows that the aligned structure of Eq.~\eqref{full-yuk} is not sufficient to avoid tree level effects, since a diagonalisation of the Yukawas may still leave the set of four-fermion operators misaligned with respect to the mass basis. Notice also that, despite the four-fermion operators of Eq.~\eqref{4fermi} originate from different effects than (pseudo)scalar-mediated FCNCs, they are not less harmful and, more importantly, they are present also in the $C_2$ symmetric case.  

A mechanism to generate an approximate alignment between the flavour structures of Eqs \eqref{full-yuk} and \eqref{4fermi} is possible under the working assumption of a flavour universal composite sector. In this case the latter  enjoys a symmetry $\mathcal{G}_F^5$ that commutes with $\mathcal{H}$ while the elementary sector has the usual ${\rm U}(3)^5$ flavour symmetry.

Only by the explicit breaking of $\mathcal{G}_F^5$ and ${\rm U}(3)^5$ down to baryon number the
CKM structure of the Yukawa sector can be reproduced: in absence of this breaking the SM fermions are in fact  massless. The couplings $y, \tilde y$ in Eq.~\eqref{Lmixing} break explicitly this symmetry and, depending on their structure in flavour space, they may lead to misaligned (with respect to the mass basis) flavour interactions. 

Several possibilities have been considered in the literature to prevent large tree level flavour violations in composite models \cite{Redi:2011zi,Barbieri:2012tu}.
In particular,  it is worth mentioning the following. 
\begin{enumerate}
\item Higgs-mediated tree level FCNCs are absent when there is only one flavour structure per SM representation.  This means that the invariants in $\mathscr{L}_{\rm mixing}$ given in Eq.~\eqref{Lmixing} need to be aligned in flavour space (e.g., $y_L^{ij} \propto \tilde y_L^{ij}$). For any form of $y^{ij}$ in flavour space then Higgs-mediated FCNCs are zero at tree level and appear only at loop order.
\item Tree level effects in the four-fermion operators of Eq.~\eqref{4fermi} are much suppressed when, in addition to the assumption in point 1, one also realises a partial alignment of $y^{ij}$ with the CKM matrix.
\end{enumerate}
We will work under these assumptions in order to realise a flavour symmetric composite sector. 

\section{An explicit model}\label{sec:model}
When parameterising a composite sector one is faced with a few practical approximations that are needed to capture its main features. Since we would like to focus on the connection between Higgs phenomenology and the spectrum of heavier resonances, we adopt a description of the composite sector based on a 2-site model, as a generalisation of \cite {DeCurtis:2011yx}, as already intimated.

We consider here a simplified picture that includes the minimal amount of new resonances that allow for a calculable Higgs potential \cite{DeCurtis:2011yx}. Here, we focus on the gauge sector.
Despite we are interested in the full coset structure
\be
\frac{\mathcal{G}}{\mathcal{H}}=\frac{{\rm SU}(3)_c \times {\rm SO}(6)\times {\rm U}(1)_X}{{\rm SU}(3)_c \times {\rm SO}(4)\times {\rm SO}(2)\times {\rm U}(1)_X},
\ee
the consistent inclusion of spin-1 resonances requires additional (gauged) symmetries as typical of 2 and 3-site models \cite{Panico:2011pw}. In principle, we should expect any type of resonances classified accordingly to the unbroken group $\mathcal{H}$, however, since we are mainly interested in deriving the Higgs potential, only spin-1 resonances associated to 
${\rm SO}(4)\times {\rm SO}(2)$ would play a major role\footnote{Coloured spin-1 resonances affect the Higgs potential only at two loop order, while resonances associated to {\rm U}(1)$_X$ would give a subdominant contribution to the Higgs potential. }. 

\begin{figure}
\begin{center}
\includegraphics[width=0.50\textwidth]{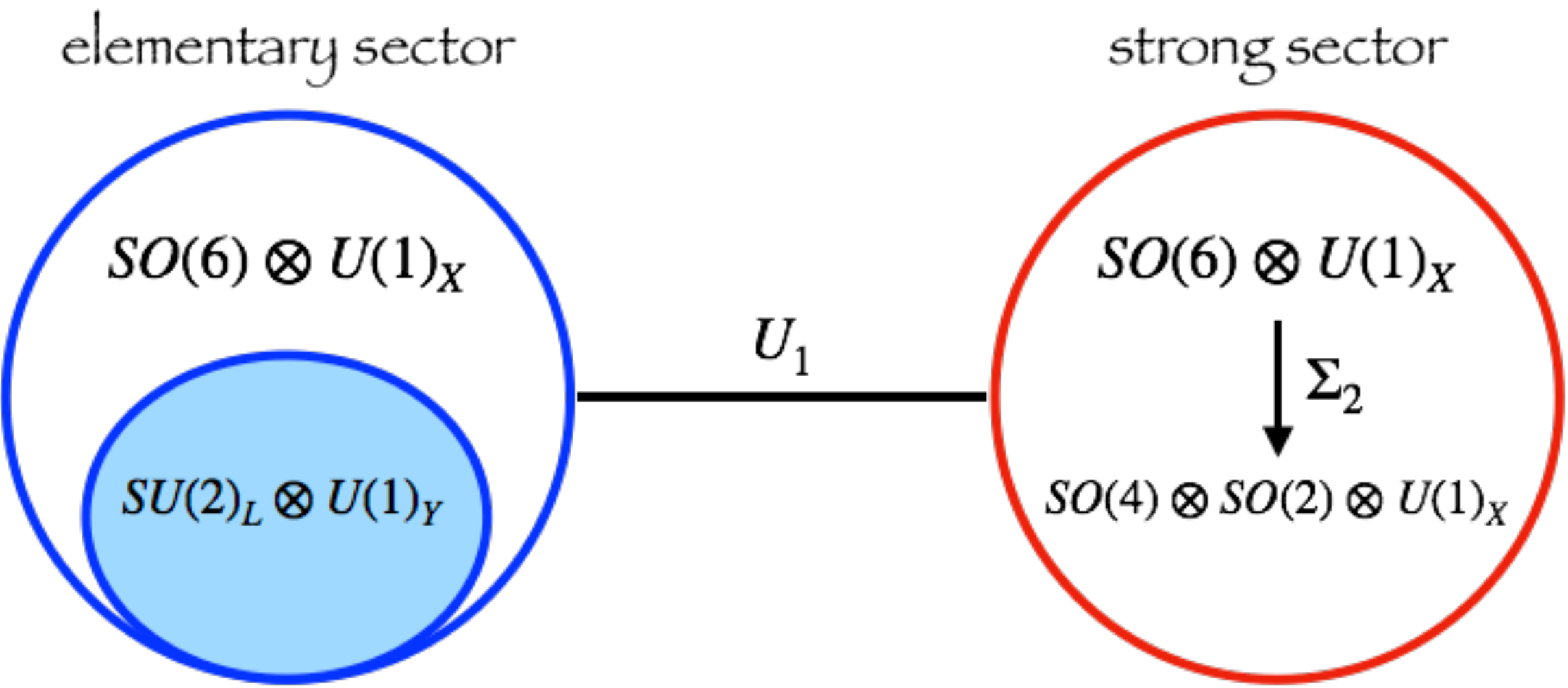}
\caption{\label{fig:2site} The 2-site construction in the gauge sector based on the ${\rm SO}(6)/{\rm SO}(4) \times {\rm SO}(2)$ coset. The first site is the elementary sector while the second one is the composite sector with $\rm SO(6) \times U(1)_X$ heavy resonances.}
\end{center}
\end{figure}

In the 2-site scenario, the Lagrangian of the gauge sector can be written as
\begin{align}
{\cal L}^{\text{gauge}}_{\rm C2HDM} &= 
\frac{f_1^2}{4}\text{Tr}|D_\mu U_1|^2 +\frac{f_2^2}{4}\text{Tr}|D_\mu \Sigma_2|^2 - \frac{1}{4g_\rho^2}(\rho^A)_{\mu\nu}(\rho^A)^{\mu\nu}  - \frac{1}{4g_{\rho_X}^2}(\rho^X)_{\mu\nu}(\rho^X)^{\mu\nu}   \nonumber \\
&- \frac{1}{4g_A^2}(A^A)_{\mu\nu}(A^A)^{\mu\nu}  -  \frac{1}{4g_X^2} X_{\mu\nu} X^{\mu\nu},
\label{4dlag}
\end{align}
where two copies, $G_{1,2}$, of the symmetry group $G = {\rm SO}(6) \times {\rm U}(1)_X$ characterise the two sites. Here, 
$G_2$ is a local group and describes the spin-1 resonances through  the gauge fields $\rho^X_\mu$ and $\rho^A_\mu$, with $A \in \textrm{Adj}({\rm SO}(6))$. 
Further, $G_1$ is global with only the ${\rm SU}(2)_L \times {\rm U}(1)_Y$ component lifted to a local subgroup. 
The corresponding elementary SM gauge fields are conveniently embedded into spurions, $A^A_\mu$ and $X_\mu$, of the adjoint of $G_1$, and $g_A$, $g_X$ are the corresponding gauge couplings.
The $U_1$ field in Eq.~(\ref{4dlag}) is commonly dubbed {\it link} due to its transformation properties under both symmetry groups of the two sites it connects, namely, $U_1 \rightarrow g_1 U_1 g_2^\dag$.
Notice that $g_1$ is an element of the global $G_1$ group in the first site while $g_2$ belongs to the local $G_2$ in the second  site. 
By virtue of the EW gauging, the ${\rm SU}(2)_L \times {\rm U}(1)_Y$ component of the global symmetry on the first site is promoted to a local one. 
The $\Sigma_2$ field defined on the second site transforms, instead, under the local group $G_2$. 
Having specified the transformation properties of the fields introduced in Eq.~(\ref{4dlag}), the covariant derivatives are easily worked out to be  
\begin{align}
D_\mu U_1 & = \partial_\mu U_1 -iA_\mu U_1  +i U_1 \rho_\mu, \nonumber \\
D_\mu \Sigma_2 &= \partial_\mu \Sigma_2 - i[\rho_\mu, \Sigma_2] ,
\end{align}
where $A_\mu \equiv A_\mu^A T^A + X_\mu T^X$ and $\rho_\mu \equiv \rho_\mu^A T^A + \rho^X_\mu T^X$, with $T^A$ and $T^X$ being the generators of ${\rm SO}(6)$ and ${\rm U}(1)_X$, respectively.
The link field $U_1$ realises the spontaneous symmetry breaking of  $G_1 \times G_2$ to the diagonal component $G$ while the VEV of $\Sigma_2$ accounts for the breaking to ${\rm SO}(4) \times {\rm SO}(2) \times {\rm U}(1)_X$. The 2-site construction is schematically depicted in Fig.~\ref{fig:2site}.
This breaking pattern provides 24 NGBs, 16 of which are reabsorbed in the longitudinal components of the gauge fields, while the remaining 8 can be identified with Higgs fields.
In the unitary gauge, where the physical degrees of freedom are clearly evident, the $U_1$ and $\Sigma_2$ are given by $U_i = \exp {i\frac{f}{f_i^2}\Pi}$ and $\Sigma_2 = U_2 \Sigma_0 U_2^T$ 
with $\Pi$ being the usual NGB matrix given in Eq.~\eqref{Pi} , $\Sigma_0 = i \sqrt{2} T_S$ where $T_S= -{i}/{\sqrt{2}}(\delta_I^5\delta_J^6-\delta_J^5\delta_I^6)$ with $I,J=1,.,6$ and $f^{-2} = f_1^{-2} + f_2^{-2}$, following from the canonical normalisation of the Higgs kinetic term.  

Before  EW gauging, the model possesses an unbroken ${\rm SO}(4) \times {\rm SO}(2) \times {\rm U}(1)_X$ symmetry and the masses of the spin-1 resonances are described by the following relations:
\begin{align}
\label{eq:rhomass}
& m_\rho^2 = \frac{g_\rho^2 f_1^2}{2} \,, \qquad m_{\rho_X}^2 = \frac{g_{\rho_X}^2 f_1^2}{2} \,, \qquad  m_{\hat \rho}^2 = \frac{g_\rho^2 (f_1^2+f_2^2)}{2},
\end{align}
where the first two characterise the resonances spanning the unbroken group while the last one is related to the broken sector. 
The gauging of the EW subgroup explicitly breaks ${\rm SO}(4) \times {\rm SO}(2) \times {\rm U}(1)_X$ and induces a mixing between the elementary and composite fields as well as 
corrections to the masses defined above.

Upon integrating out the spin-1 resonances, we obtain the following effective Lagrangian in momentum space  up to quadratic terms:
\begin{align}
{\mathcal L}_{\text{Composite}}^{\text{gauge}}  
= -\frac{(P_T)^{\mu\nu}}{2}&\Big[ q^2\tilde{\Pi}_0(q^2) A^A_\mu A^A_\nu + q^2\tilde{\Pi}_X(q^2)X_\mu X_\nu \notag\\
& + f^2\tilde{\Pi}_1(q^2) A^A_\mu A^B_\nu \text{Tr} (\Sigma T^A T^B \Sigma) + f^2\tilde{\Pi}_2(q^2) A^A_\mu A^B_\nu \text{Tr}(T^A \Sigma T^B \Sigma) \Big],\label{eff6} 
\end{align}  
where $\Sigma = U_1 \Sigma_2 U_1^T$ and $P^T_{\mu\nu}$ is the projection operator $P^T_{\mu\nu} = \eta_{\mu\nu} - q_\mu q_\nu / q^2$. 
The form factors $\tilde \Pi$ are determined by the parameters of the strong sector, namely, by the masses and couplings of the resonances, 
with  their explicit expressions  given in Appendix \ref{sec:formfactors}. 
At $q^2 = 0$, the $\tilde{\Pi}_1$ form factor can be fixed in terms of the SM gauge masses. 
Indeed, keeping only the CP-even components of the $\Sigma$ matrix and removing the non-dynamical fields, the $W$ and $Z$ gauge boson masses are given by
\begin{align}
\label{eq:bosonmasses}
m_W^2 = - \frac{\Pi_W(0)}{4}f^2\sin^2 \frac{v}{f},\quad 
m_Z^2 = - \frac{\Pi_W(0)}{4}f^2\sin^2 \frac{v}{f}(1+ \tan^2 \theta_W ),
\end{align}
where $\theta_W$ is the Weinberg angle and we recall that  $v^2 = v_1^2 + v_2^2$, with $v_{1,2}$ the VEVs of the two CP-even Higgs boson components.
The form factor $\Pi_W$ is normalised in order to correctly reproduce the canonical kinetic term of the SM gauge fields.
From Eq.~(\ref{eq:bosonmasses}) we can finally identify 
\begin{align}
v_{\text{SM}}^2 = f^2\sin^2 \frac{v}{f},  \qquad  g^2 = - \Pi_W(0),
\end{align}
where $v_{\text{SM}} \simeq 246$ GeV is the SM VEV and $g$ is the ${\rm SU}(2)_L$ gauge coupling. As usual,  the corrections, with respect to the SM, will be parameterised by $\xi=v_{\text{SM}}^2/f^2$.

Differently from the gauge sector, which is model independent and fixed only by the symmetry group of the strong dynamics, 
the fermion sector is not uniquely determined due to the possibility to choose different group representations for the fermionic fields. 
In this work we opt for the simplest one in which the SM fermions are embedded into the fundamental of ${\rm SO}(6)$. 
Another scenario, for instance, envisages the {\bf 20} representation \cite{Mrazek:2011iu}. 

\begin{figure}
\begin{center}
\includegraphics[width=0.3\textwidth]{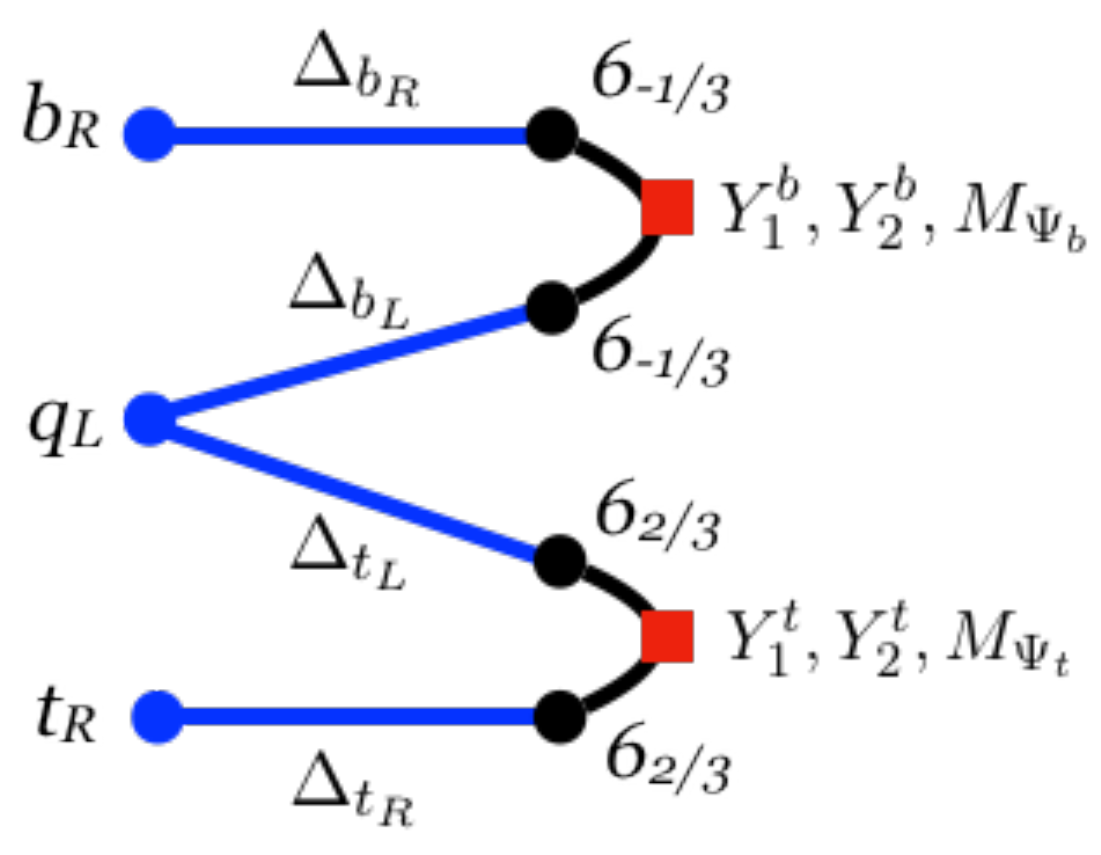}
\caption{\label{fig:fermions} The 2-site construction in the fermion sector with the Left-Right (LR) structure based on the ${\rm SO}(6)/{\rm SO}(4) \times {\rm SO}(2)$ coset. }
\end{center}
\end{figure}

For the sake of simplicity, we consider only the third generation and focus ourself on the top quark contributions. 
Indeed, all the other SM quarks provide only sub-leading corrections to the Higgs effective potential.
Needless to say, all the other fermions can be included, if necessary, by simply extending the formalism described below. 

In order to construct a ${\rm SO}(6) \times {\rm U}(1)_X$ invariant Lagrangian, it is useful to embed the top quark using the spurion method into a complete representation of ${\rm SO}(6)$ with $X = 2/3$. 
More precisely, the $\bf 6$ of ${\rm SO}(6)$ decomposes into the $(\bf 4,\bf 1) \oplus (\bf 1, \bf 2)$ of ${\rm SO}(4) \times {\rm SO}(2)$. 
The L-hand top quark doublet $q_L$ has a unique embedding into the $({\bf 4},{\bf 1})_{{ 2/3}}$  
while for the R-handed component of the top quark $t_R$, described by the $({\bf 1}, {\bf 2})_{{2/3}}$, an extra angle $\theta_t$ parameterises the ambiguity of the embedding in the $\bf 6$, 
since the fundamental representation contains two ${\rm SU}(2)_L$ singlets. 
The R-handed component of the bottom quark $b_R$ is coupled to the $({\bf 1, \bf 2})_{{-1/3}}$ of ${\bf 6}_{{-1/3}}$ and, due to  ${\rm U}(1)_X$ invariance, 
a second embedding for $q_L$ (in another ${\bf 6}_{{-1/3}}$) is needed in order to generate the bottom mass. 
The embedding of the $\tau$ lepton follows the same line of reasoning of the bottom quark with $X = -1$.
In particular, the L-hand (doublet) and  R-handed (singlet) components are promoted to spurions
\begin{align}
&(q_{L}^{\bm{6}})_t^A = q_L^\alpha (\Upsilon^t_{L})^{\alpha A}, \qquad  (t_R^{\bm{6}})^A = t_R^{} (\Upsilon^t_R)^A,  \nonumber \\
&(q_{L}^{\bm{6}})_b^A = q_L^\alpha (\Upsilon^b_{L})^{\alpha A}, \qquad  (b_R^{\bm{6}})^A = b_R^{} (\Upsilon^b_R)^A,  \nonumber \\
&(l_{L}^{\bm{6}})_\tau^A = l_L^\alpha (\Upsilon^\tau_{L})^{\alpha A}, \qquad  (\tau_R^{\bm{6}})^A = \tau_R^{} (\Upsilon^\tau_R)^A, 
\end{align}
with $\alpha$ being the $SU(2)_L$ index and the spurion VEVs defined as
\begin{align}
\langle \Upsilon^t_{L} \rangle= \frac{1}{\sqrt{2}} \left( \begin{matrix}
0 & 0 & 1 & i & 0 &0 \\
1 & -i & 0 & 0 & 0 & 0
\end{matrix} \right) ,\qquad
\langle\Upsilon^t_R \rangle=(0,0,0,0,\cos\theta_t, i\sin\theta_t) \,, \nonumber \\
\langle \Upsilon^b_{L} \rangle= \frac{1}{\sqrt{2}} \left( \begin{matrix}
1 & i & 0 & 0 & 0 &0 \\
0 & 0 & -1 & i & 0 & 0
\end{matrix} \right) ,\qquad
\langle\Upsilon^b_R \rangle=(0,0,0,0,\cos\theta_b, i\sin\theta_b) 
\label{spurion61}
\end{align}
and  $\langle \Upsilon^\tau_{L,R} \rangle = \langle \Upsilon^b_{L,R} \rangle$ with $b \rightarrow \tau$. 

The spin-$1/2$ resonances of the top quark are described by ${\bm 6}$-plets $\Psi$ with $X=2/3$. 
As $\bold 6_{{2/3} } = (\bold 4, \bold 1)_{{2/3}} \oplus (\bold 1, \bold 2)_{{2/3}}$, the former delivers two ${\rm SU}(2)_L$ doublets with hypercharge $7/6$ and $1/6$, respectively, while the latter delivers two ${\rm SU}(2)_L$ singlets with hypercharge $2/3$. 
After EWSB we count four top partners with electric charge $Q=2/3$, one bottom partner with $Q=-1/3$ and one exotic fermion with $Q=5/3$.
For the down type quarks and charged leptons, we can analogously introduce other spin-1/2 resonances with $X=-1/3$ and $-1$, respectively. 
The Lagrangian of the composite fermion sector is
\begin{align}
{\cal L}_{\text{C2HDM}} ^ {\text{fermion}}& = 
(\bar q_{L}^{\bm{6}})_t  iD\hspace{-2.7mm}/\hspace{1mm} ( q_{L}^{\bm{6}})_t   + (\bar t_{R}^{\bm{6}})  iD\hspace{-2.7mm}/\hspace{1mm} ( t_{R}^{\bm{6}})  +    \bar{\Psi}_t^I iD\hspace{-2.7mm}/\hspace{1mm} \Psi_t^I  
 - \bar{\Psi}_{t}^I (M_{\Psi_t})_{IJ} P_R \Psi_t^J  - \bar{\Psi}_t^I [ (Y_1^t)_{IJ} \Sigma_2 + (Y_2^t)_{IJ} \Sigma_2^2] P_R \Psi_t^J    \nonumber \\
&  +  (\Delta^t_L)_{I}(\bar{q}_L^{\bm{6}})_t U_1 P_R \Psi_t^I +  (\Delta^t_R)_{I}(\bar{t}_R^{\bm{6}}) U_1 P_L \Psi_t^I     + (t \rightarrow b, \tau) + \text{h.c.} \,,
\label{eq:lag_ferm0}
\end{align}
where the covariant derivatives of the elementary fermions include the interactions with the elementary gauge bosons while the covariant derivative of the resonance $\Psi$ provides the couplings to the spin-1 resonances introduced above. 
In the following, we will restrict ourselves to a realisation with $I=1,2$ fermionic resonances. 
The dimensionful parameters $\Delta_L$ and $\Delta_R$ induce a mixing between the elementary and composite fermions and, as such, explicitly break the ${\rm SO}(4) \times {\rm SO}(2) \times {\rm U}(1)_X$ symmetry.
All the parameters in the above Lagrangian are taken to be real in order to realise a CP invariant scenario.
Moreover, we assume the LR structure of the fermionic Lagrangian discussed in \cite{DeCurtis:2011yx} which allows us to simplify the parameterisation of the spin-$1/2$ resonances and to reduce the number of free couplings. This construction is sketched in Fig.~\ref{fig:fermions}. This represents the minimal choice able to generate the SM Yukawa interactions and to guarantee the Ultra-Violet (UV) finiteness of the CW potential. Notice that 
the LR assumption requires $\Delta_L^2 = \Delta_R^1 = M_\Psi^{21} = Y_1^{11} = Y_1^{22} = Y_2^{11} = Y_2^{22}  = Y_1^{21} = Y_2^{21} = 0$.

Upon integrating out the heavy resonances $\Psi^I$, the effective Lagrangian takes the form
\begin{align}
{\cal L}_{\text{Composite}} ^ {\text{fermion}}&= 
(\bar{q}_{L}^{\bm{6}})_t \,\, q \hspace{-2mm}/[\tilde{\Pi}_0^{q_t}(q^2) + \tilde{\Pi}_1^{q_t} (q^2)\Sigma + \tilde{\Pi}_2^{q_t} (q^2)\Sigma^2]  (q_{L}^{\bm{6}})_t \notag\\
&+ \bar{t}_R^{\bm{6}} \, q\hspace{-2mm}/[\tilde{\Pi}_0^t(q^2) + \tilde{\Pi}_1^t(q^2)\Sigma + \tilde{\Pi}_2^t(q^2)\Sigma^2] t_R^{\bm{6}} \notag\\
&+ (\bar{q}_{L}^{\bm{6}})_t  [\tilde{M}_0^t(q^2) + \tilde{M}_1^t(q^2)\Sigma  + \tilde{M}_2^t(q^2)\Sigma^2] t_R^{\bm{6}} + (t \rightarrow b, \tau)  + \text{h.c.} \,,
\label{fermion6}
\end{align}
where the explicit expression of the form factors is given in Appendix \ref{sec:formfactors}.
Due to $\Sigma^\dag = - \Sigma$, with $\Sigma = U_1 \Sigma_2 U_1^T$, the hermeticity of the Lagrangian implies that the form factors $\tilde{\Pi}_1^{q_t}$ and $\tilde{\Pi}_1^t$ are purely imaginary while  $\tilde{\Pi}_0^{{q_t},t}$ and $\tilde{\Pi}_2^{{q_t},t}$ are real. The form factors $\tilde M_1^t$ and $\tilde M_2^t$ can be complex but we will restrict them to real values, as a consequence of the reality of the strong sector parameters. 

It is important to notice that the above Lagrangian has non-canonically normalised fields, therefore, when computing a given quantity, one has to take this fact into account. The form factors in Eq.~(\ref{fermion6}) are listed in Appendix \ref{sec:formfactors}.

The mass spectrum of the top-partners can be extracted from the poles and zeros of the form factors of Eq.~(\ref{fermion6}). For instance, before EWSB the spectrum of the heavy top-partner resonances is given by
\begin{itemize}
\item two $\bold 2_{ 7/6}$ with masses $m_4, \tilde m_4$ from the poles of $\tilde{\Pi}_0^{q_t}$,
\item two $\bold 2_{ 1/6}$ with masses $m_Q, \tilde m_Q$ from the zeros of $\tilde{\Pi}_0^{q_t}$,
\item two $\bold 1_{ 2/3}$ with masses $m_T, \tilde m_T$ from the zeros of $\tilde{\Pi}_0^{t}$,
\item two $\bold 1_{ 2/3}$ with masses $m_1, \tilde m_1$ from the poles of $\tilde{M}_1^t$,
\end{itemize}
where, e.g., $\bold 2_{ 7/6}$ denotes a $SU(2)_L$ doublet with hypercharge $7/6$. The masses listed above are functions of the fundamental parameters and, in particular, only $m_Q, \tilde m_Q$ and $m_T, \tilde m_T$ get corrections from the elementary/composite mixings $\Delta_{L,R}$.

\section{The Higgs potential}\label{sec:CW}

As already pointed out, the elementary sector is defined by the SM fermions and the gauge fields that linearly couple to operators of the strong sector and explicitly break its symmetry. 
As a result, the NGB symmetry becomes only approximate and the scalar potential for the Higgs is radiatively generated together with the SM gauge boson and fermion masses. 
Once the symmetries of the strong sector are fixed and the representations of the fermion fields are chosen, the computation of the effective potential can be carried out without a complete knowledge of all the details of the strong UV dynamics. Indeed, it can be entirely expressed in terms of the form factors introduced in Eqs.~(\ref{eff6}) and (\ref{fermion6}) which have been obtained 
after the integration of the heavy resonances. The CW effective potential, at one-loop order in perturbation theory and up to the fourth power in $1/f$, is formally written as
\begin{align}
iV_{\text{1-loop}}  &= \frac{1}{f^4}\int\frac{d^4k}{(2\pi)^4}\left[\frac{3}{2}V_G(H_1, H_2)
-2N_cV_F(H_1, H_2)\right]+ {\cal O}\left( \frac{1}{f^{6}} \right), 
\end{align}
where $V_G(H_1, H_2)$ and $V_F(H_1, H_2)$ show the same structure of the Higgs potential in the renormalisable E2HDM. 
The normalisation of the parameters has been fixed in Eq.~(\ref{2HDM-potential}) and they read as 
\begin{align}
m_i^2 &= \frac{-i}{f^2}\int \frac{d^4k}{(2\pi)^4}\left[\frac{3}{2}(m_{i}^G)^2 - 2N_c(m_{i}^F)^2\right], \quad 
\lambda_j = \frac{-i}{f^4}\int \frac{d^4k}{(2\pi)^4}\left[\frac{3}{2}\lambda_{j}^G - 2N_c\lambda_j^F\right], \notag\\
& i=1,\dots,3,~j=1,\dots,7,
\end{align}
where their explicit dependence on the form factors is given below.  
Here we focus only on the leading top quark and gauge contributions as well as on the CP-conserving scenario while we allow, at the same time, for an explicit $C_2$ breaking (differently from \cite{Mrazek:2011iu} where the $C_2$ symmetry is enforced).
The former is realised if the parameters of the strong sector are real and the embedding of the R-handed top is aligned in the $\theta_t = 0$ direction.
In this configuration, all the form factors are real and, in particular, $\tilde \Pi_{1}^{t,q_t} = 0$. 

The contribution of the gauge bosons is given by
\begin{align}
& \frac{(m_1^G)^2}{f^2} =\frac{(m_2^G)^2}{f^2}= \frac{3\overline{\Pi}_W + \overline{\Pi}_B}{2}, \quad  \frac{(m_3^G)^2}{f^2}   =  0, \nonumber \\
& \lambda_1^G =\lambda_2^G = -\frac{2}{3}(3\overline{\Pi}_W+\overline{\Pi}_B) -\frac{1}{4}(\overline{\Pi}_B^2 + 2\overline{\Pi}_W\overline{\Pi}_B + 3\overline{\Pi}_W^2), \quad \nonumber \\
&\lambda_3^G  = -2\overline{\Pi}_W -\frac{1}{4}(\overline{\Pi}_B^2 - 2\overline{\Pi}_W\overline{\Pi}_B + 3\overline{\Pi}_W^2), \nonumber \\
& \lambda_4^G  = -\frac{4}{3}\overline{\Pi}_B  -\overline{\Pi}_W\overline{\Pi}_B, \quad
\lambda_5^G  = \frac{2}{3}\overline{\Pi}_B , \quad
\lambda_6^G  = \lambda_7^G =  0.
\label{eq:parameters_g}
\end{align}
The relation between the form factors appearing in the previous equations and those from Eq. (\ref{eff6}) is worked out in Appendix \ref{sec:formfactors}. 
The fermion contribution to the parameters of the scalar potential is 
\begin{align}
\label{eq:parameters_f}
& \frac{(m_1^F)^2}{f^2} = -\Pi_2^{q_t} + 2\Pi_2^t -\frac{(M_2^t)^2}{k^2}, \quad
\frac{(m_2^F)^2}{f^2}  = -\Pi_2^{q_t} - \frac{(M_1^t)^2}{k^2}, \quad
\frac{(m_3^F)^2}{f^2}  =  \frac{(M_1^tM_2^t)}{k^2}, \nonumber \\
& \lambda_1^F = \frac{4}{3}(\Pi_2^{q_t} -2\Pi_2^t  ) -(\Pi_2^{q_t})^2 - 4(\Pi_2^t)^2 + \frac{16(M_2^t)^2}{3k^2},    \quad
\lambda_2^F = \frac{4}{3}\Pi_2^{q_t}   -(\Pi_2^{q_t})^2  + \frac{4(M_1^t)^2}{3k^2},   \nonumber \\
& \lambda_3^F  =   \frac{2(M_1^t)^2}{k^2}, \quad
 \lambda_4^F  = \frac{1}{3}\left[2\Pi_2^{q_t}  -2\Pi_2^t  -3(\Pi_2^{q_t})^2 \right]  - \frac{2}{3k^2}[(M_1^t)^2-2(M_2^t)^2], \nonumber \\
& \lambda_5^F  = \frac{2}{3}\left(\Pi_2^{q_t}-\Pi_2^t \right) -\frac{2}{3k^2}[(M_1^t)^2 -2(M_2^t)^2], \quad
 \lambda_6^F  = \lambda_7^F =  \frac{5(M_1^tM_2^t )}{3k^2}. 
\end{align}
The quadratic as well as quartic parameters and, therefore, the masses and  couplings of the Higgs bosons are completely predicted by the strong sector.
Notice also that in our construction $\lambda_6 = \lambda_7 = (5/3) m_3^2/f^2$.

\subsection{Structure of the scalar potential}
In order to understand the relevance of the various terms in the potential we can organise our discussion based on the presence of accidental symmetries and on the amount of breaking of the shift symmetry due to the elementary couplings.

\subsubsection*{Gauge contribution}
From the above discussion we notice that the gauge contribution to the Higgs potential respects the SO(2) symmetry that rotates the two scalar four-plets and is both CP and $C_2$ conserving for any value of the parameters.
Moreover, the explicit expressions of $\bar\Pi_{W,B}$ are proportional to the ratio $g^2/g_\rho^2$ and, by construction,  give a positive contribution to the mass terms of the two Higgs doublets, which are proportional to
\be
m_1^2|_{\rm gauge}=m_2^2|_{\rm gauge}\sim \frac{9 g^2}{32\pi^2} m_\rho^2,
\ee
and grows with the mass of the spin-1 resonances.
In isolation, the gauge radiative corrections are not sufficient to break the EW symmetry.

\subsubsection*{Fermionic contribution}

The effect of the fermion sector on the Higgs potential is more complicated due to the fact that the elementary-composite mixings of the L-handed doublet $q_L$ and R-handed $t_R$ top break different symmetries. In order to treat the fermionic contribution on the same footing as the gauge one, it is convenient to introduce the dimensionless couplings $y_{L,R}=\Delta_{L,R}/f$. The presence of a $C_2$ breaking in the composite sector generates a non-vanishing $m_3^2$ and $\lambda_{6,7}$. At the approximation we are working, these contributions are proportional to each other. It is important to stress that the effects of $C_2$ breaking appear at quartic order in the couplings $y_{L,R}$, as the scaling is given by
\be
m_3^2|_{\rm fermion} \sim \frac{N_c }{16\pi^2} y_L^2 y_R^2 f^2, 
\ee
while the contributions to the mass terms $m_{1,2}^2$ start at the quadratic order. Accidentally, we also notice that the fermionic contribution to $\lambda_3$ vanishes when the composite sector displays  a $C_2$ symmetry.

\subsubsection*{Comparison with other studies}
The presence of  the $C_2$ breaking terms represents the main difference with respect to the scenario discussed in \cite{Mrazek:2011iu} which, instead, focused on the CP and $C_2$ invariant configuration.
Incidentally, the parities under CP and $C_2$ of each of the operators that can be generated at one loop, and that have been classified in \cite{Mrazek:2011iu}, are the same, with the only exception of one contribution proportional to $\Delta_L^2 \Delta_R^2 (\Upsilon^{t \, \dag}_{L})^{i \alpha} (\Upsilon^t_{L})^{\alpha j} (\Upsilon^{t\,\dag}_{R})^{m \beta} (\Upsilon^t_{R})^{\beta n} \delta_{ij} \epsilon_{mn}$, where $\delta_{ij}$ and $\epsilon_{mn}$ sum, respectively, over the ${\rm SO}(4)$ and ${\rm SO}(2)$ indices. This operator provides the possibility to construct a CP invariant model that is not $C_2$ symmetric and to eventually realise the scenario that we have considered in this work. 

\subsubsection{Parameters of the model}

Before moving to the characterisation of the scalar spectrum, we stress again that the effective potential may be, in general, UV divergent and that an explicit realisation of the strong sector requires, at least, two heavy fermions. Among the possible structures of the parameters controlling the strong dynamics, the LR one presented in \cite{DeCurtis:2011yx} and discussed above provides the most economical condition for a calculable potential. In contrast, the gauge contributions do not give rise to UV singularities at one-loop level. 
Moreover, for the sake of simplicity, we can also require $f_1 = f_2$ and $g_\rho = g_{\rho_X}$, which gives us eight free input parameters to the Higgs sector of the C2HDM, namely, 
\begin{equation}\label{eq:C2HDM-params}
f, \quad Y_1^{12}, \quad Y_2^{12},\quad  \Delta_L^1, \quad \Delta_R^2, \quad M_\Psi^{11}, \quad M_\Psi^{22}, \quad M_\Psi^{12}, \quad g_\rho.
\end{equation}
Under these assumptions, all the results presented below have been obtained with parameters scanned in the ranges $600 \, \textrm{GeV} \le f \le 3000 ~\textrm{GeV}$, $0 \le X \le 10 f$, with $X = Y_1^{12}, Y_2^{12}, \Delta_L^1, \Delta_R^2, M_\Psi^{11}, M_\Psi^{22}, M_\Psi^{12}$ and $2 \le g_\rho \le 10$.

\begin{figure}
\centering
\subfloat{\includegraphics[scale=0.35]{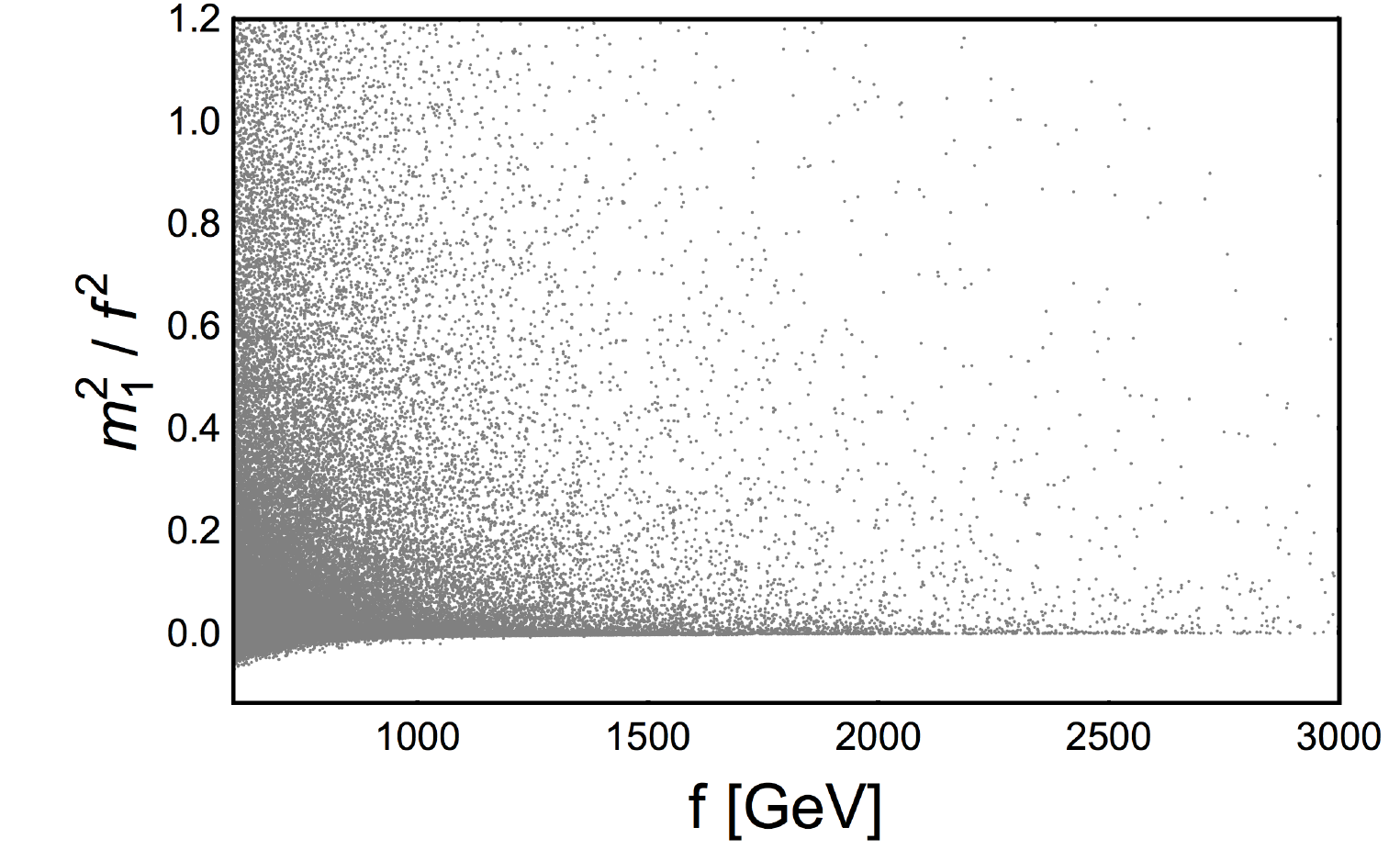}}
\subfloat{\includegraphics[scale=0.35]{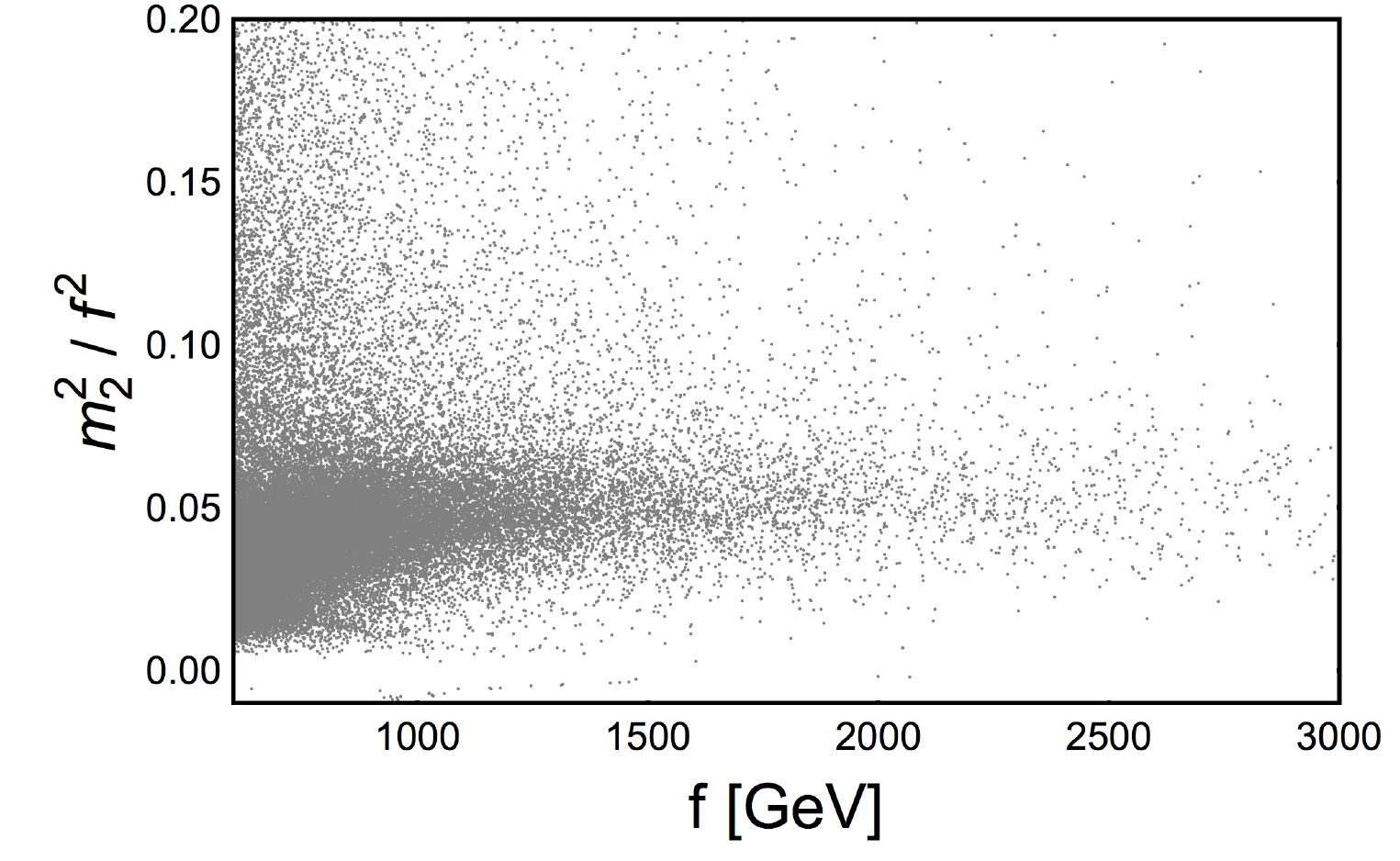}}
\subfloat{\includegraphics[scale=0.35]{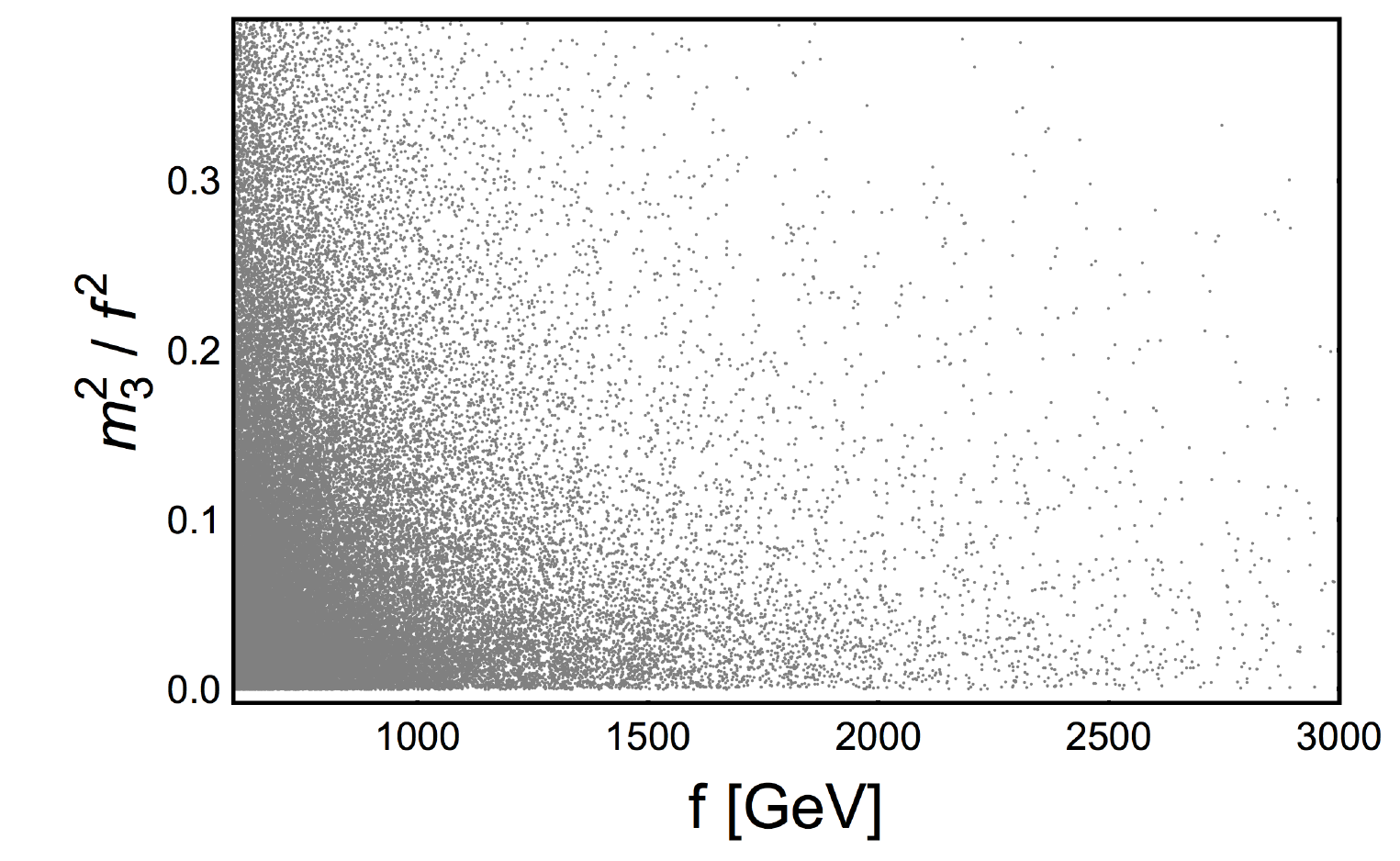}}\\
\subfloat{\includegraphics[scale=0.35]{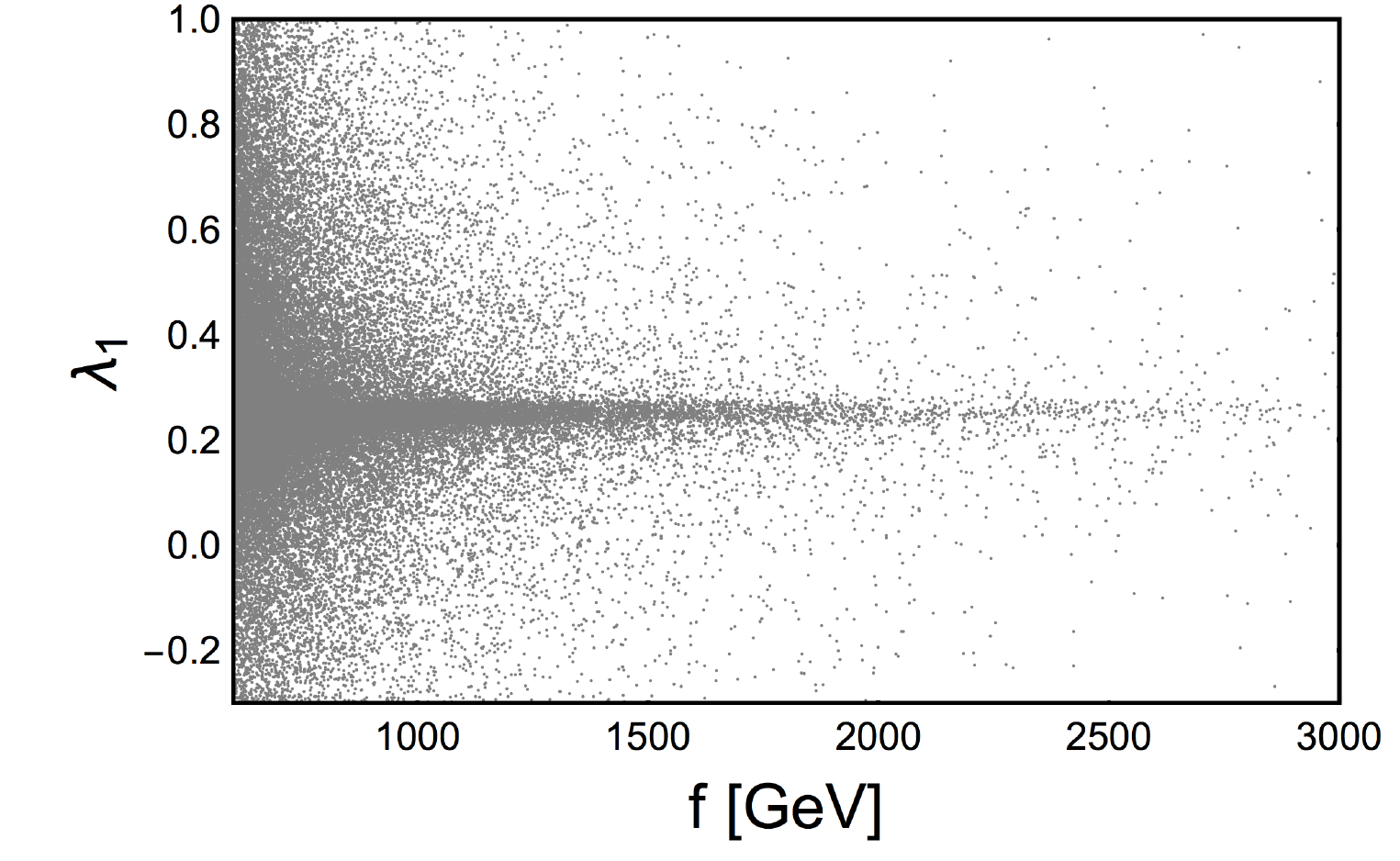}}
\subfloat{\includegraphics[scale=0.35]{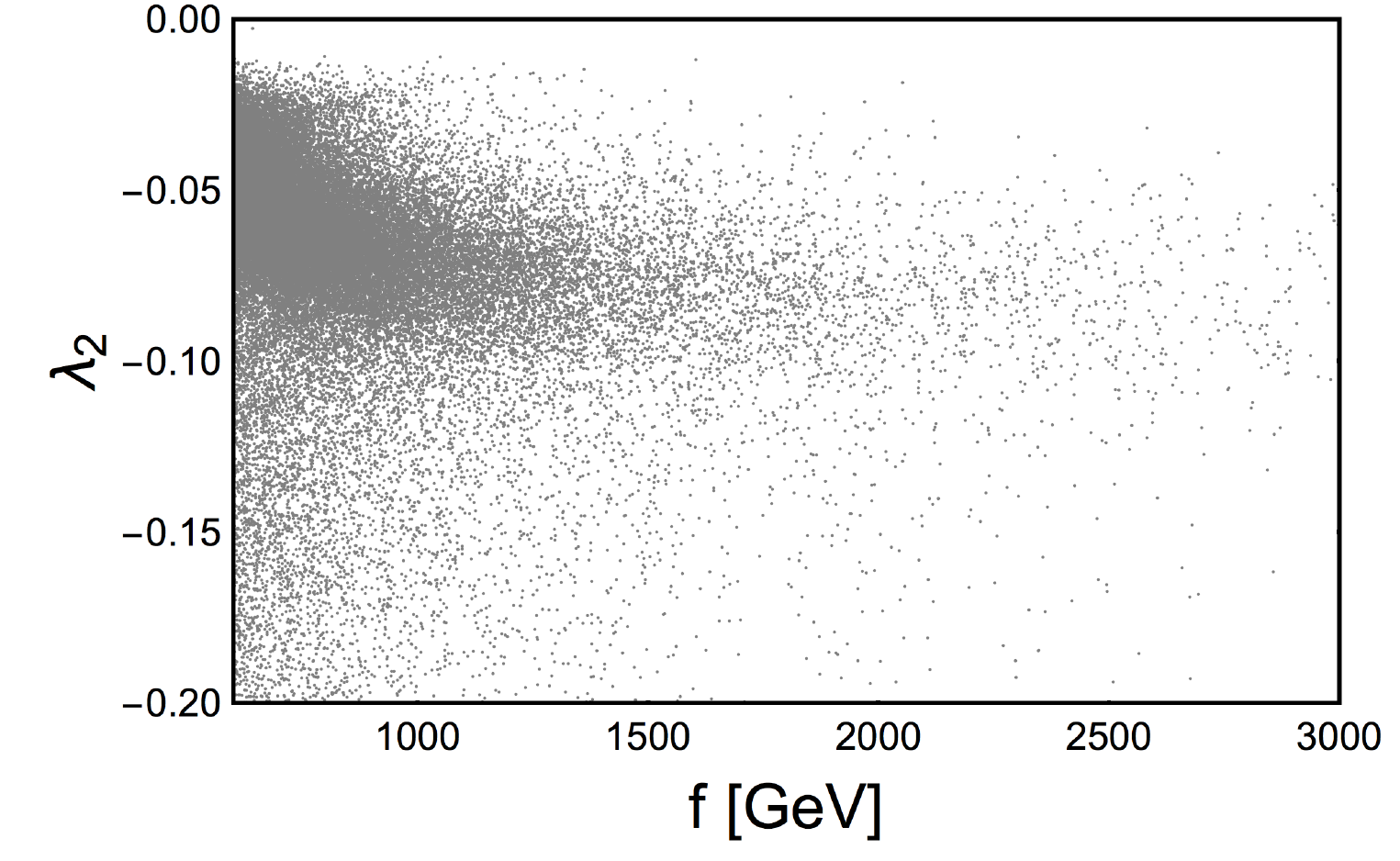}}
\subfloat{\includegraphics[scale=0.35]{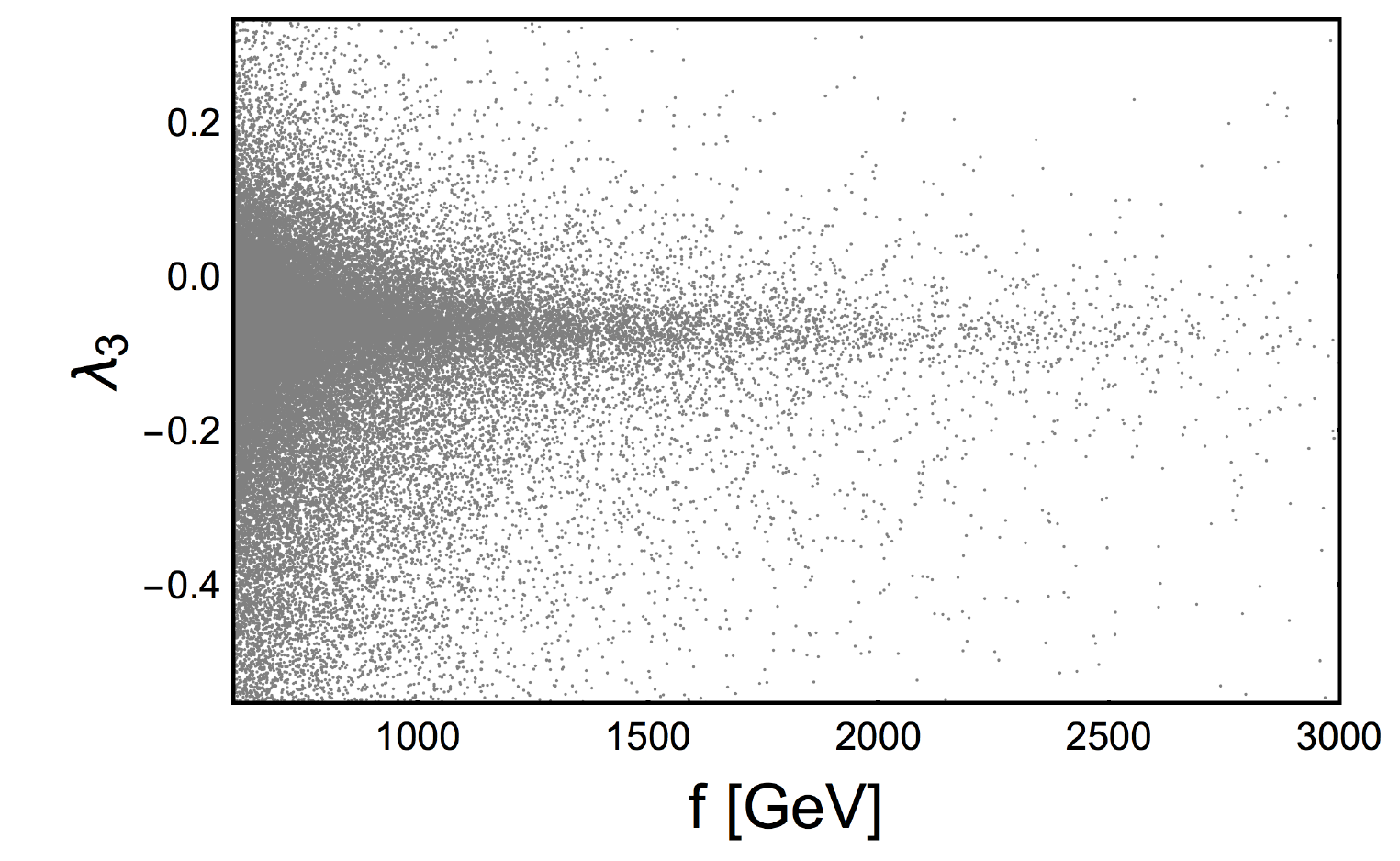}}\\
\subfloat{\includegraphics[scale=0.35]{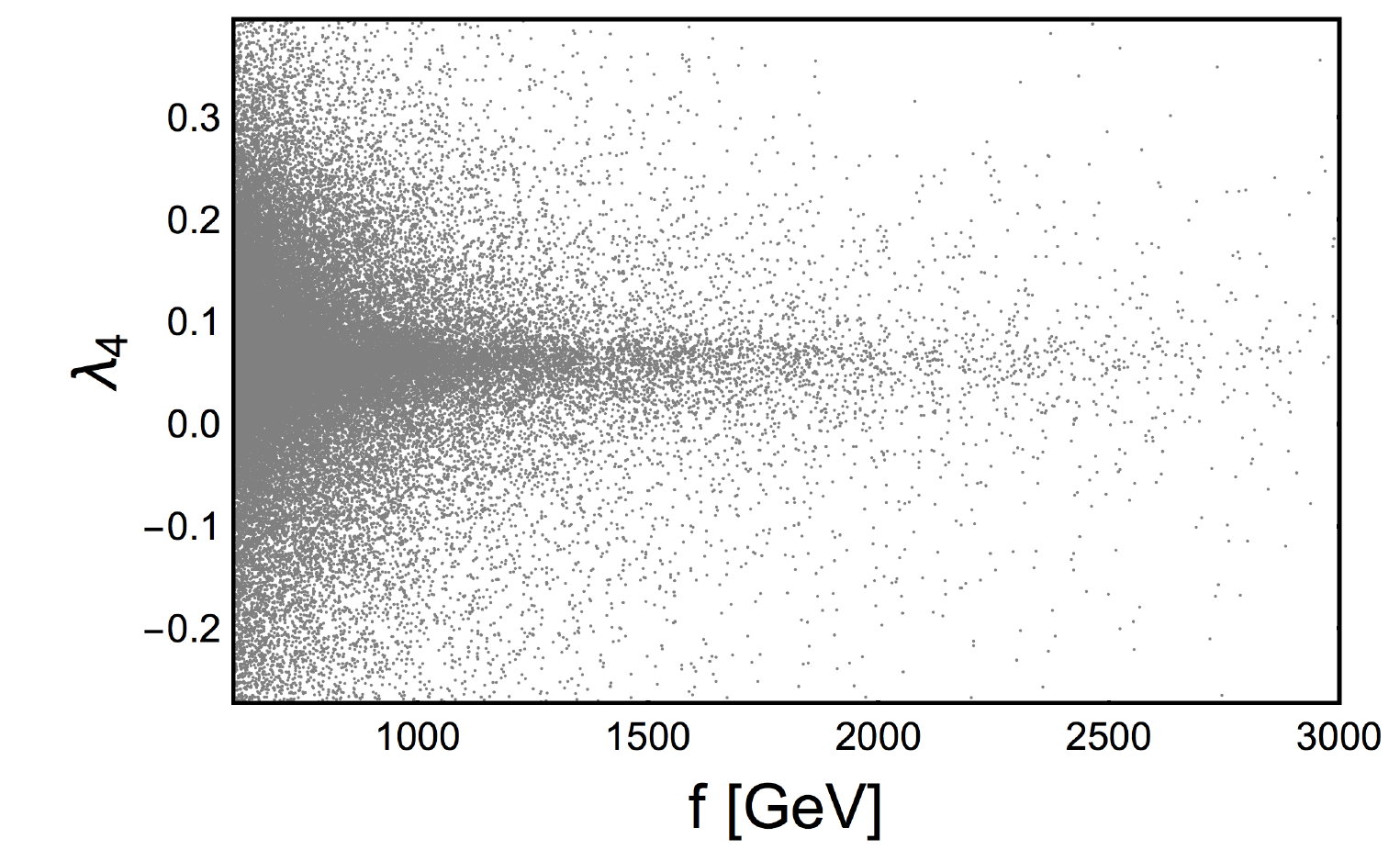}}
\subfloat{\includegraphics[scale=0.35]{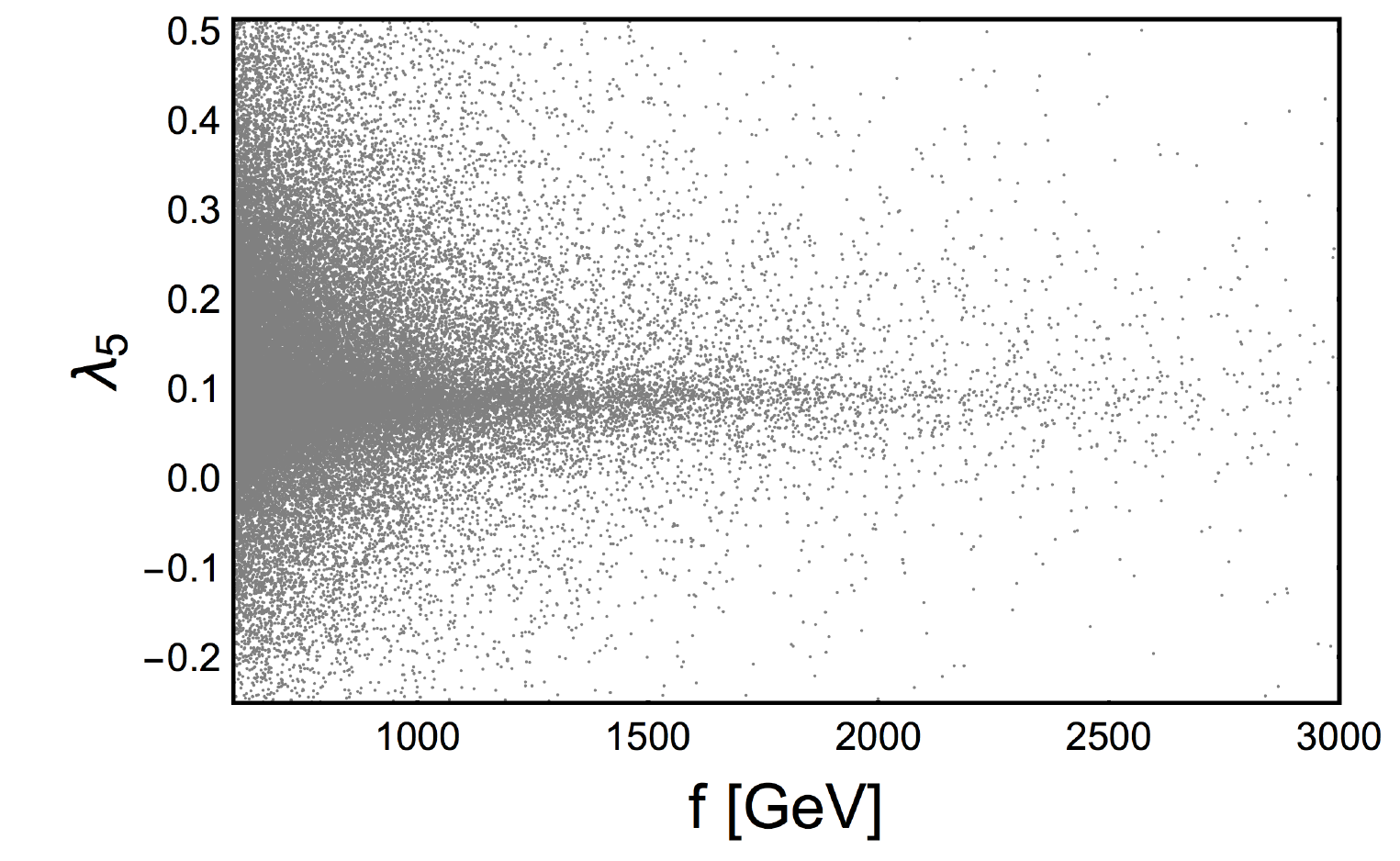}}
\subfloat{\includegraphics[scale=0.35]{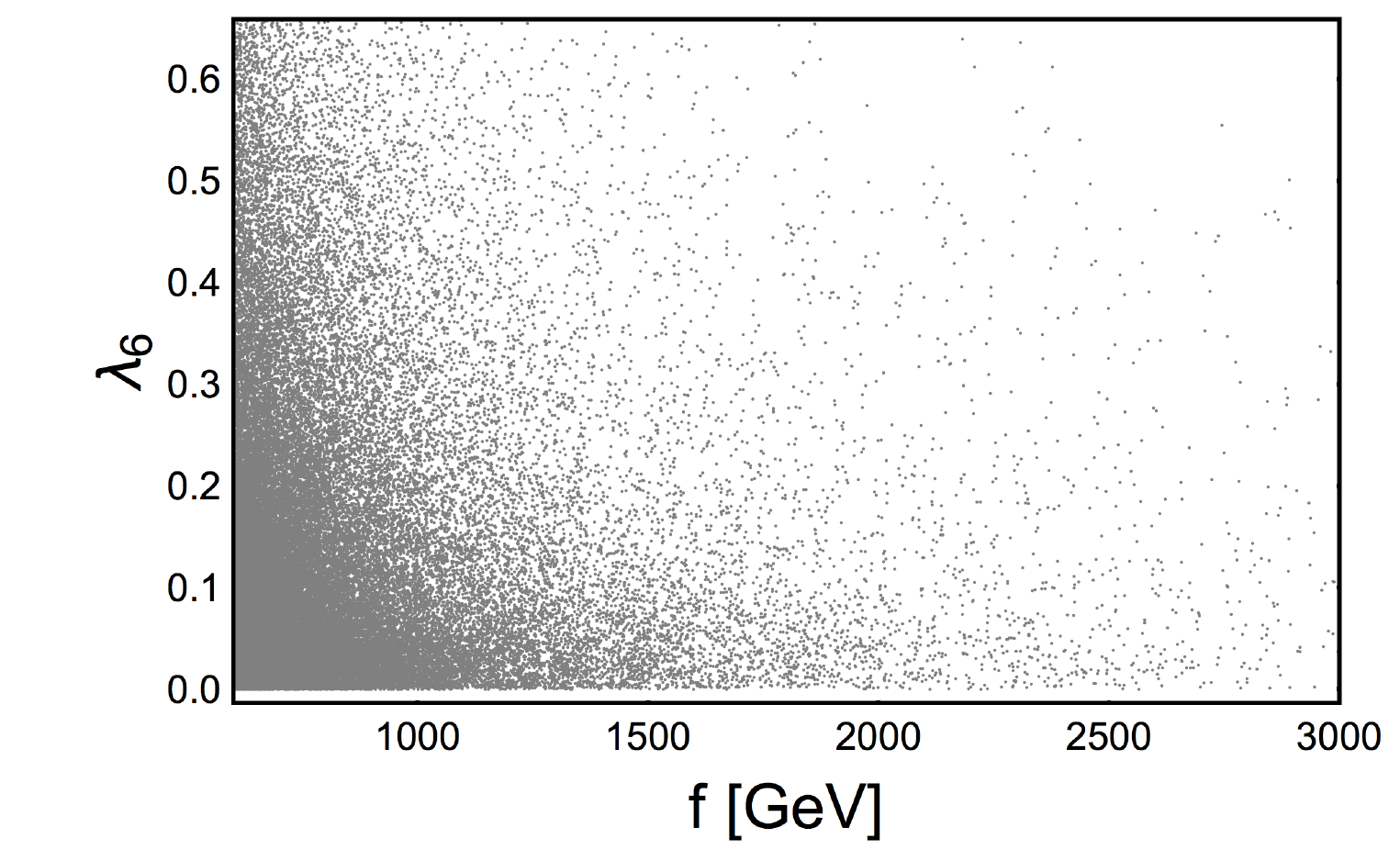}}
\caption{The mass parameters and the quartic couplings of the scalar potential as functions of $f$. Notice that $\lambda_6 = \lambda_7$.
\label{fig:parameters}
}
\end{figure}
\begin{figure}
\centering
\includegraphics[scale=0.4]{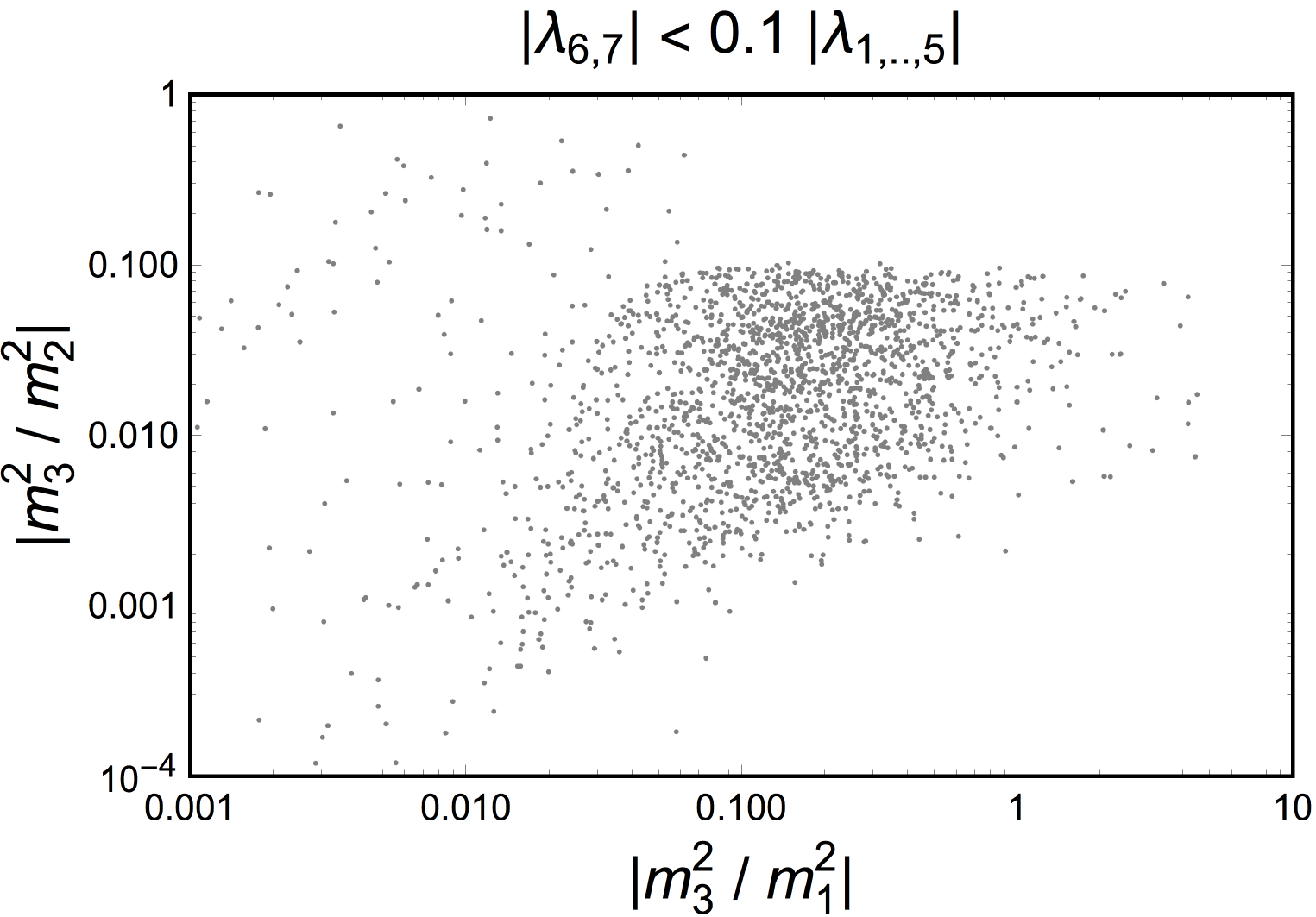}
\caption{The relative size of the mass parameters of the scalar potential satisfying the
condition  $\lambda_{6,7}<0.1\; \lambda_{1,2,3,4,5}$.
\label{fig:contact}
}
\end{figure}
Before moving to the study of the EWSB dynamics and the 
characterisation of the spectrum of the composite Higgs bosons, we show in Fig.~\ref{fig:parameters} the parameters of the scalar potential in the general basis as functions of $f$ that represent the main outcome of the constraints imposed by the strong dynamics. Of some relevance here is to establish a contact with the results of  Refs. \cite{DeCurtis:2016scv}, \cite{DeCurtis:2016tsm} and \cite{DeCurtis:2017gzi},  where the following parameter region of the C2HDM scalar potential 
was investigated: $\lambda_{6,7}\ll \lambda_{1,2,3,4,5}$, mimicking the one normally exploited in investigations of E2HDMs  with a softly broken $Z_2$ symmetry, namely $\lambda_6 = \lambda_7 = 0$ and $m_3^2 \neq 0$. To this end, we show in Fig.~\ref{fig:contact} the population of points, extracted from those in 
Fig.~\ref{fig:parameters}, which satisfy the  requirements 
 $\lambda_{6,7}<0.1\; \lambda_{1,2,3,4,5}$,  plotted over the plane $(m_3^2/m_1^2, m_3^2/m_2^2)$. As the values of
the mass parameters were in  Refs. \cite{DeCurtis:2016scv}, \cite{DeCurtis:2016tsm} and \cite{DeCurtis:2017gzi}
 taken within an order of magnitude of each other, over the corresponding region of parameter space of Fig.~\ref{fig:contact},   the same results found therein can be adopted for our C2HDM construct as well.

\subsection{EWSB and the significance of $\tan\beta$}

Differently from an E2HDM scenario in which all the Lagrangian parameters $(m_i^2,\lambda_j)$ appearing in Eq.~\eqref{2HDM-potential} can be taken as free\footnote{Unless an additional overarching symmetry is imposed, e.g., Supersymmetry: see Ref.~\cite{DeCurtis:2018iqd} for a comparison between Supersymmetry and Compositeness realisations embedding 2HDMs.},
all the masses and couplings in the Higgs potential of the C2HDM are predicted by the strong dynamics. Therefore,  achieving EWSB successfully in the C2HDM is not straightforward and a given amount of tuning is always necessary. 
Moreover, the potential of a CP-conserving 2HDM can allow, in general, two separate minima \cite{Barroso:2013awa} and one has to make sure that the EW one corresponds to a stable configuration.
In our analysis we have explicitly checked that, if this particular configuration is realised, the EW vacuum always corresponds to the global minimum.
In addition,  we have further demanded to reconstruct the observed Higgs and top masses in the intervals (120, 130) GeV and $(165, 175)$ GeV, respectively.
As a result of these constraints and of the implications of the strong dynamics, the distributions of the allowed points imply strong correlations among the physical observables
that will be investigated in the following. 

\begin{figure}[t]
\centering
\includegraphics[width=0.35\textwidth]{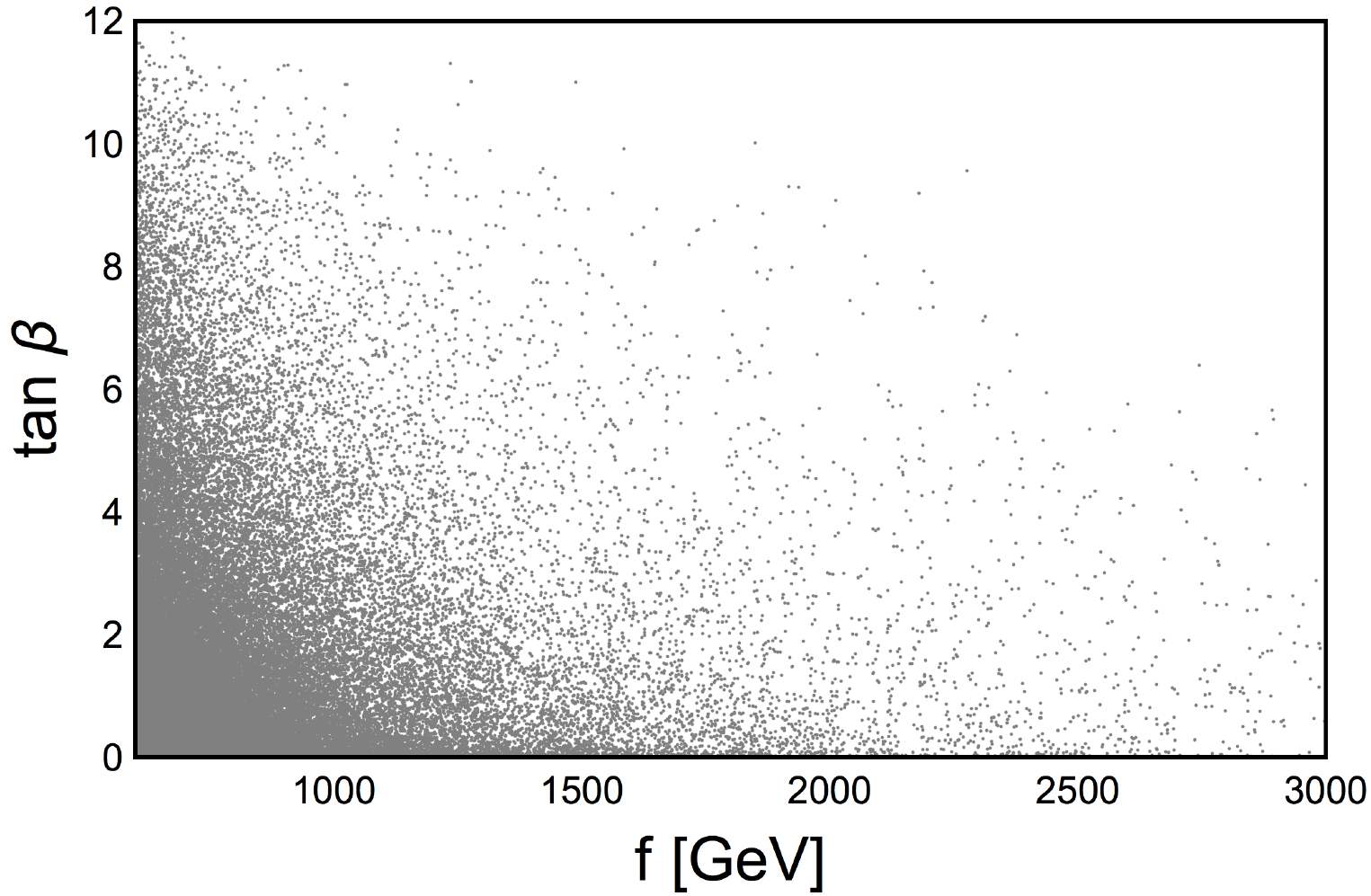} \qquad\qquad
\includegraphics[width=0.35\textwidth]{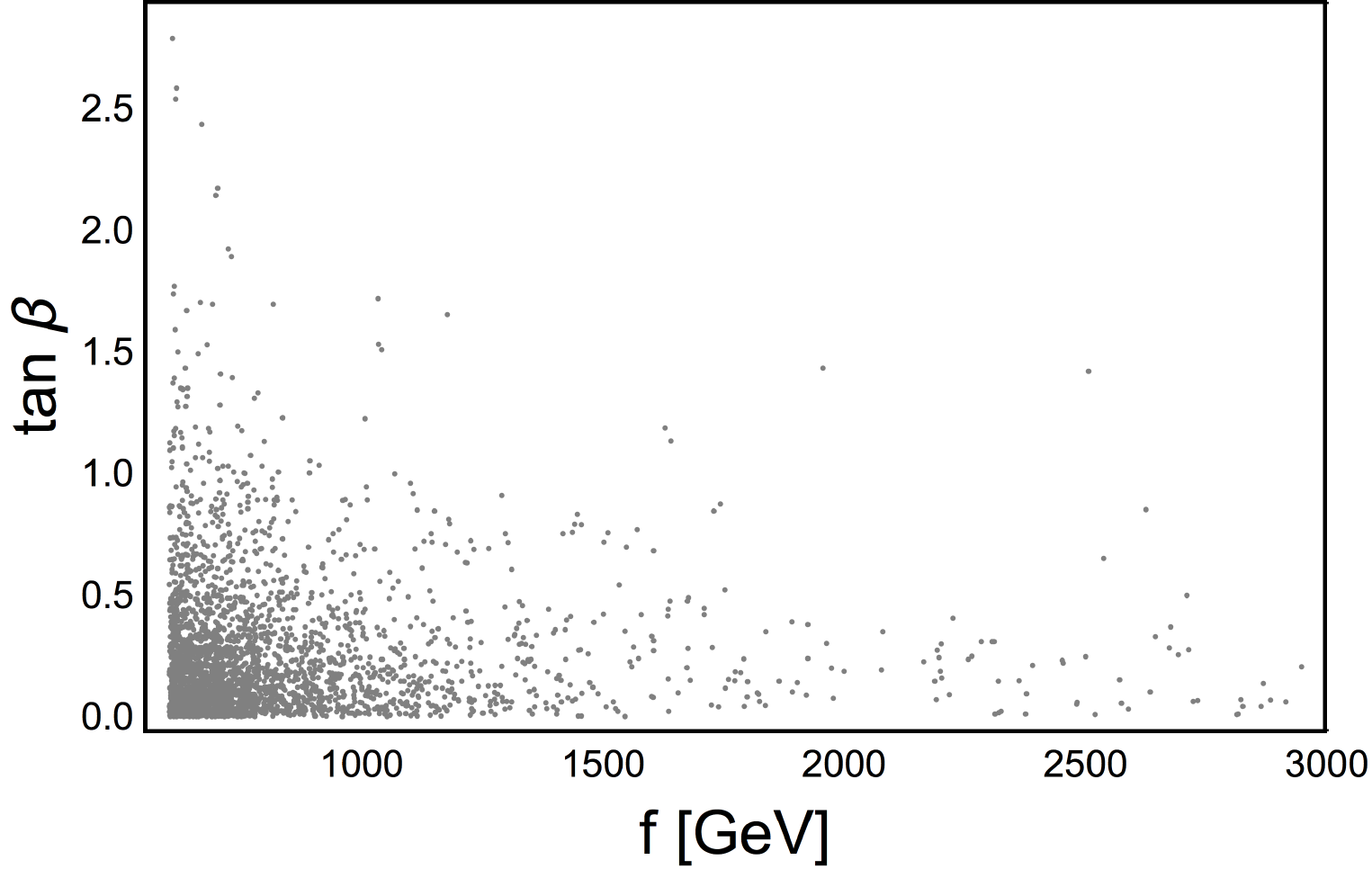}
\caption{Correlation between $\tan \beta$ and $f$. Left: full potential. Right: potential without the gauge contribution. \label{fig:f-tanbeta}}
\end{figure}

The existence of a non-trivial vacuum is secured by a careful solution of the two tadpole equations which provide values for the VEV and $\tan \beta$.
While $v_\textrm{SM}$ is fixed to 246 GeV, $\tan \beta$ could  potentially be  unconstrained if it were not for the requirements on $m_h$ and $m_{t}$, which select $\tan \beta \sim O(1-10)$ among all its possible values. 
The surviving range of $\tan \beta$, mapped against $f$, after imposing the two aforementioned constraints is illustrated in Fig.~\ref{fig:f-tanbeta} (left), from where it is clear that $\tan \beta$ never exceeds $\sim 10$.

\begin{figure}[t]
\begin{center}

\includegraphics[width=0.35\textwidth]{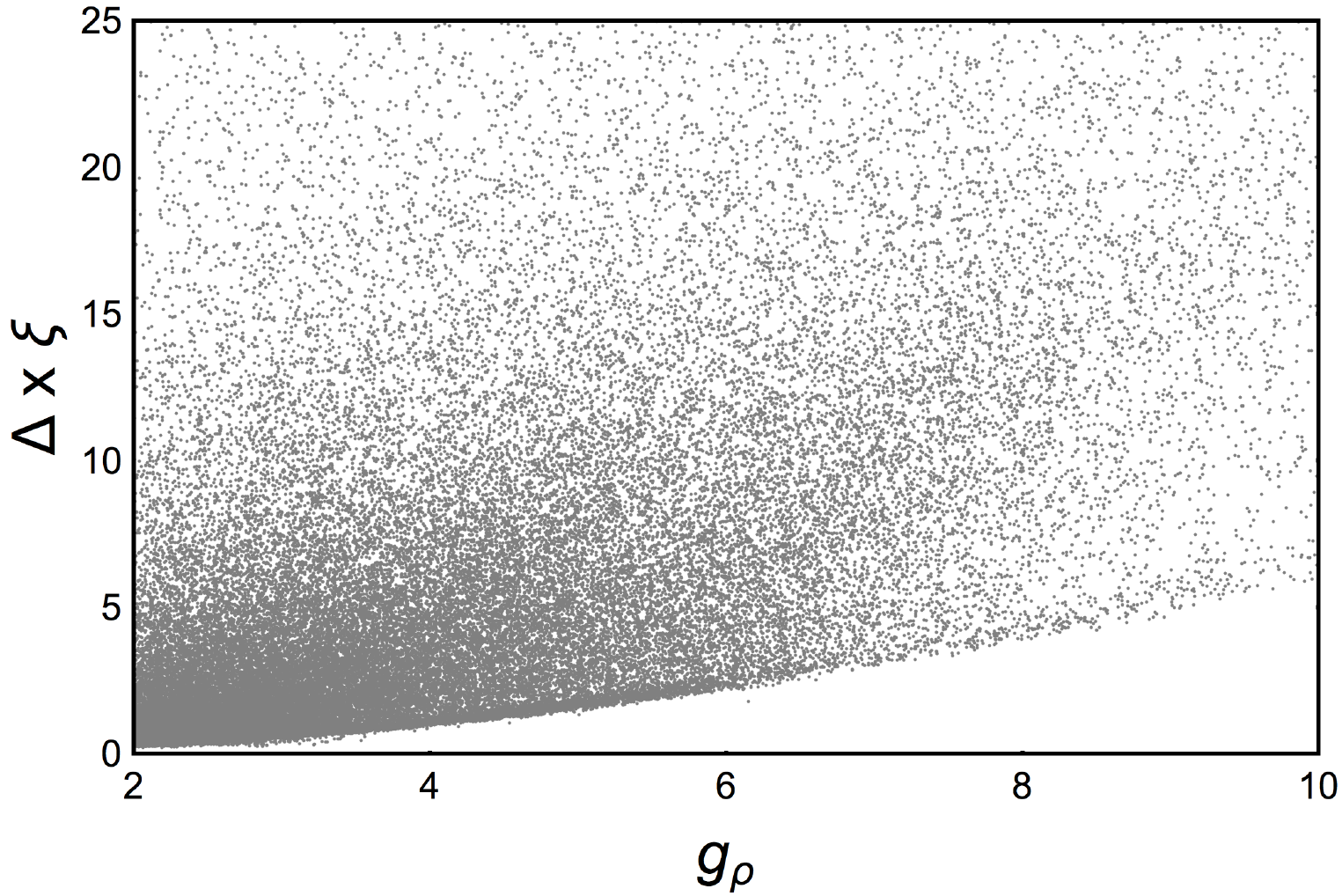}\qquad \qquad
\includegraphics[width=0.35\textwidth]{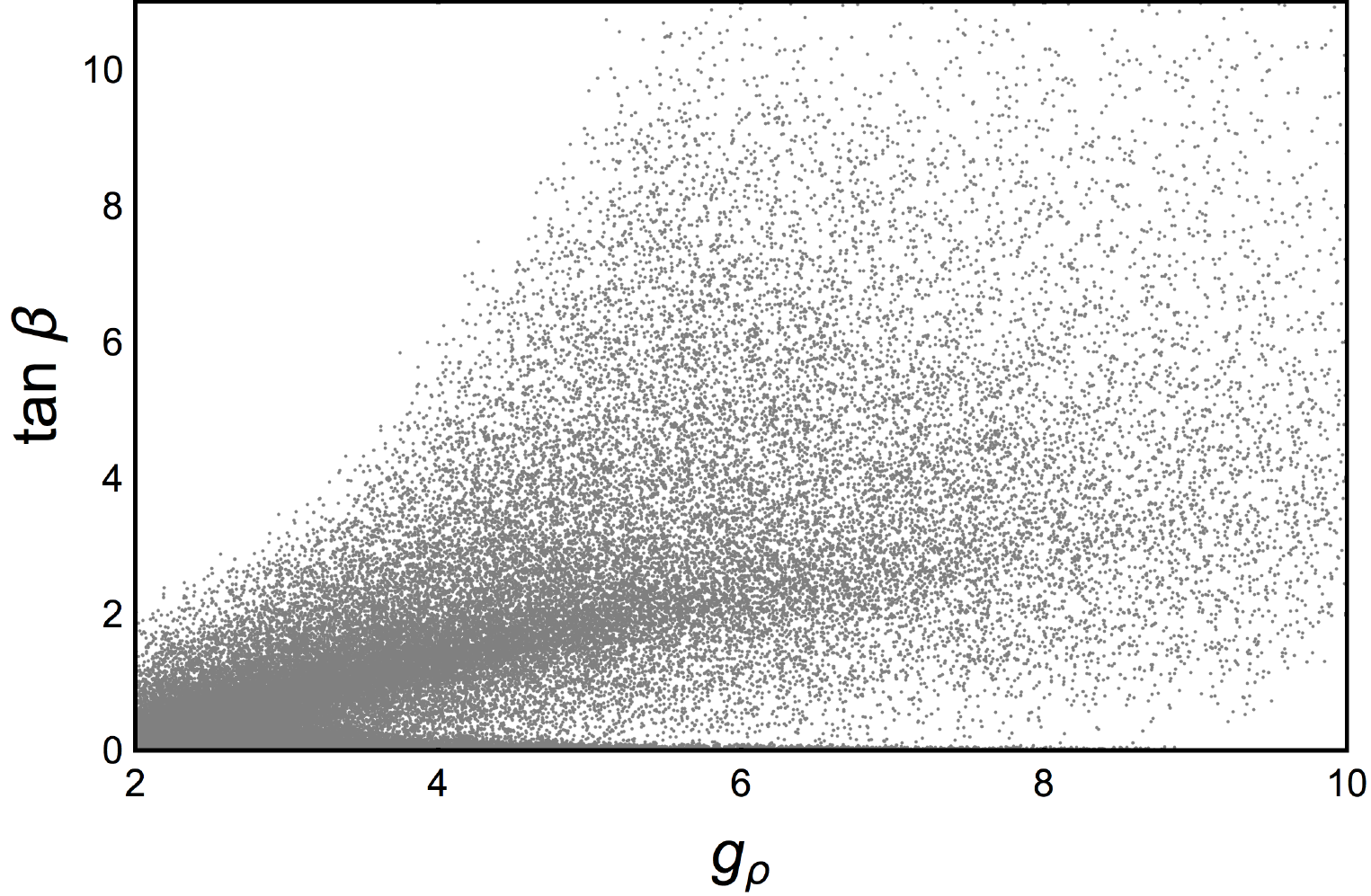}
\caption{ Left: Fine-tuning $\Delta \times \xi$ as a function of the coupling $g_\rho$. Right: Correlation between $\tan\beta$ and $g_\rho$.  \label{fig:finetuning}}
\end{center}
\end{figure}

Before moving to the characterisation of the scalar spectrum we comment on the significance of $\tan \beta$ for 2HDMs, both elementary and composite. 
In general, $\tan \beta$ is not a physical parameter of the 2HDM since it is not basis-independent \cite{Davidson:2005cw,Haber:2006ue}. 
This is, for instance, the case of an E2HDM in which the $Z_2$ transformation properties of the two Higgs doublets are not specified.
By imposing specific discrete symmetries, one selects a particular basis (the one in which such symmetries are manifest) where $\tan \beta$ can be uniquely identified, thus promoting it to a physical observable. Notable examples are the type-I and type-II E2HDMs \cite{Branco:2011iw}. 
The composite scenario described in this work does not have a $C_2$ invariance in the strong sector and thus cannot be related to any of the well know $Z_2$ realisations of the E2HDM.
Nevertheless, the requirement of CP conservation ($\theta_t = 0$) in the mixing between the composite states and the elementary R-handed quark automatically selects the $C_2$ invariant embedding of the latter. This choice eventually picks up a special basis, thus a special $\tan \beta$ among all possible basis-dependent definitions. Interestingly, this is not the case for the E2HDM in which the absence of a $Z_2$ symmetry prevents  the identification of a particular $\tan \beta$.
As a final remark, then,  it is worth noticing that one should pay attention in comparing C2HDMs vs E2HDMs for fixed values of $\tan \beta$ as the procedure is not meaningful unless the realisation of the 2HDMs is the same. 
Indeed, even though the two definitions of $\tan \beta$ may be both physical in the two models, if they belong to different bases, the observables they describe are not the same.

\subsection{The tuning of the Higgs potential}

Another important message learnt from Fig.~\ref{fig:f-tanbeta} (left) is that the density of points becomes smaller at large $f$, as naively expected from fine-tuning arguments.
Indeed, when $f$ is far from $v_\textrm{SM}$, severe cancellations among the parameters of the strong sector are necessary to satisfy the tadpole conditions. 
In general, a single tuning of the order $1/\xi = f^2/v_\textrm{SM}^2$ is not sufficient to depart from the natural solution of the tadpole equations, $v_\textrm{SM} \sim f$, and other cancellations, which depend on the fermionic and gauge embeddings in the global symmetry group, must also be advocated. 
In order to understand what are the most natural regions of the model, we compute the tuning $\Delta$ associated with the EW scale using the measure \cite{Barbieri:1987fn}
\begin{equation}
\Delta = \max_i \left |\frac{\partial \log m_Z}{\partial \log x_i} \right |,
\end{equation}
where the $x_i$'s span the parameters of the strong sector, i.e., those in Eq.~(\ref{eq:C2HDM-params}).
As stated above, $\Delta \sim \xi^{-1}$ represents only the minimal and unavoidable tuning necessary to trigger EWSB and bounds from below the distribution of points. 
Indeed, for a given $\xi$, the actual tuning may vary over different orders of magnitude. This situation, which is also manifest in ${\rm SO}(5)/{\rm SO}(4)$ CHMs with fermions in the fundamental representations, is dubbed \emph{double tuning} and has been extensively discussed in \cite{Panico:2012uw}, where the role of the fermion sector has been emphasised. It has been shown that
the parametrically leading quadratic contributions $y_{L,R}^2$, with $y_{L,R} = \Delta_{L,R}/f$, in the mass terms cannot be made arbitrarily small without reducing at the same time 
the quartic couplings. This makes the potential not tunable at the order $O(y_{L,R}^2)$. 
The presence of  higher order corrections is therefore crucial and one is obliged to firstly demand the leading $y_{L,R}^2$ contributions to be of the same order of the subleading $y_{L,R}^4$ ones before invoking the second cancellation which finally generates the order $\xi$ hierarchy between the EW vacuum and the scale of compositeness. 

Here we focus, instead, on the impact of the gauge sector which is most of the time overlooked and neglected with respect to the top quark one due to the smallness of the weak gauge couplings. 
This is not always the case as one can naively see from Fig. \ref{fig:f-tanbeta} (right), where we have considered the limiting case when $g,g'\to 0$, i.e.,  the unphysical regime where there is no gauge contribution. The distribution of $\tan \beta$ is clearly very different and limited to much more small values with respect to that of Fig. \ref{fig:f-tanbeta} (left), where all the gauge and fermionic contributions are correctly taken into account. 

In Fig. \ref{fig:finetuning} (left), we notice that the minimum amount of tuning increases in the region where the coupling $g_\rho$ (and, thus, the mass of the spin-1 resonances) becomes large. Indeed, a rather model independent gauge contribution to $m_{1,2}^2$ is proportional to $(9/32\pi^2) g^2 m_\rho^2$, which is $C_2$ symmetric and tends to prevent the breaking of the EW symmetry. As such, a larger cancellation between the fermionic and gauge contributions must be advocated.
As the gauge contributions become large, the fermionic ones are forced to increase as well and, being intrinsically $C_2$ breaking, drive the model into a region of the parameter space which deviates substantially from the inert case. This expectation finds confirmation in Fig.~\ref{fig:finetuning} (right), where we show a correlation between large values of $g_\rho$ and sizeable values of $\tan\beta$ (we remind the reader that the inert case implies $\tan \beta = 0$).

\section{Higgs boson masses and couplings}
\label{sec:higgscouplings}

We start here by recalling that the physical Higgs states of the C2HDM are the same as those of the renormalisable E2HDM, namely, the two CP-even scalars $h$ and $H$ (in the C2HDM, $h$ is always the SM-like Higgs with mass around 125 GeV), the pseudoscalar $A$ and the charged Higgs $H^\pm$. These are easily identified in the Higgs basis (see Appendix \ref{sec:Higgsbasis}) in which only one of the two doublets provides a VEV and it is obtained from Eq.~(\ref{2HDM-potential}) after a rotation by an angle $\beta$. The correspondence between parameters in the Higgs and  general basis is worked out, for instance, in \cite{Davidson:2005cw}. 
The mass matrix for the CP-even states is 
\begin{eqnarray}
\mathcal M^2 = \left(
\begin{array}{cc}
\mathcal M_{11}^2 &  \mathcal  M_{12}^2 \\
 \mathcal  M_{12}^2 & \mathcal M_{22}^2 
\end{array}
\right) 
\qquad  \textrm{with} \qquad
\left\{
\begin{array}{l}
\mathcal M_{11}^2 =  - 2 M_{11}^2 \\
\mathcal M_{12}^2 =  2  \frac{v}{v_\textrm{SM}} M_{12}^2 \\
\mathcal M_{22}^2 =  \frac{v^2}{v_\textrm{SM}^2} \left[ M_{22}^2 + \frac{v^2}{2} \left( \Lambda_3 + \Lambda_4 + \Lambda_5 \right) \right]
\end{array}
\right.
\,,
\end{eqnarray}
where the prefactor $v/v_\textrm{SM}$ arises from the canonical normalisation of the kinetic terms and the parameters $M^2_{ij}$ and $\Lambda_i$ which are given in Appendix \ref{sec:Higgsbasis}. The diagonalisation of the mass matrix provides the masses of the physical CP-even Higgses
\begin{align}
m_h^2 &= c_\theta^2 \mathcal M_{11}^2   +   s_\theta^2 \mathcal M_{22}^2   + s_{2\theta} \mathcal M_{12}^2 \,, \nonumber \\
m_H^2 &= s_\theta^2 \mathcal M_{11}^2   +   c_\theta^2 \mathcal M_{22}^2   - s_{2\theta} \mathcal M_{12}^2 \,, \nonumber \\
\tan 2 \theta &= 2 \frac{\mathcal M_{12}^2}{\mathcal M_{11}^2 - \mathcal M_{22}^2} \,.
\end{align}
Interestingly, the minimum conditions in the Higgs basis are
\begin{equation}
M_{11}^2 = - \frac{1}{2} \Lambda_1 \, v^2 \,, \qquad M_{12}^2 =  \frac{1}{2} \Lambda_6 \, v^2,
\end{equation}
which involve only $M_{11}^2$ and $M_{12}^2$ while the value of $M_{22}^2$ is unconstrained. Therefore, while $M_{11}$ and $M_{12}$ are tuned to $v_\textrm{SM}$ (or, equivalently, $\mathcal M_{11}$ and $\mathcal M_{12}$), thus providing a physical Higgs state with $m_h \simeq \mathcal M_{11}$ much lighter than the scale of compositeness and a small mixing angle $\theta$ between the two CP-even scalars, the mass of the heavy Higgses is naturally of order $m_H \simeq \mathcal M_{22} \simeq f$, modulo corrections induced by the mixing $\theta$. 
For the latter we have
\begin{equation}
\label{eq:scalarmixing}
\tan 2 \theta = 2 \frac{\mathcal M_{12}^2}{\mathcal M_{11}^2 - \mathcal M_{22}^2} \simeq - 2 \frac{\mathcal M_{12}^2}{\mathcal M_{22}^2} \simeq - 2 \Lambda_6 \xi \,.
\end{equation}
The dependence of $\theta$ from $f$ is depicted in Fig.~\ref{fig:scalarmixing}. Further details on the behaviour of the masses and  mixing angle are discussed in the following sections. 

The masses of the CP-odd and charged Higgs states are, instead, given by
\begin{align}
m_A^2 &= M_{22}^2 + \frac{v^2}{2} \left( \Lambda_3 + \Lambda_4 - \Lambda_5 \right), \nonumber \\
m_{H^\pm}^2 &= M_{22}^2 + \frac{v^2}{2} \left( \Lambda_3 - \Lambda_4 \right),
\end{align}
and, as $m_H$, are naturally of order $f$.

We now present explicitly in analytical form the couplings of all Higgs states of the C2HDM to both fermions and gauge bosons of the the SM as well those among themselves which are relevant for LHC phenomenology. We shall do so in three separate sub-sections.

\subsection{Couplings to fermions} 
Assuming  flavour alignment in order to guarantee the absence of FCNCs at tree level, the leading couplings of the scalars to the fermions are extracted from Eq.~(\ref{fermion6}) at first order in $\xi$ and can be described by the Yukawa Lagrangian
\begin{align}\label{eq:Hff}
- \mathcal L_\textrm{Yukawa} &=  \sum_{f = u,d,l} \frac{m_f}{v_\textrm{SM}} \bar f \left[ \xi_h^f \, h   + \xi_H^f \, H  - 2 i I_f \xi_A^f \, A \gamma^5 \right] f \nonumber \\
 & + \frac{\sqrt{2}}{v_\textrm{SM}} \left[ V_{ud} \, \bar u \left( - \xi^u_A m_u P_L + \xi^d_A m_d P_R \right) d \, H^+ + \xi^l_A \, m_l \, \bar \nu P_R l \, H^+\right] + \textrm{h.c.},
\end{align}
where $I_f = 1/2 (-1/2)$ for $f = u \, (d,l)$ and the $\xi^f$ coefficients are
\begin{align}
\xi^f_h &= (1 + c^h_f \, \xi) \cos \theta + (\zeta_f + c^H_f \, \xi)  \sin \theta   \,, \quad
\xi^f_H = - (1 + c^h_f \, \xi) \sin \theta + (\zeta_f + c^H_f \, \xi) \cos \theta    \,, \nonumber  \\
\xi^f_A &=   \zeta_f    +    \xi \left[ - \frac{\tan \beta}{2}  \frac{1 + \bar \zeta_t^2}{(1 + \bar \zeta_f \, \tan \beta)^2} \right],
\label{eq:fermion-coupling-0}
\end{align}
with
\begin{align}\label{eq:fermion-coupling}
c_f^h   &=  - \frac{1}{2} \frac{3 + \bar \zeta_f \, \tan \beta}{1 + \bar \zeta_f \, \tan \beta} \,, \quad
c_f^H   = \frac{1}{2} \frac{\bar \zeta_f (1 + \tan^2 \beta)}{(1 + \bar \zeta_f \, \tan \beta)^2}  \,, \nonumber \\
\zeta_f &= \frac{\bar \zeta_f - \tan \beta}{1 + \bar \zeta_f \, \tan \beta} \,, \qquad \qquad \bar \zeta_f = - \frac{Y_1^f}{Y_2^f} \,.
\end{align}
As mentioned, the parameter $\theta$ denotes the mixing between the two CP-even states while $\zeta_f$ represents the normalised coupling to the fermion $f$ of the CP-even scalar that does not acquire a VEV in the Higgs basis. Since $\theta$ is predicted to be small \cite{Mrazek:2011iu,DeCurtis:2018iqd}, $\zeta_f$ controls the interactions of the Higgs states $H,A, H^\pm$ at the zeroth order in $\xi$.
At that order, the structure of the Yukawa Lagrangian is the same as the E2HDM one in which the alignment in the flavour sector has been enforced. 
The crucial difference is that in the E2HDM $\zeta_f$ is a free parameter while in the C2HDM is fixed by the strong dynamics and  correlated 
to other physical observables. Some of these correlations will be explored in the following sections. 

The mass of fermions are also predicted quantities in CHMs. In particular, exploiting the explicit expressions of the form factors in Eq.~(\ref{fermion6}) (and listed in Appendix \ref{sec:formfactors}), we read 
the top mass
\be
\label{eq:topmass_pred}
m_t =\frac{v_\textrm{SM}}{\sqrt{2}}\frac{\Delta_L\Delta_R}{m_Q m_T}\frac{M_\Psi^2}{\tilde m_Q \tilde m_T}\frac{Y_1 s_\beta + Y_2 c_\beta}{f}\big[1 + O(\frac{\Delta^2}{M_\Psi^2}\xi)\big] \equiv \frac{y_t}{\sqrt{2}} v_\textrm{SM}
\ee
where  $m_{Q,T}$ and $\tilde m_{Q,T}$ are the physical masses of the top partners coupled to $q_L$ and $t_R$, respectively, and, for simplicity, we dropped the superscript $t$ from the parameters of Eq.~(\ref{eq:lag_ferm0}).
 When the mixing parameters are such that $\Delta\sim M_\Psi$ the subleading corrections can be numerically relevant at small $f$. The expression of the bottom mass is totally analogue and better approximated even at smaller $f$.

\subsection{Couplings to gauge bosons} 
The trilinear interaction vertices between the physical Higgses and the SM gauge bosons can  easily be extracted from the kinetic Lagrangian of the pNGB matrix in Eq. (\ref{4dlag}) up to the order $\xi$. 
As typical in CHMs, due to the non-linearities of the derivative interactions, the kinetic terms of the NGBs must be rescaled in order to be canonical. This introduces corrections of order $\xi$ in the gauge couplings.
The relevant interaction terms and corresponding coefficients have been computed in \cite{DeCurtis:2016tsm} and are listed in Tab.~\ref{tab:3vert} as functions of the mixing angle $\theta$.
In particular, the couplings of the SM-like Higgs to the EW gauge bosons, $h VV$ with $V = W, Z$, get modified by the usual mixing angle $\theta$, as in every realisation of E2HDMs, but also by corrections of order $\xi$.
A convenient way to parameterise these couplings is to recast them in terms of the so called $\kappa_i$ `modifiers'  of Ref. 
\cite{LHCHiggsCrossSectionWorkingGroup:2012nn} which are the couplings of the SM-like Higgs boson normalised to the corresponding SM prediction 
\begin{equation}
\kappa_V = \left(1-\frac{\xi}{2} \right)\cos\theta, 
\label{eq:kappaV}
\end{equation}
where $\theta \to 0$ with $f \to \infty$ corresponds to the alignment limit, i.e., the couplings of $h$ to SM particles become the same as those of the SM-like Higgs at tree level.  
\begin{table}
\centering
\begin{tabular}{|c|c||c|c|}
\hline
vertex & coupling 														& vertex & coupling \\ \hline \hline
$h W^+_\mu W^{- \, \mu}$  & $ g \, m_W (1- \xi/2) \cos \theta$ 						& $H W^+_\mu W^{- \, \mu}$  & $- g \, m_W (1- \xi/2) \sin \theta$ \\
$h Z_\mu Z^{\mu}$  & $g_Z \, m_Z/2 \, (1- \xi/2) \cos \theta$ 						& $H Z_\mu Z^{\mu}$  & $- g_Z \, m_Z/2 \, (1- \xi/2) \sin \theta$ \\
$H^\pm \partial_\mu h W^{\mp \, \mu}$ & $\mp i g/2 \, (1 - 5 \xi/6) \sin \theta$			& $H^\pm \partial_\mu H W^{\mp \, \mu}$ & $\mp i g/2 \, (1 - 5 \xi/6) \cos \theta$\\
$h \partial_\mu H^\pm W^{\mp \, \mu}$ & $\pm i g/2 \, (1 -  \xi/6) \sin \theta$			& $H \partial_\mu H^\pm W^{\mp \, \mu}$ & $\pm i g/2 \, (1 -  \xi/6) \cos \theta$\\
$A \partial_\mu h Z^\mu$	& $-g_Z/2 \, (1 - 5 \xi/6) \sin \theta$						& $A \partial_\mu H Z^\mu$	& $-g_Z/2 \, (1 - 5 \xi/6) \cos \theta$\\
$h \partial_\mu A Z^\mu$	& $g_Z/2 \, (1 -  \xi/6) \sin \theta$						& $H \partial_\mu A Z^\mu$	& $g_Z/2 \, (1 -  \xi/6) \cos \theta$\\
$H^+ \overset{\leftrightarrow}{\partial}_\mu H^- Z^\mu$ & $-i g_Z/2 \cos(2 \theta_W)$		& $H^+ \overset{\leftrightarrow}{\partial}_\mu H^- A^\mu$ & $-i e$\\
$H^\pm \overset{\leftrightarrow}{\partial}_\mu A W^{\mp\,\mu}$	& $g/2$					&	&	\\
\hline
\end{tabular}
\caption{Trilinear couplings between (pseudo)scalars and SM gauge bosons. 
\label{tab:3vert}}
\label{tab:trilinear}
\end{table} \\ \\
\subsection{Higgs boson self-couplings} 

The scalar trilinear couplings are extracted from the cubic part of the potential, $V_3 = \sum_{i,j,k} \lambda_{\phi_i \phi_j \phi_k} \phi_i \phi_j \phi_k$. 
Here we only show the most relevant ones to the phenomenological studies that will be carried out below, i.e.,  \begin{eqnarray}
\lambda_{h H^+ H^-} &=& v_\textrm{SM} \left[  \left( c_\theta \Lambda_3 + s_\theta \Lambda_7 \right)   + \frac{\xi}{6}  \left(  c_\theta \Lambda_3 + 2 s_\theta \Lambda_7 \right)    \right]  \,, \nonumber \\
\lambda_{H A A} &=& \frac{v_\textrm{SM}}{2} \left[ \left( c_\theta \Lambda_7 - s_\theta ( \Lambda_{3} + \Lambda_{4} - \Lambda_{5}) \right)  
				+ \frac{\xi}{6} \left( 2 c_\theta \Lambda_7 - s_\theta ( \Lambda_{3} + \Lambda_{4} - \Lambda_{5}) \right) \right] \,, \nonumber \\
\lambda_{H H^+ H^-} &=& v_\textrm{SM} \left[  \left( c_\theta \Lambda_7 - s_\theta \Lambda_3 \right)   + \frac{\xi}{6}  \left( 2 c_\theta \Lambda_7 - s_\theta \Lambda_3 \right)    \right]  \,, \nonumber \\
\lambda_{H h h} &=& \frac{v_\textrm{SM}}{2} \left[ 
\left(  - s_\theta^3 \Lambda_{345} + 3 s_\theta^2 c_\theta  (\Lambda_7  -2 \Lambda_6 ) +  s_\theta c_\theta^2  (2 \Lambda_{345} - 3 \Lambda_1)  + 3 c_\theta^3 \Lambda_6 \right) \right. \nonumber  \\
&+&
\left. \frac{\xi}{2} \left(  - s_\theta^3 \Lambda_{345} + 4 s_\theta^2 c_\theta  (\Lambda_7  - \Lambda_6 ) +  s_\theta c_\theta^2  (2 \Lambda_{345} -  \Lambda_1)  + 2 c_\theta^3 \Lambda_6 \right)
\right] ,
\label{eq:HHH}
\end{eqnarray}
where the quartic couplings $\Lambda_i$ are defined in the Higgs basis and explicitly given in Appendix \ref{sec:Higgsbasis} and $\Lambda_{345} = \Lambda_3 + \Lambda_4 + \Lambda_5$. Finally, due to $\theta\sim\xi$ (see Eq.~(\ref{eq:scalarmixing})), the terms $\propto s_\theta^n$ for $n>1$ in Eq.~\eqref{eq:HHH} can safely be omitted at the order ${O(\xi})$. We note that other terms at the first order in $\xi$ arise from the non-linearities of the derivative terms of the NGB Lagrangian. For the sake of simplicity, these are not shown here but correctly taken into account in all our numerical computations.

\section{Phenomenology of the C2HDM Higgs bosons}
\label{sec:pheno}
We here list the expression for the scalar masses and  mixing angle once the parameters are fixed to reproduce the correct EW VEV.
In particular, we have the following prediction for the mass of the Higgs states and the rotation angle $\theta$ from the Higgs basis to the mass basis
\begin{eqnarray}\label{estimate-MH2}
\frac{m_{h}^2}{v^2} &=& \frac{1}{(1+t_\beta^2)^2} \big(\lambda_1 +t_\beta^4 \lambda_2 + 2 t_\beta^2 \lambda_{345}+4\lambda_6 (1+t_\beta^2)t_\beta \big) + O(\xi),\\
m_{H}^2 &=& m_3^2 \frac{1+t_\beta^2}{t_\beta} + v^2 \frac{t_\beta^2}{(1+t_\beta^2)^2} (\lambda_{1}+\lambda_{2}-2\lambda_{345} - \frac{1}{2} \lambda_6 \frac{(1+t_\beta^2)^3}{t_\beta^3}) + O(\xi),\\
m_A^2 &=& m_3^2 \frac{1+t_\beta^2}{t_\beta} - \frac{v^2}{2} (2 \lambda_5 +  \lambda_6  \frac{(1+t_\beta^2)}{t_\beta} )   ,   \label{eq:mAmass}   \\
m_{H^\pm}^2 &=& m_3^2 \frac{1+t_\beta^2}{t_\beta} - \frac{v^2}{2} (\lambda_4 + \lambda_5 +  \lambda_6   \frac{(1+t_\beta^2)}{t_\beta} )   , \\
\theta &=& -\frac{m_h^2}{t_\beta m_H^2}+\frac{\lambda_6 v^2}{m_H^2}\frac{3+t_\beta^2}{1+t_\beta^2}+\frac{\lambda_1 v^2}{t_\beta (1+t_\beta^2) m_H^2} +\frac{\lambda_{345}v^2 t_\beta}{(1+t_\beta^2)m_H^2}+ O(\xi^2).
\end{eqnarray}
By numerical evaluation we can show that indeed the $O(\xi)$ corrections are negligible in the determination of the mass of the heavy Higgses. In the remainder of this section we are going to discuss in more detail the phenomenological impact of the above formulas trying also to correlate (when possible) the Higgs properties  with the parameters of the composite sector.

\subsection{The mass of the heavy Higgs bosons}
\begin{figure}
\begin{center}
\includegraphics[width=0.45\textwidth,height=0.3\textwidth]{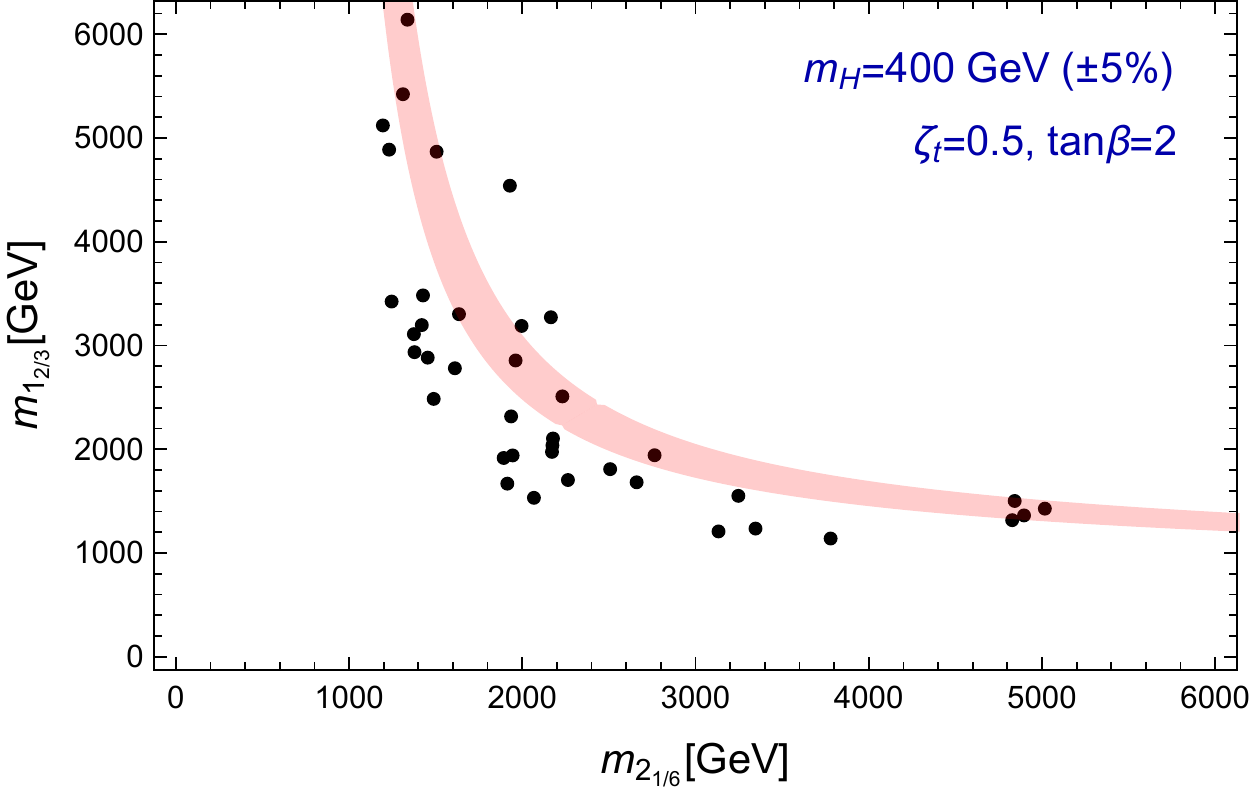}
\includegraphics[width=0.45\textwidth,height=0.3\textwidth]{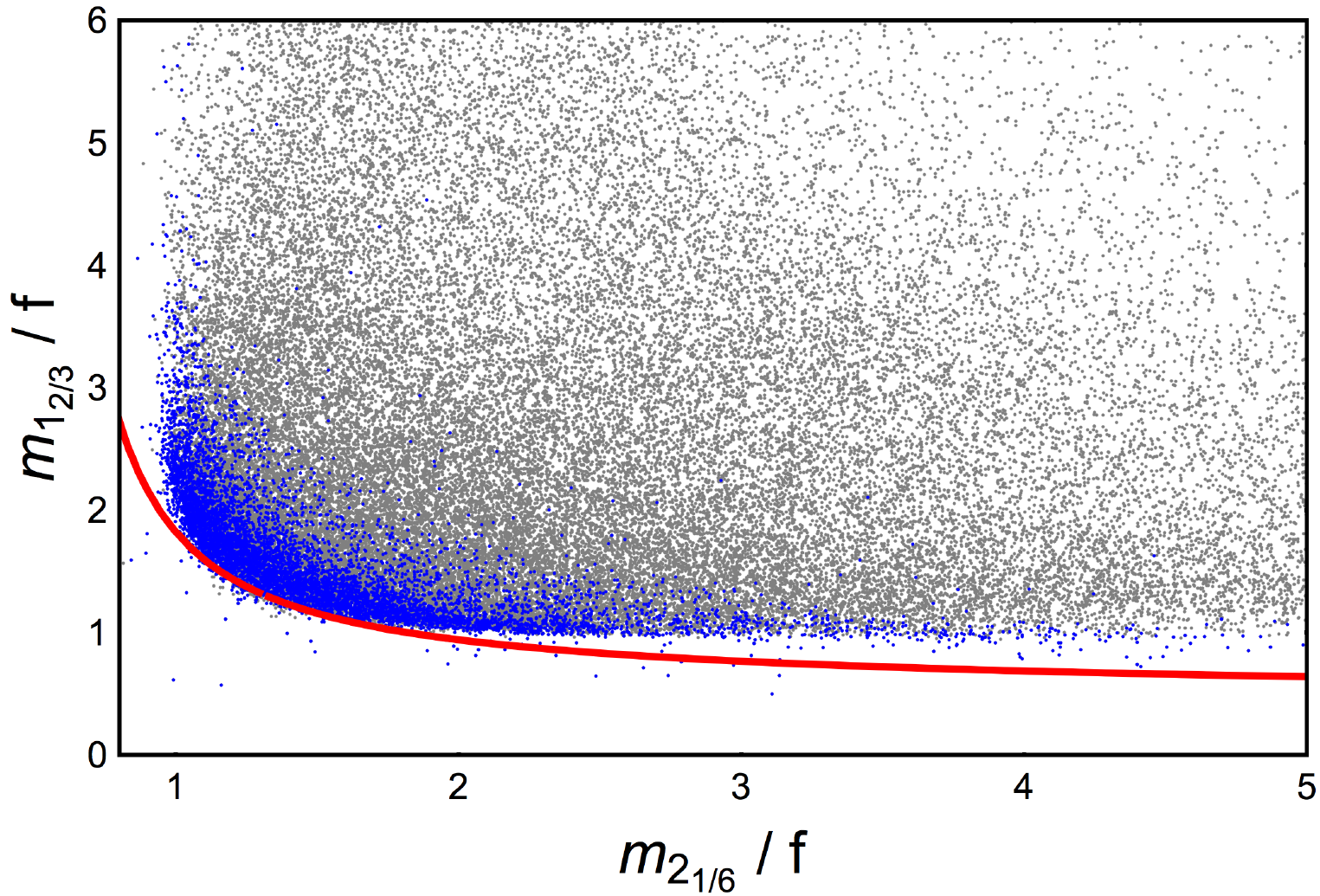}
\caption{\label{fig:scalingMH} Distribution of the masses of the lightest top-partners. (Left) Zoomed region of a selected mass window for the heavy CP-even Higgs boson. The red shaded region represents the analytic formula for $m_H$ at $O(\xi)$. (Right): the blue points are subject to the constraint $\tan\beta<0.1$ and the red line is the theoretical expectation for the Higgs mass in the minimal CHM.}
\end{center}
\end{figure}

Up to corrections  of order $\xi$, the mass of the heavy Higgs bosons is mainly set by the $m_3^2$ parameter, which is associated to the explicit breaking of $C_2$. In order to correlate $m_H$, or equivalently $m_A$ and $m_{H^\pm}$, to the fermionic resonances of the composite sector, we can inspect the form factors from which $m_3^2$ originates. Given the expression of $m_3^2$ in Eq.~(\ref{eq:parameters_f}) and exploiting the poles and the zeros of the form factors listed in Appendix \ref{sec:formfactors}, in particular of the product $M_1^t M_2^t$ in Eq.~(\ref{eq:M1M2ff}), we can write (after requiring the top mass in Eq.~(\ref{eq:topmass_pred})) the following expression for $m_3^2$
\be\label{m3}
m_3^2= \frac{N_c \, y_t^2}{8\pi^2}  \frac{\bar\zeta_t(1+t_\beta^2)}{(t_\beta \bar \zeta_t +1)^2} \int d k^2 \frac{m_{Q}^2 \tilde m_{Q}^2 m_T^2 \tilde m_T^2}{(k^2 - m_Q^2)(k^2 - \tilde m_Q^2)(k^2 - m_T^2)(k^2 - \tilde m_T^2)}F(k^2;\{m\}),
\ee
where the masses $m_Q, \tilde m_Q, m_T, \tilde m_T$ refer to the top-partners in the $2_{1/6}$ and $1_{2/3}$ representations while $F(k^2;\{m\})$ is a ratio of quartic polynomials in $k$ 
and functions of the masses $m_4, \tilde m_4, m_1,\tilde m_1$ introduced at the end of Sec.~\ref{sec:model}. Due to the large degeneracy between $m_4$ and $m_1$ and between $\tilde m_4$ and $\tilde m_1$, $F(k^2;\{m\})$ can be well approximated by $F(k^2;\{m\}) \simeq 1$ for any value of $k$. As such, the dependence of $m_3^2$ on the top-partner masses can be recasted in a simple analytic expression obtained from the integral in Eq.~(\ref{m3}) and only through $m_Q, \tilde m_Q, m_T, \tilde m_T$. Such a dependence further simplifies when a given hierarchy is established among the fermion masses. In that case, we expect $m_3^2$ to be approximated by
\be\label{estimate-m3}
m_3^2\approx \frac{N_c \, y_t^2}{8\pi^2} \frac{\bar\zeta_t(1+t_\beta^2)}{(1+t_\beta \bar\zeta_t)^2} \frac{m_{l_1}^2 m_{l_2}^2}{ (m_{l_1}^2 - m_{l_2}^2)} \log(\frac{m_{l_1}^2}{m_{l_2}^2}),
\ee
where $m_{l_{1,2}}$ are the masses of the lightest and next-to-lightest top-partners, respectively, among $m_Q, \tilde m_Q$, $m_T, \tilde m_T$. Notice that between the two lightest states there is always a fermion in the $2_{1/6}$ and one in the $1_{2/3}$. A comparison with the numerical computations is done in the left panel of Fig. \ref{fig:scalingMH}. By requiring the next-to-lightest top partners to be much lighter than $m_{l_3}$ we recover the expected behaviour, while the estimate in Eq.~\eqref{estimate-m3} is violated as soon as the condition $m_{l_2}\ll m_{l_3}$ is not satisfied, as the full dependence on the four top-partner masses becomes important in the determination of $m_3^2$. Notice also that we recover the generic result that the terms proportional to quartic powers of the elementary composite mixings are UV finite and calculable even with just two top-partners. 

Finally, in order to make a comparison with previous studies, we plot in Fig. \ref{fig:scalingMH} (right) the correlation between the two lightest states in the spectrum of the top-partners, respectively,  in the $2_{1/6}$ and $1_{2/3}$ representations and we contrast it with the prediction of minimal CHMs based on the coset $\rm SO(5)/SO(4)$. In this case, the correlation is usually constrained by the mass of the SM-like Higgs as the aforementioned top-partners dominantly contribute to $m_h$ (shown with a red line). In the present scenario, however, the presence of a non-vanishing $\tan\beta$ allows for a larger region of parameter space.

\subsection{Scalar mixing and mass splittings}

\begin{figure}[t]
\begin{center}
\includegraphics[width=0.45\textwidth]{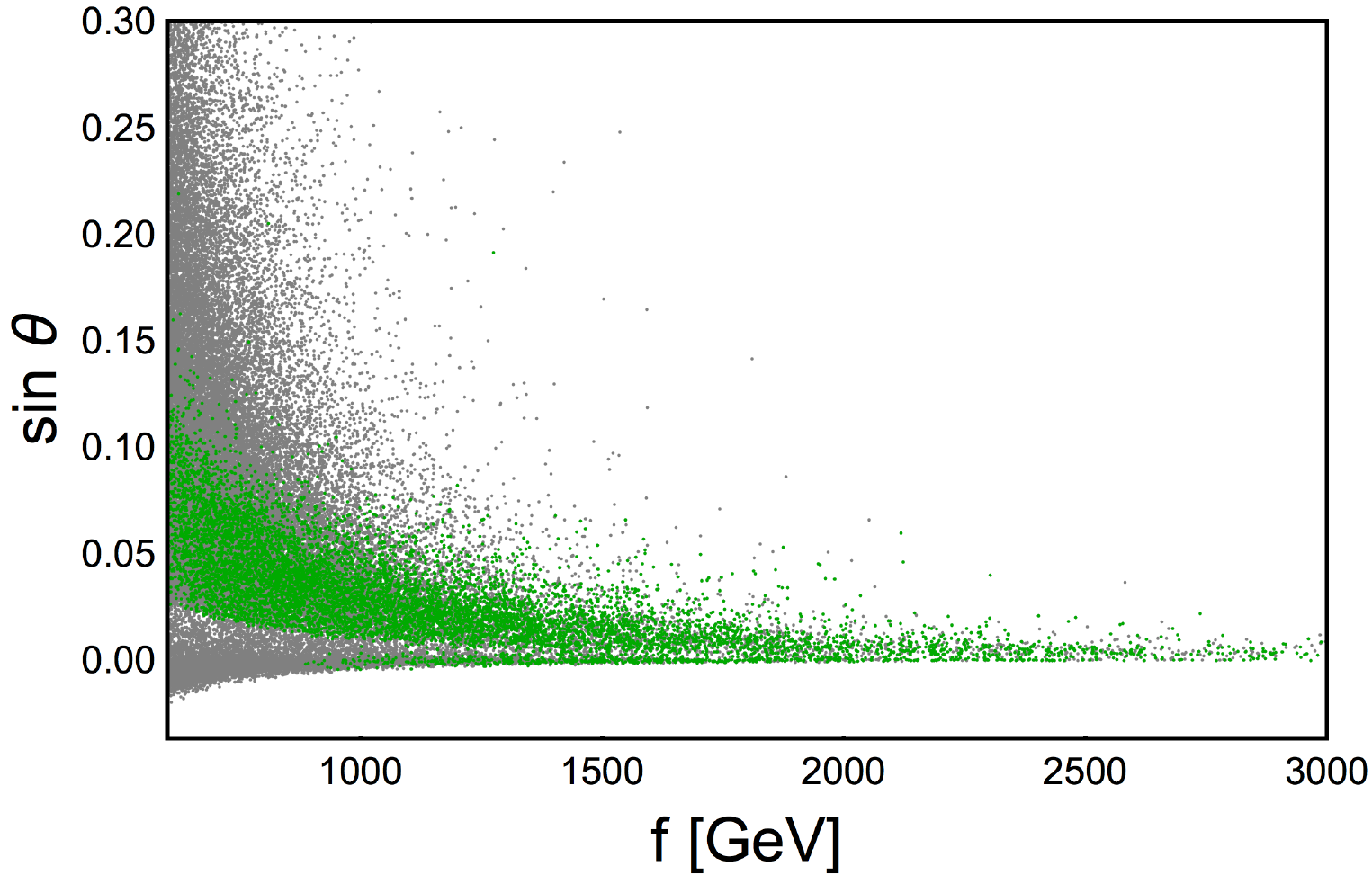}
\includegraphics[width=0.45\textwidth]{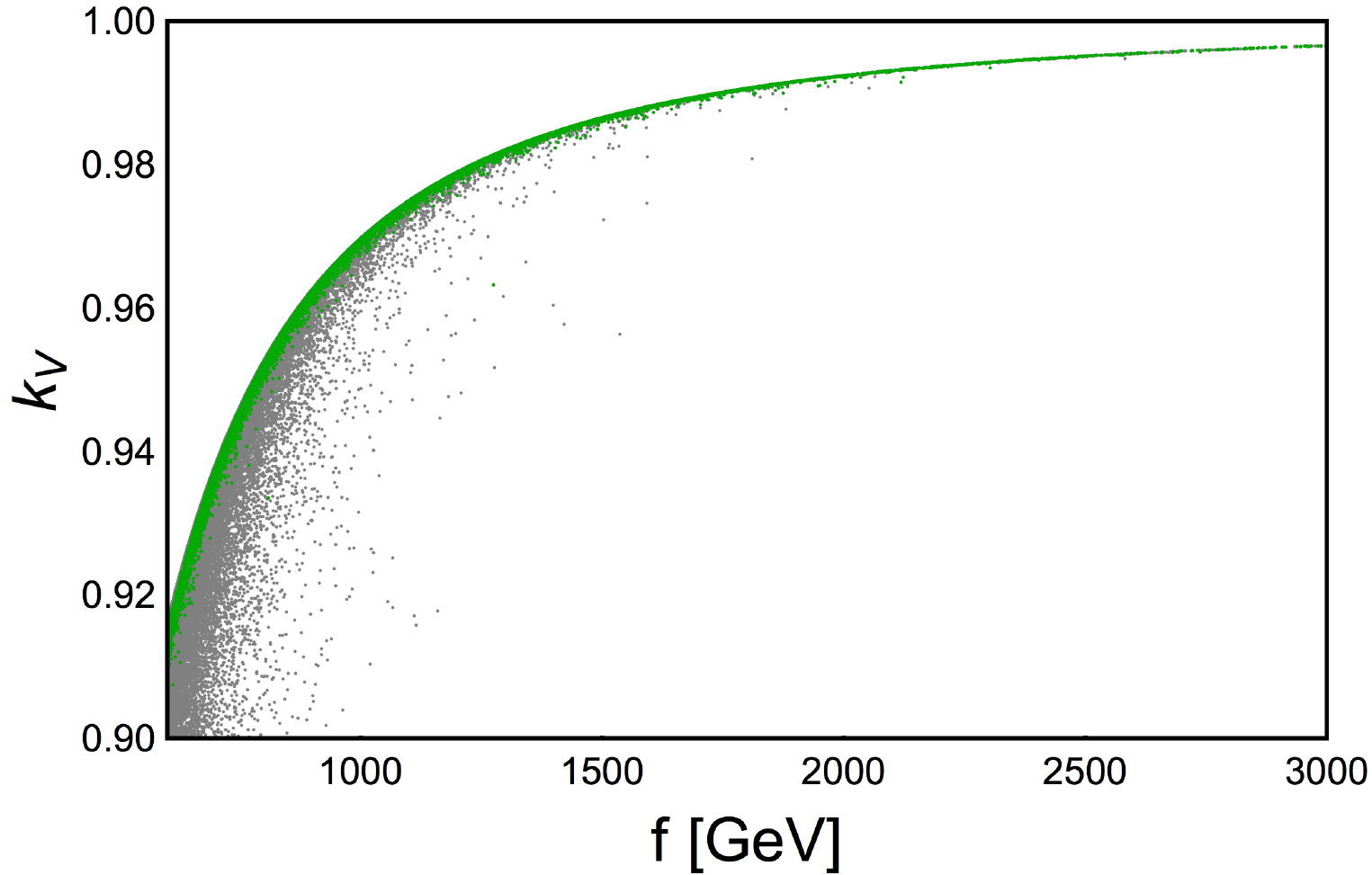}
\caption{ Here, we show $\sin \theta$ (left) and $k_V$ (right) as a function of $f$. The green(gray) points (do not) satisfy the constraints from current indirect and direct Higgs searches. \label{fig:scalarmixing}}
\end{center}
\end{figure}
\begin{figure}[t]
\begin{center}
\includegraphics[width=0.45\textwidth]{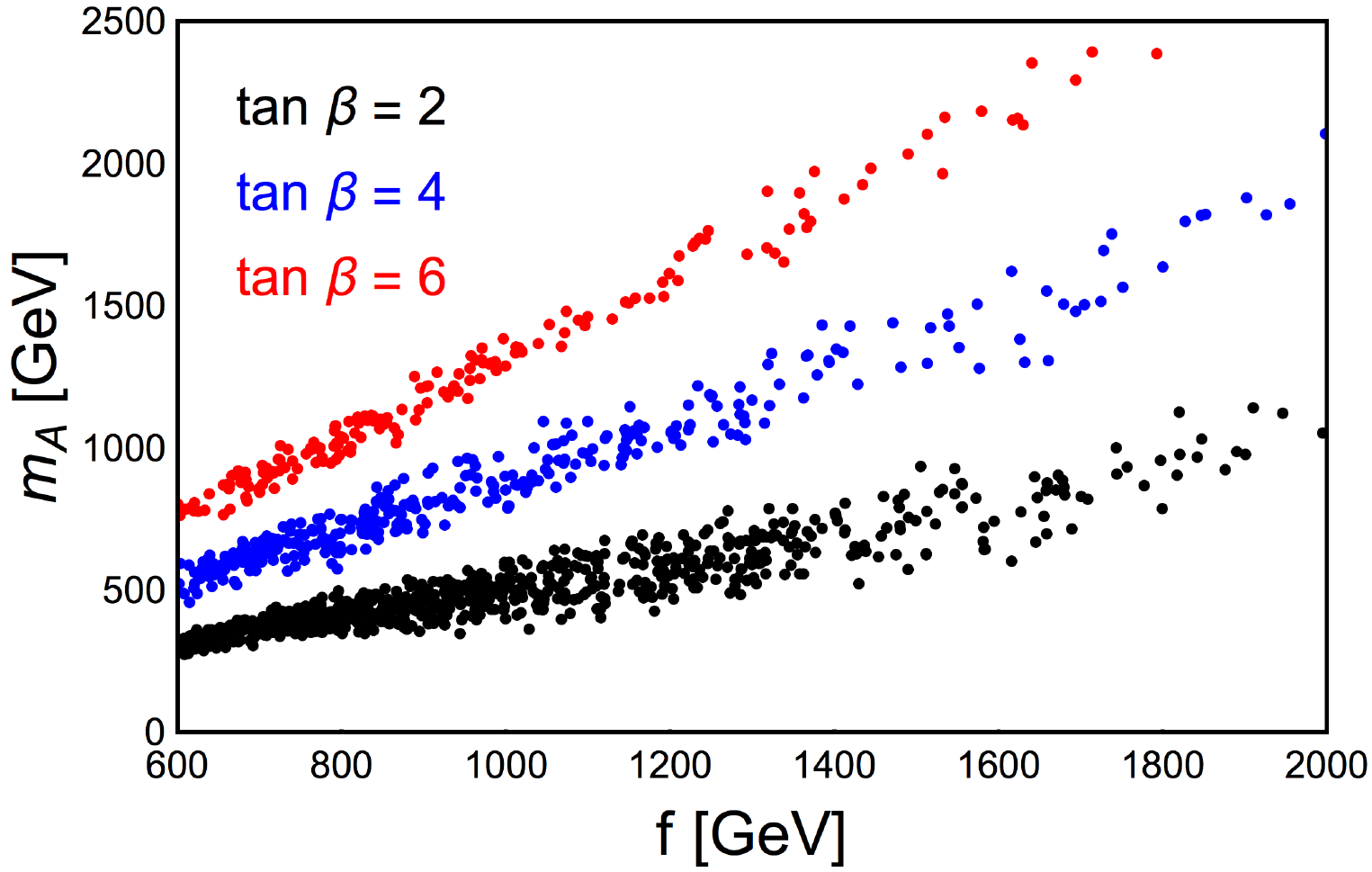}
\includegraphics[width=0.45\textwidth]{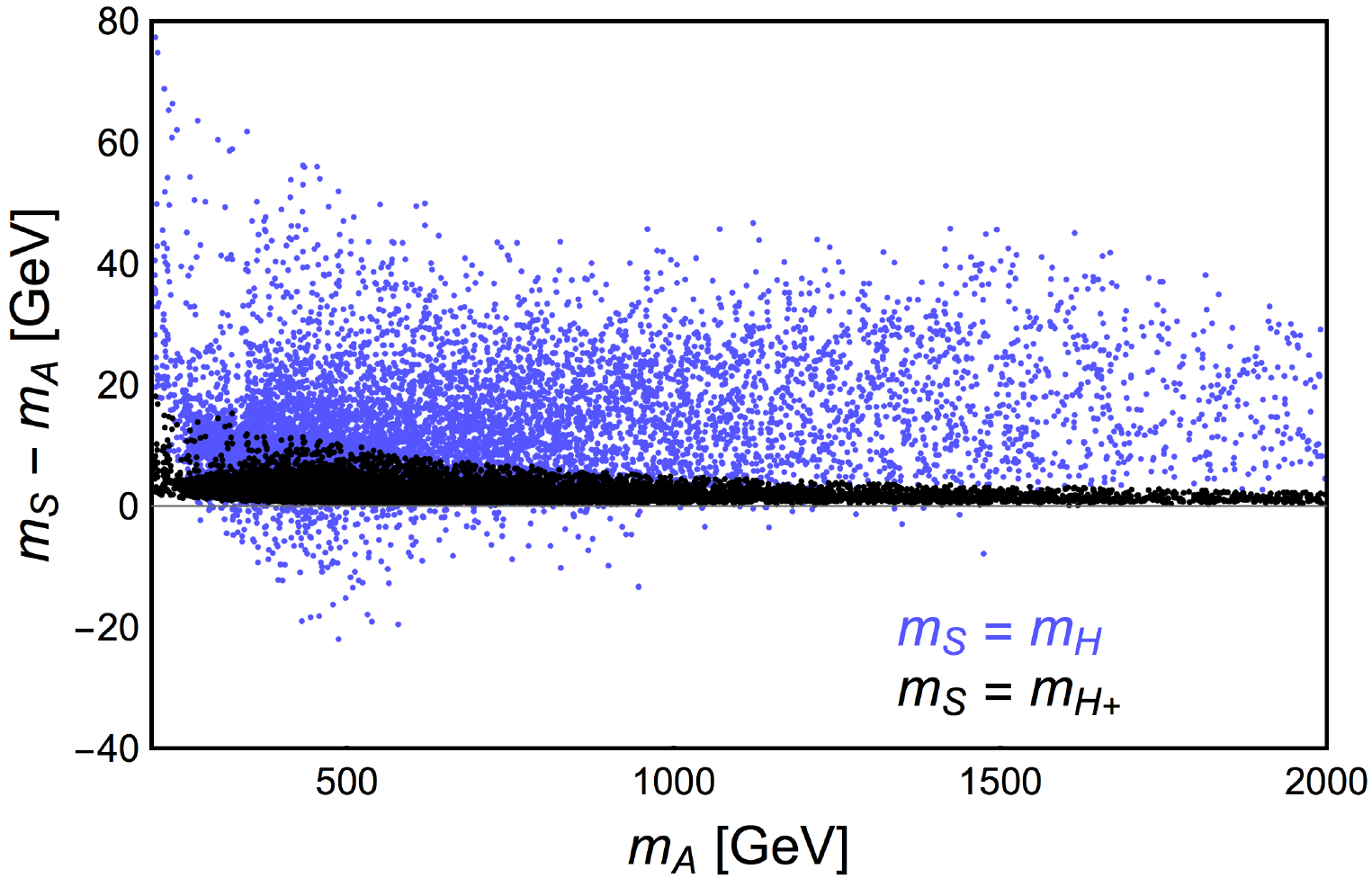}
\caption{\label{figure:splittings}(Left) The CP-odd scalar mass $m_A$ as a function of $f$ for three values of $\tan \beta$. (Right) Mass splittings $m_H - m_A$ (blue) and $m_{H^+} - m_A$ (black) versus $m_A$. \label{fig:spectrum}}
\end{center}
\end{figure}

We now proceed to show some phenomenological results of relevance to Higgs physics at the LHC, concerning both precision measurements of the SM-like Higgs boson and masses, coupling plus decay properties of its C2HDM companions. However, in order to do so, we first ought to extract the viable regions of C2HDM parameter space following the latest experimental constraints.

The points generated from the scan are tested against void experimental searches for extra Higgs bosons, through HiggsBounds \cite{Bechtle:2013wla}, and signal strength measurements of the discovered Higgs state, via HiggsSignals \cite{Bechtle:2013xfa}. Acceptable regions of the C2HDM parameter space are determined according to the exclusion limits computed at 95\% Confidence Level (CL) and then further sifted if the corresponding $\chi^2$ lied within $2\sigma$ from the best fit point. 
The aforementioned tools have been fed with the normalised (with respect to the SM values) Higgs couplings to SM fermions and bosons and with the neutral and charged Higgs Branching Ratios (BRs) without SM equivalents. The Higgs production and decay rates are computed including $O(\xi)$ corrections with the Feynman rules listed in Sec. \ref{sec:higgscouplings}. 
For the Higgs couplings and masses, as well as for the top quark mass, we used the numerical values predicted, for each data point, by the strong dynamics. 
For the sake of definiteness, we assumed that the $C_2$ symmetry is broken with the same strength in all the three sectors of the third generation, namely, the top, the bottom and the $\tau$ lepton. 
This simplified assumption implies $\bar \zeta_b = \bar \zeta_\tau = \bar \zeta_t$ or, equivalently, $\zeta_b = \zeta_\tau = \zeta_t$. Even though this particular choice may have an impact on the couplings of the heavy scalars to SM fermions, 
the predictions for the parameters of the Higgs potential remain unaffected since they are sensitive only to the top sector and thus to $\zeta_t$.
Besides the extra scalars discussed above, the spectrum of the model is characterised by vector and fermion resonances. Their impact has been taken into account in the computation of the gauge and fermionic form factors that define the scalar potential.
Direct searches in di-boson final states, see for instance \cite{Aaboud:2017eta}, effectively constrain the masses and couplings of the vector resonances of composite Higgs models. Our scenario can be mapped, for what concerns the heavy spin-1 states, into the model of the $SU(2)$ vector triplet considered in \cite{Aaboud:2017eta}. We will take into account the corresponding exclusion bound at $2\sigma$ level which was obtained by assuming a branching ratio into SM gauge bosons of $\sim 50\%$ and the narrow width approximation.  As such, these bounds are more than conservative since, as shown in \cite{Barducci:2012kk,Accomando:2016mvz}, the decay modes into SM gauge bosons can be suppressed and the total width substantially enlarged as soon as the heavy fermion decay channels open. 
In the sector of the heavy fermion resonances, we will require for the lightest state, which usually corresponds to the exotic resonance $X_{5/3}$, $M_{X_{5/3}} \gtrsim 1$ TeV in order to comply with the exclusion bound at $2\sigma$ level extracted in \cite{Aaboud:2018uek} under the assumption $\textrm{BR}(X_{5/3} \to W t) = 1$. Nevertheless, we expect that this constraint will be relaxed in our model compared to the minimal scenario examined in \cite{Aaboud:2018uek} as the presence of an extra Higgs doublet also allows for the decay mode $X_{5/3} \to H^+ t$.
\\
Hereafter, all the results discussed in the text and figures are taken to satisfy the constraints from direct and indirect searches described above.

The first consequence of these bounds is on the Higgs couplings. In particular we can constrain the mixing angle $\theta$ as depicted in the left panel of Fig. \ref{fig:scalarmixing} in which we show its dependence on the compositeness scale, in agreement with the expectation that $\theta \sim \xi$ for large $f$, see Eq.~(\ref{eq:scalarmixing}). The green points are all those satisfying the bounds discussed above while the gray ones represent those failing these. For smaller $f$, the mixing angle can vary in principle over a wide range of values but is at present bounded by Higgs coupling measurements to be $|\sin\theta|\lesssim 0.15$. (We also notice that, in the C2HDM, $\sin\theta$  is predicted to be predominantly of positive sign.) The shape makes clear that the more aligned with the SM-like Higgs predictions the LHC measurements are, the larger $f$ ought to be. Indeed, for  
 $f \gtrsim 1$ TeV, all the $\sin \theta$ predicted values do pass the indirect and direct constraints.  Conversely, if significant deviations from $\sin\theta=0$ are eventually established, this implies that $f$ might well be at the sub-TeV scale, in turn hinting at the existence of other C2HDM states in the LHC regime. 
The plot on the right of Fig. \ref{fig:scalarmixing} shows, instead, the Higgs coupling modifier $k_V$ introduced in Eq.~(\ref{eq:kappaV}). With respect to the E2HDM, $\kappa_V$ in the C2HDM approaches the alignment limit more slowly, as evident  from the negative ${\cal O}(\xi)$ corrections, as seen in Eq.~(\ref{eq:kappaV}) and exemplified by the upper edge of the distribution presented. However,  $\sin\theta$ also feeds into this distribution, so that its spread seen on the left-hand side of  Fig. \ref{fig:scalarmixing} is responsible for the departures from the $(1-\xi/2)$ behaviour on the right-hand side. 
We finally note that values of $\kappa_V \gtrsim 0.9$ are currently compatible with LHC data at 1$\sigma$  level~\cite{Khachatryan:2016vau}, hence allowing for several C2HDM solutions at small $f$. 
Finally, concerning the ability to distinguish between the C2HDM hypothesis and the E2HDM one, as stressed in \cite{DeCurtis:2016tsm}, once equipped with a measurement of $\kappa_V$, one  can look for differences in the correlation of possible deviations in $\kappa_E^{}$ and $\kappa_D^{}$, where  $E$ and $D$ represent a charged lepton (e.g., a $\tau$) and a down-type quark (e.g., a $b$), respectively.

The size of the mass of the CP-odd scalar is shown in Fig.~\ref{fig:spectrum}(a) for three specific values of $\tan \beta$ and, as expected, grows linearly in $f$. 
Indeed, as shown in Eq.~(\ref{eq:mAmass}), the mass of the pseudoscalar, as well as that of $H^\pm$ and $H$, is controlled by $m_3$ which is not constrained to the EW scale by the minimisation conditions of the potential.
From the same equation it is also possible to extract the dependence of the mass on $\tan \beta$. In particular, as $m_3^2$ grows linearly in $\tan \beta$, one  finds $m_A^2 \propto f^2 (1+\tan^2 \beta)$. 
The splitting between the heavy CP-even state and the CP-odd scalar (or, equivalently, the charged Higgs) is shown in Fig.~\ref{fig:spectrum}(b). The mass difference $m_H - m_A$ spans from $-20$ GeV to 60 GeV while a quite definite prediction exists  for the splitting between $A$ and $H^\pm$, indeed of high degeneracy, since $m_{H^\pm}^2-m_{A}^2=(\lambda_5-\lambda_4)v^2/2$ is mainly controlled by the gauge contribution and scales like $g'^2/(16\pi^2) g_\rho^2$. Hence, the ability to establish $A\to Z^*H$ or $H\to Z^*A$ signals, respectively, at the LHC will be a strong hint towards a C2HDM dynamics for EWSB, especially if accompanied by the absence of $A\to W^{\pm *}H^\mp$ and $ H^\pm \to W^{\pm *} A$ decays. Clearly, also  $H\to W^{\pm *}H^\mp$ or $ H^\pm \to W^{\pm *} H$ decays would  simultaneously be possible in the C2HDM.  

However, similar decay patterns may also emerge in the E2HDM\footnote{Albeit not in its Supersymmetric incarnation \cite{DeCurtis:2018iqd}.}. Notwithstanding this, though, an intriguing situation \cite{DeCurtis:2016tsm} could occur when, e.g.,
in the presence of an established deviation (of, say, a few percents) from the SM prediction for the $hVV$  ($V=W^\pm,Z$)
 coupling, the E2HDM would require
the mixing between the $h$ and $H$ states to be non-zero whereas in the C2HDM compliance with such a measurement could be achieved also for the zero mixing case. Hence, in this situation, the $H\to W^+W^-$ and $ZZ$ decays would be forbidden in the composite case, while still being allowed in the elementary one. (Similarly, Higgs-strahlung and vector-boson-fusion would be nullified in the C2DHM scenario, unlike in the E2HDM, while potentially large differences  would also appear in the case
of gluon-gluon fusion and associated production with $b\bar b$ pairs.) Clearly, also intermediate situations can be
realised. Therefore, a close scrutiny of the aforementioned signatures of a heavy CP-even Higgs boson, $H$, would be a key to assess the viability of either model. Regarding the CP-odd Higgs state, $A$, in the case of non-zero(zero)
mixing in the E2HDM(C2HDM), again, it is the absence of a decay, i.e., $A\to Z^*h$,  in the C2HDM that would distinguish it from the E2HDM.         
In the case of the $H^\pm$ state, a similar role is played by the $H^\pm\to W^{\pm *} h$ decay. Obviously, for  both these states too, intermediate situations are again possible\footnote{As far as $A$ and $H^\pm$ production modes which are accessible at the
LHC, i.e., gluon-gluon fusion and associate production with $b\bar b$  pairs (for the $A$) 
and associated production with $b\bar t$ pairs (for the $H^+$), 
are concerned though, practically no difference appears between the two scenarios \cite{DeCurtis:2016tsm}.},  so that one is eventually forced to also investigate the fermionic decays of heavy Higgs states, chiefly, those into top quarks. (We will dwell further on all this in an upcoming section.)

\subsubsection{Comments on the exact $C_2$ symmetry scenario}
\label{sec:c2symm}
\begin{figure}
\begin{center}
\includegraphics[width=0.45\textwidth]{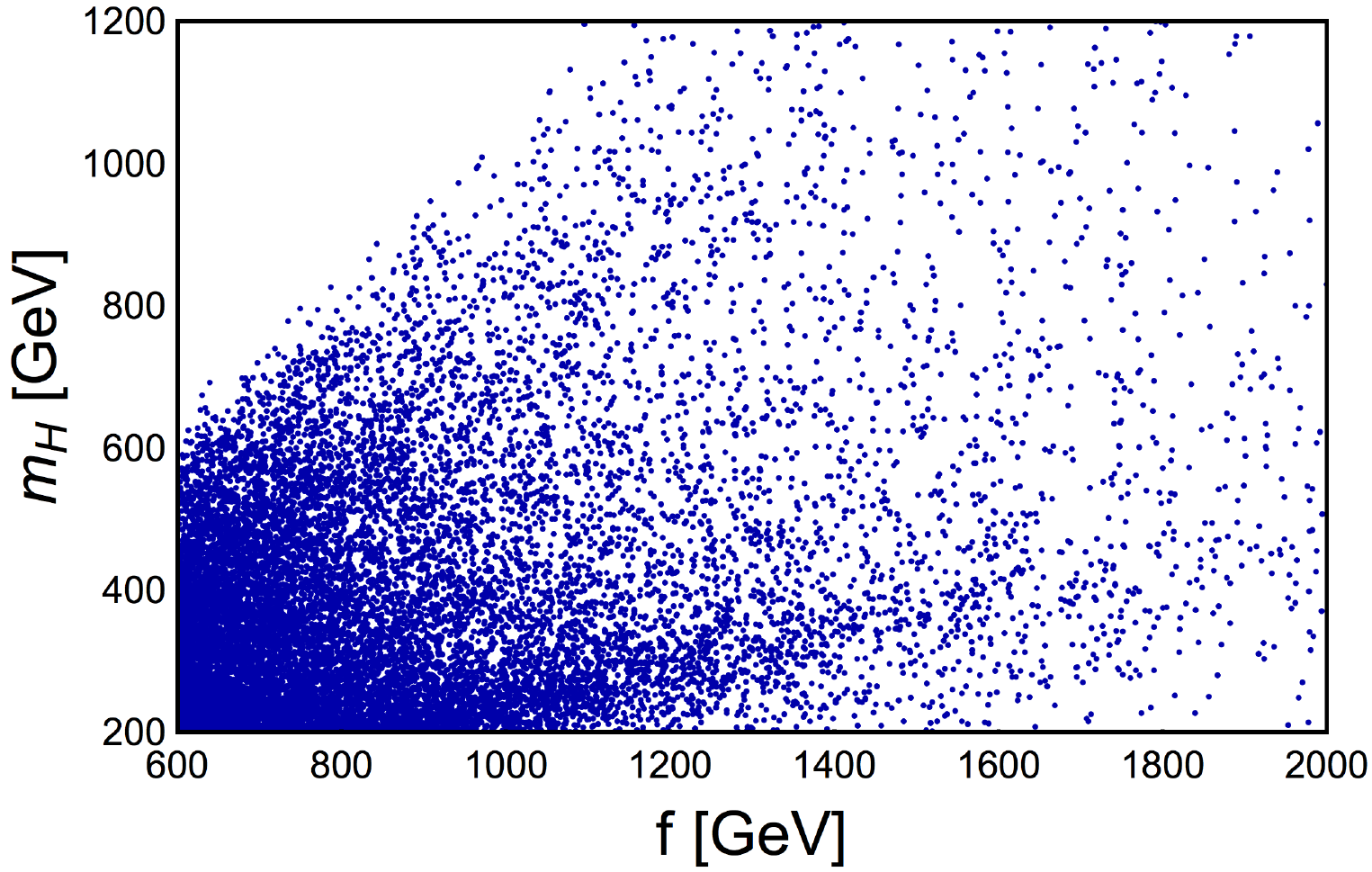}
\includegraphics[width=0.45\textwidth]{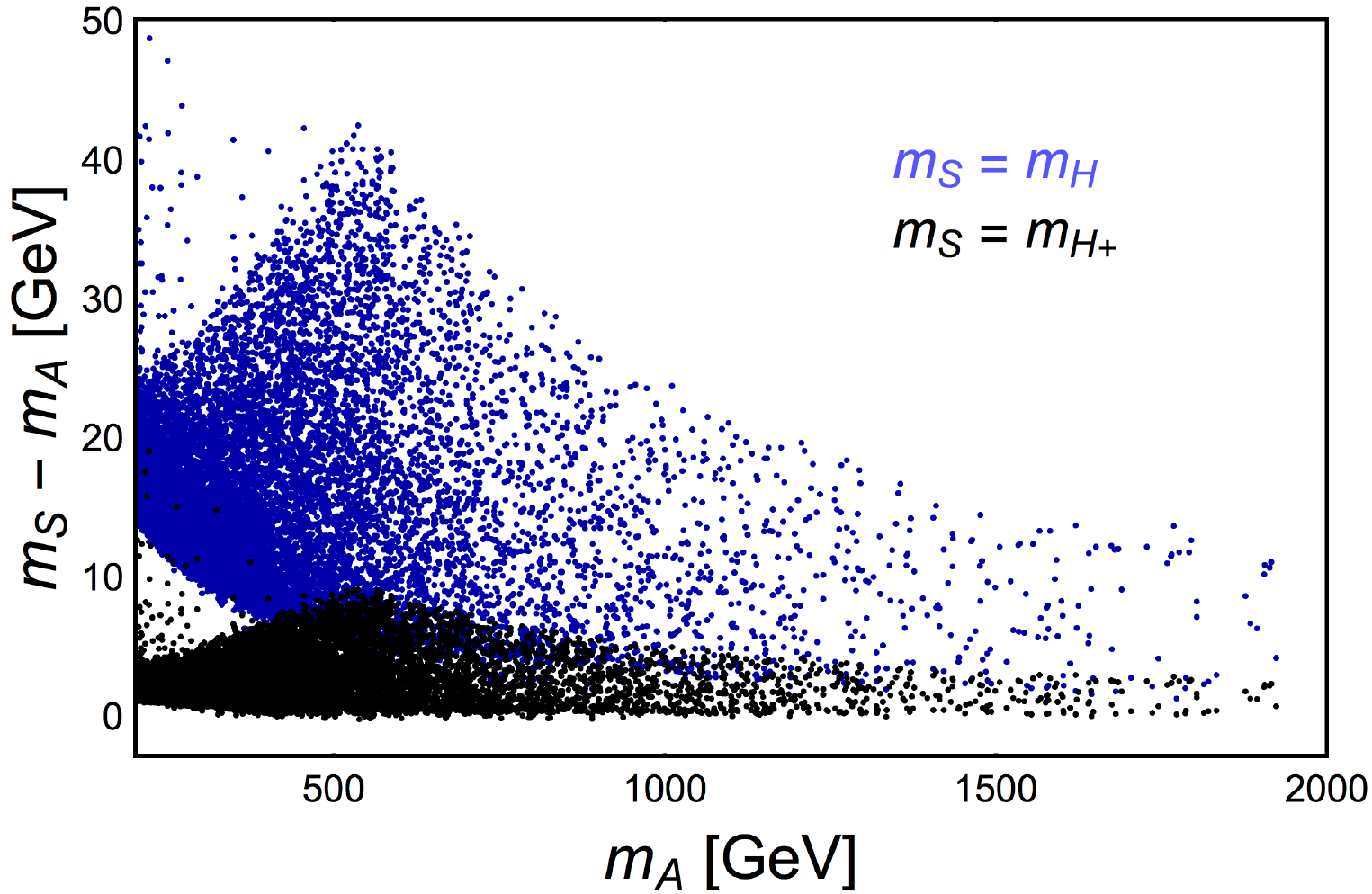}
\caption{\label{fig:C2simm} (a) Correlation between $m_H$ and $f$ and (b) mass splittings in the case of an exact $C_2$ symmetric scenario.}
\end{center}
\end{figure}

An interesting limit of our model is $\bar\zeta_t=0$ ($Y_1 = 0$) which corresponds to a restored $C_2$ symmetry. This scenario realises a composite version of the inert 2HDM. The presence of a $C_2$ symmetry is consistent with the fact that only one Higgs doublet develops a VEV.
By performing a numerical computation of the Higgs potential in the $C_2$ symmetric case we verified that $m_2^2$ gives the mass of the physical components of the second Higgs doublet. We also checked the absence of solutions providing the spontaneous breaking of $C_2$. The predicted value of the mass of the heavy CP-even Higgs is shown in Fig. \ref{fig:C2simm} where we include all the points generated by the scan without implementing direct and indirect experimental constraints. 

\begin{figure}
\begin{center}
\includegraphics[width=0.45\textwidth]{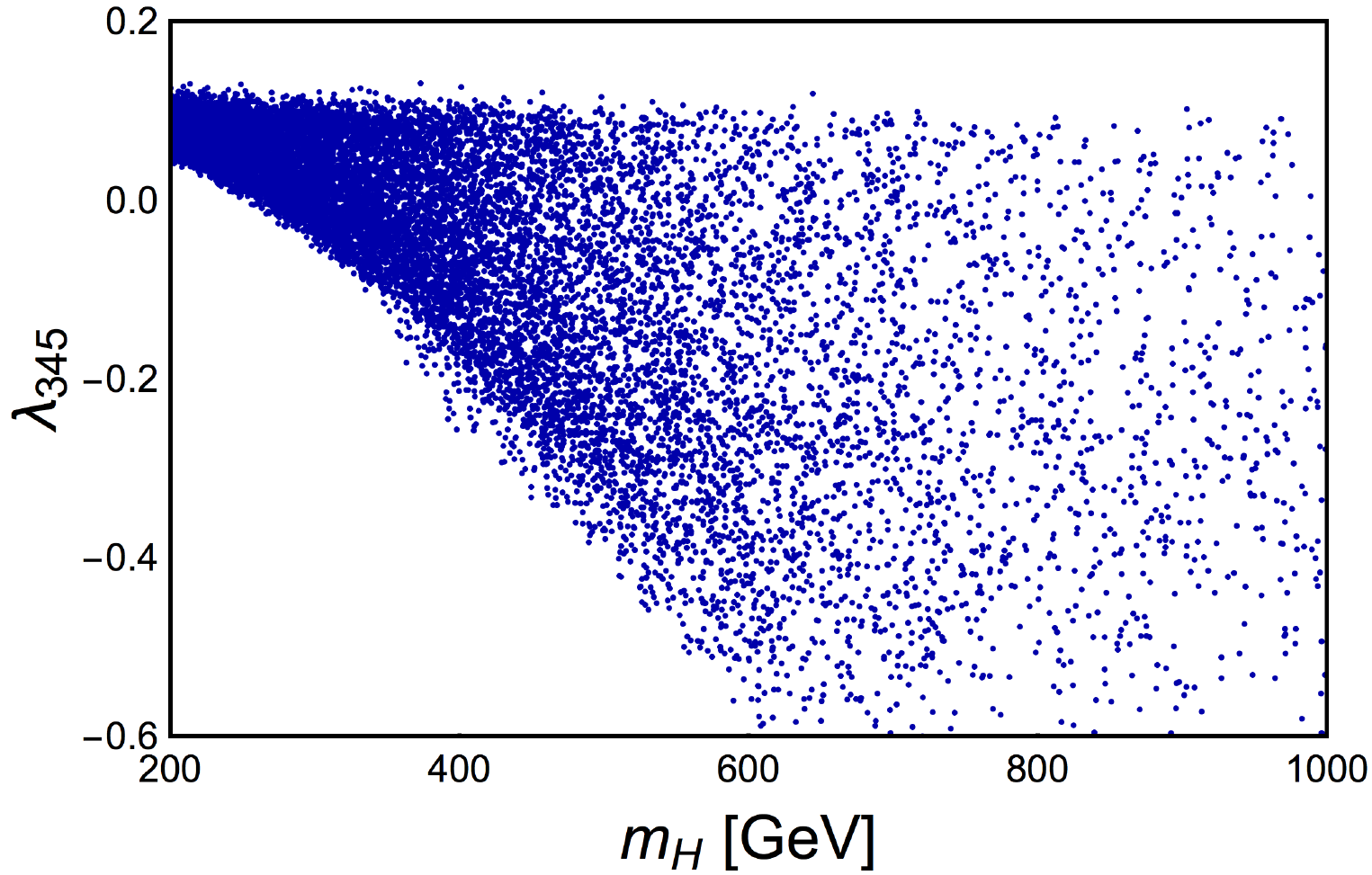}
\caption{\label{fig:l345}  Correlation between the coupling $\lambda_{345}$ and the mass of the inert doublet $m_H$ in the $C_2$ symmetric potential.}
\end{center}
\end{figure}

In the case where $C_2$ is also preserved by lighter quarks and leptons, the neutral component of the second Higgs doublet can be a Dark Matter (DM) candidate. 
For this to happen and also to avoid strong constraints, at least one  neutral component should be  lighter than $H^\pm$ (this is always the case in the parameter space explored). The possibility to have DM as the neutral component of an inert Higgs doublet has been thoroughly discussed in the literature, see for instance \cite{Belyaev:2016lok}. In this context we notice that reproducing the DM relic density $\Omega_{\rm DM}$ requires a specific value of the couplings $\lambda_{hHH,hAA}$ for any mass point. The same couplings are also important for direct detection which occurs via tree level Higgs-exchange and loops of $W^\pm$'s. The tree level contribution is a direct test of the quartic coupling $\lambda_{345}$, which then receives an upper bound from direct detection experiments, $\lambda_{345} \lesssim 1$ for $m_{H,A} \gtrsim 200$ GeV \cite{Belyaev:2016lok}. As we can see from the size of the coupling $\lambda_{345}$ presented in Fig. \ref{fig:l345}, the model could allow for a DM candidate, providing the observed value of the relic abundance, while complying with direct detection bounds, for $m_{H,A} \gtrsim 800$ GeV \cite{Belyaev:2016lok}. 

\subsection{Flavour constraints}

In this model, even though we assume a flavour symmetric composite sector, we find modifications to rare flavour transitions in the SM from the exchange of  pNGBs. 
As already stressed previously, in CHMs, there are several effects in flavour physics, depending on which composite resonance is integrated out at low energy.
In the literature large attention has been given to vector mediators, while in this section we would only consider effects originating from the scalar sector since this is strongly correlated to the phenomenology discussed in this work.

Under the assumption of a flavour symmetric composite sector, the heavy Higgs bosons can only mediate tree level effects in charged current processes and loop effects in the neutral ones. Since, by virtue of the flavour symmetries, the SM-like Higgs and the heavy companions have interactions with the fermions aligned in flavour space, all the flavour constraints are due to a rescaling of the corresponding SM rates. Therefore, the bounds arise because of the relative precision of the SM observables, which roughly ranges from 1 to 10\% accuracy. We review in turn the most stringent ones for our construction. 
The relevant interactions have been detailed in Eq.~(\ref{eq:Hff}) and involve, in particular, the couplings of the charged Higgs with SM quarks and leptons which are proportional to the $\xi_A^f$ parameter given in Eq.~(\ref{eq:fermion-coupling-0}).

\begin{figure}
\begin{center}
\includegraphics[width=0.45\textwidth]{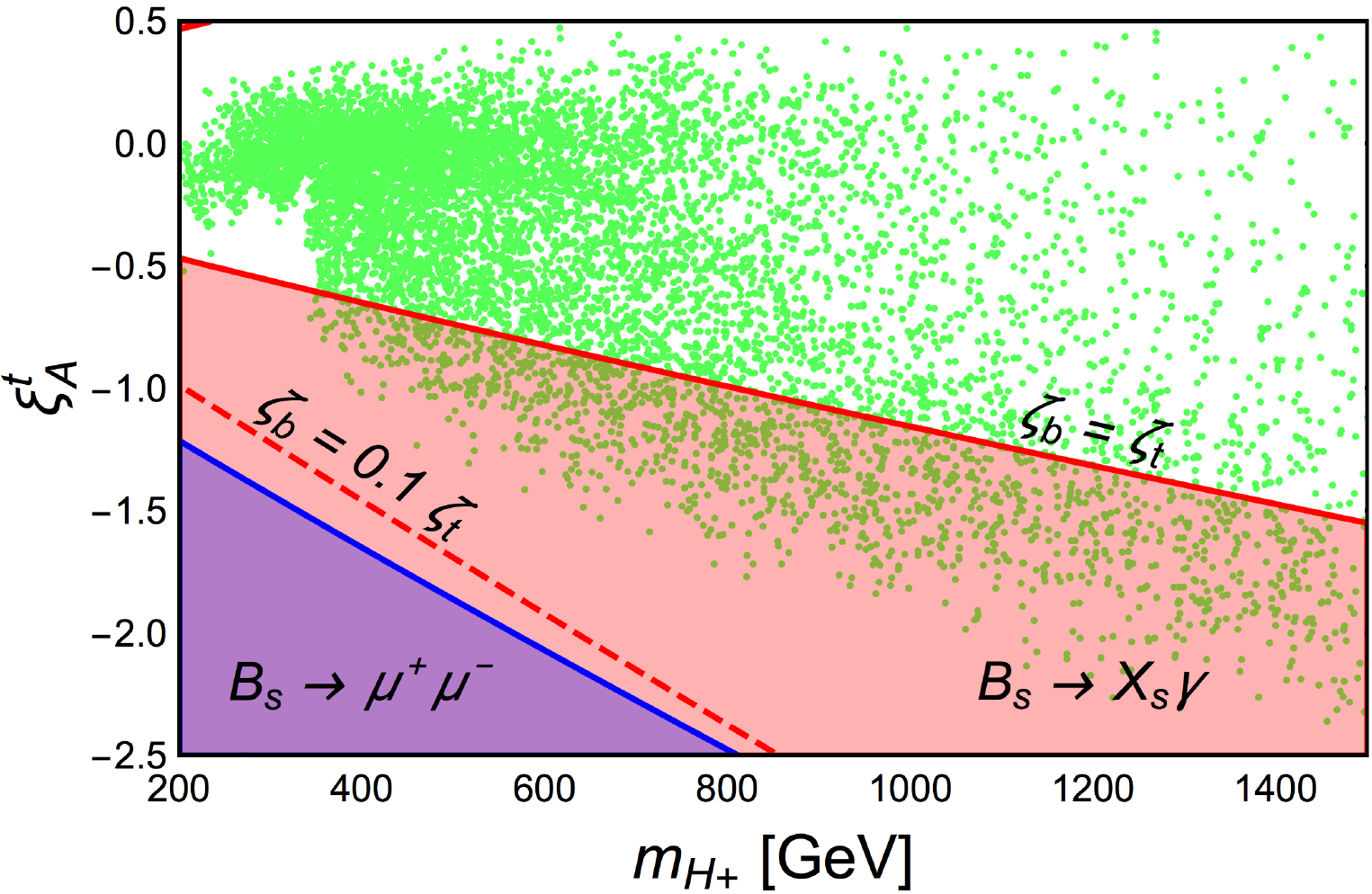}
\caption{\label{fig:flavour} Correlation between the mass of the charged Higgs $m_{H^+}$ and its coupling to the top quark, $\xi_{H^+}^t = \xi_A^t$. Green points are allowed by current direct and indirect searches at the LHC. Red and purple shaded regions are excluded at $2\sigma$ level by measurements of the $B \rightarrow X_s \gamma$ rate under the assumption $\zeta_b = \zeta_t$ and of $B_s \rightarrow \mu^+ \mu^-$ transitions, respectively.}
\end{center}
\end{figure}

\begin{itemize}
\item Meson decay $M\to \ell \nu$. The charged Higgs $H^\pm$ can mediate charged current processes aligned in flavour space alongside the $W^\pm$ mediated decay of pseudoscalar mesons. 
The expectation value of the scalar operator $\bar u d$, between the vacuum and the $M$ (mesonic) state, can be related to that of the divergence of the axial current $\bar u \gamma_\mu \gamma_5 d$. By means of this relation, one can then  write the relative variation of the BR  of the meson to leptons simply as
\be
\frac{\delta \mathrm{BR}(M \to \ell \nu)}{\mathrm{BR}(M \to \ell \nu)} \approx 2 \frac{m_u \xi_A^u - m_d \xi^d_A}{m_u+ m_d}\xi_A^l \, \frac{m_M^2}{m_{H^+}^2}.
\ee
For example, in the case of the $B$ meson decay $B\to \tau \nu$, the deviation is proportional to $\xi_A^d \xi_A^l m_B^2/m_{H^+}^2$. This shows that tree level charged current processes are sensitive (mainly) to composite parameters that enter the expressions for $\xi^{d,l}_A$ and $f$, since $m_{H^+}$ is linear in $f$. Notice, however, that the $\xi^{d,l}_A$'s are not directly related to the Higgs potential, since they originate from the down sector which contributes negligibly to $v$ and $m_h$, so they can be taken small enough to reduce effects in the charged currents. Furthermore, 
under the assumption of a flavour symmetric sector,  $D\to \tau \nu$ is sensitive to $\xi^u_A$ and therefore to the parameters of the Higgs potential but still suppressed by the small ratio $m_{D}^2/m_{H^+}^2$.

\item Transition $b\to s \gamma$. Among the best measured quantities are $B\to X_s \gamma$ transitions. Differently from tree level $\Delta F=1$ transitions, here, the relevant couplings are the ones of $H^\pm$ to the top, given that the main contributions arise from the Wilson coefficients  $C_{7,8}$ 
 in the weak Hamiltonian which are generated by box diagrams with two $H^\pm$'s. From loops of heavy (pseudo)scalars we get
\begin{align}
C_i = \frac{(\xi_A^t)^2}{3} G_i^1 \left(\frac{m_t^2}{m_{H^+}^2} \right) + \xi_A^t \xi_A^b G_i^2 \left(\frac{m_t^2}{m_{H^+}^2} \right) 
\end{align}
with
\begin{align}
G_7^1(x) &= \frac{y(7-5y-8y^2)}{24(y-1)^3} + \frac{y^2(3y-2)}{4(y-1)^4} \log x , \quad G_7^2(x) = \frac{y(3-5y)}{12(y-1)^2} + \frac{y(3y-2)}{6(y-1)^3} \log x \,, \nonumber \\
G_8^1(x) &= \frac{y(2+5y-y^2)}{8(y-1)^3} - \frac{3y^2}{4(y-1)^4} \log x , \quad G_8^2(x) = \frac{y(3-y)}{4(y-1)^2} - \frac{y}{2(y-1)^3} \log x \,.
\end{align}

\item Transition $B_s \to \mu^+\mu^-$. In the SM the leading contributions arise from $Z$ penguin diagrams contributing to the Wilson coefficient $C_{10}$. In this model we predict at the scale of the resonances that 
\begin{align}
C_{10} = (\xi_A^t)^2 \frac{m_t^2}{8} \left[ \frac{1}{m_{H^+}^2 - m_t^2} + \frac{m_{H^+}^2}{(m_{H^+}^2 - m_t^2)}  \log \frac{m_t^2}{m_{H^+}^2}   \right].
\end{align}
\end{itemize}
We finally depict in Fig.~\ref{fig:flavour} the impact of the flavour bounds discussed above on the parameter space of the C2HDM. The constraints at $2 \sigma$ level have been extracted from \cite{Enomoto:2015wbn,Misiak:2015xwa} and shown by the red and purple shaded regions.
The constraint from the measurement of the $M \rightarrow l \nu$ meson decay is usually important only for small charged Higgs masses and/or large couplings and, as such, does not affect the range of parameters discussed here even though $\zeta_\tau$ is taken as large as $\zeta_t$. The bound from the $B\to X_s \gamma$ transition depends on the interplay between the top and the bottom contributions and on the relative size of the $\xi_A^t$ and $\xi_A^b$ couplings. In the scenario discussed here, in which $C_2$ is broken equally in the top and bottom quark sectors ($\xi_A^t = \xi_A^b$), 
the corresponding constraint is shown in Fig.~\ref{fig:flavour} by the red shaded excluded region. 
Needless to say, in different scenarios realising $\zeta_b < \zeta_t$, the bound from the $b \to s \gamma$ transition can be greatly relaxed so that all the points survive the constraint. 
For example, the same exclusion bound computed for $\zeta_b < 0.1 \zeta_t$ lies well below the distribution of points (the latter does not sensibly change if the constraints from HiggsBounds and HiggsSignals are enforced for $\zeta_b = 0.1 \zeta_t$).
The bound from the measurement of the $B_s \to \mu^+\mu^-$ transition is, instead, more robust as it only depends on $\xi_A^t$ and not on the particular realisation of $C_2$ breaking in the bottom quark sector. Moreover, the corresponding excluded region does not overlap with the distribution of points.

\subsection{Phenomenology of the heavy scalars}

\begin{figure}[t]
\begin{center}
\includegraphics[width=0.45\textwidth]{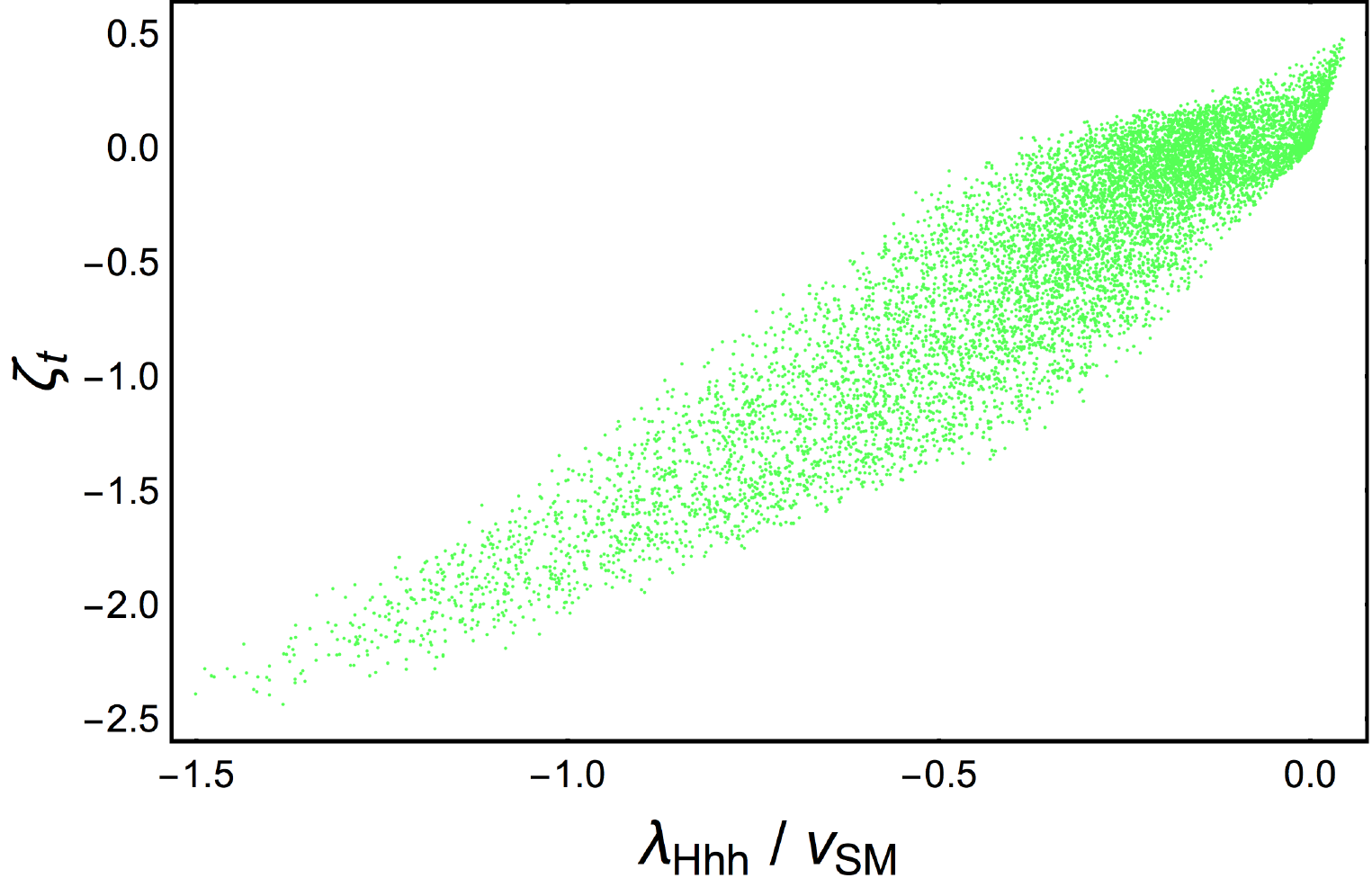}
\includegraphics[width=0.45\textwidth]{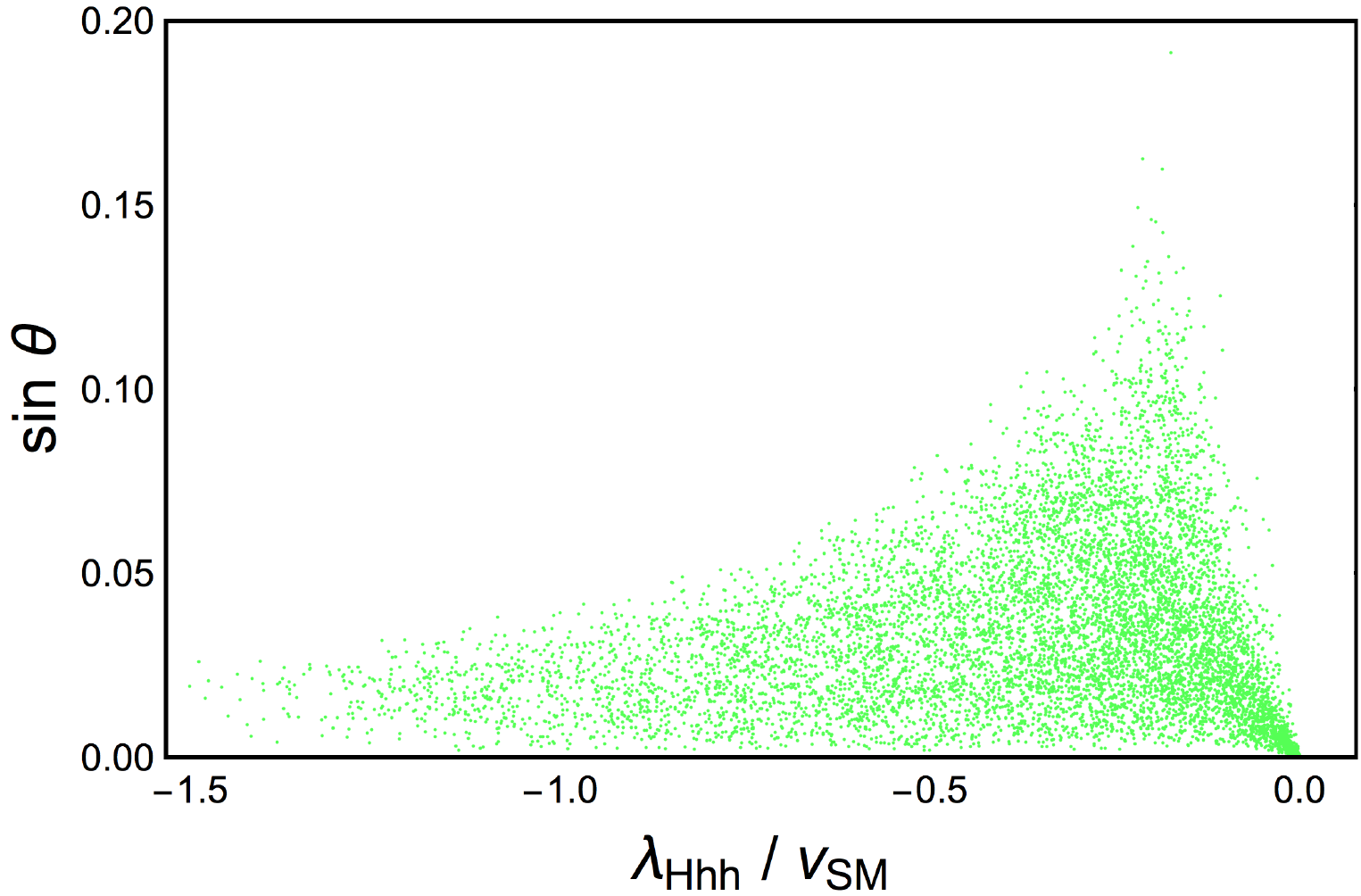}
\caption{Correlation between $\zeta_t$ and $\lambda_{Hhh}$ (left) and between $\sin \theta$ and $\lambda_{Hhh}$ (right). \label{fig:couplings-correl}}
\end{center}
\end{figure}

The LHC phenomenology of the second Higgs doublet is determined by the couplings to fermions given in Eq.~\eqref{eq:Hff} and by the trilinear couplings in Eq.~\eqref{eq:HHH} setting the decay to di-Higgs final states. As repeatedly stressed, a key feature of the C2HDM is the strong correlation between the former and the latter, since they are both generated by the breaking of the NGB shift symmetry. This correlation is exemplified in Fig.~\ref{fig:couplings-correl} where we show the parameters $\zeta_t$, controlling the couplings of the heavy scalars to the top quark, $\lambda_{Hhh}$, which is responsible for the decay channel $H \rightarrow h h$,  and $\sin \theta$, the sine of the mixing angle of the CP-even scalars which sets the size of the $H$ decay into the SM gauge bosons.  
From our numerical studies, as already mentioned, we have verified that we are always in the region where $m_H > m_h$, therefore we focus on the following three scenarios, each addressing a distinctive C2HDM phenomenology for the three heavy physical states of its spectrum, i.e., $H,A$ and $H^\pm$, respectively. 

 \begin{figure}
\centering
\includegraphics[scale=0.5]{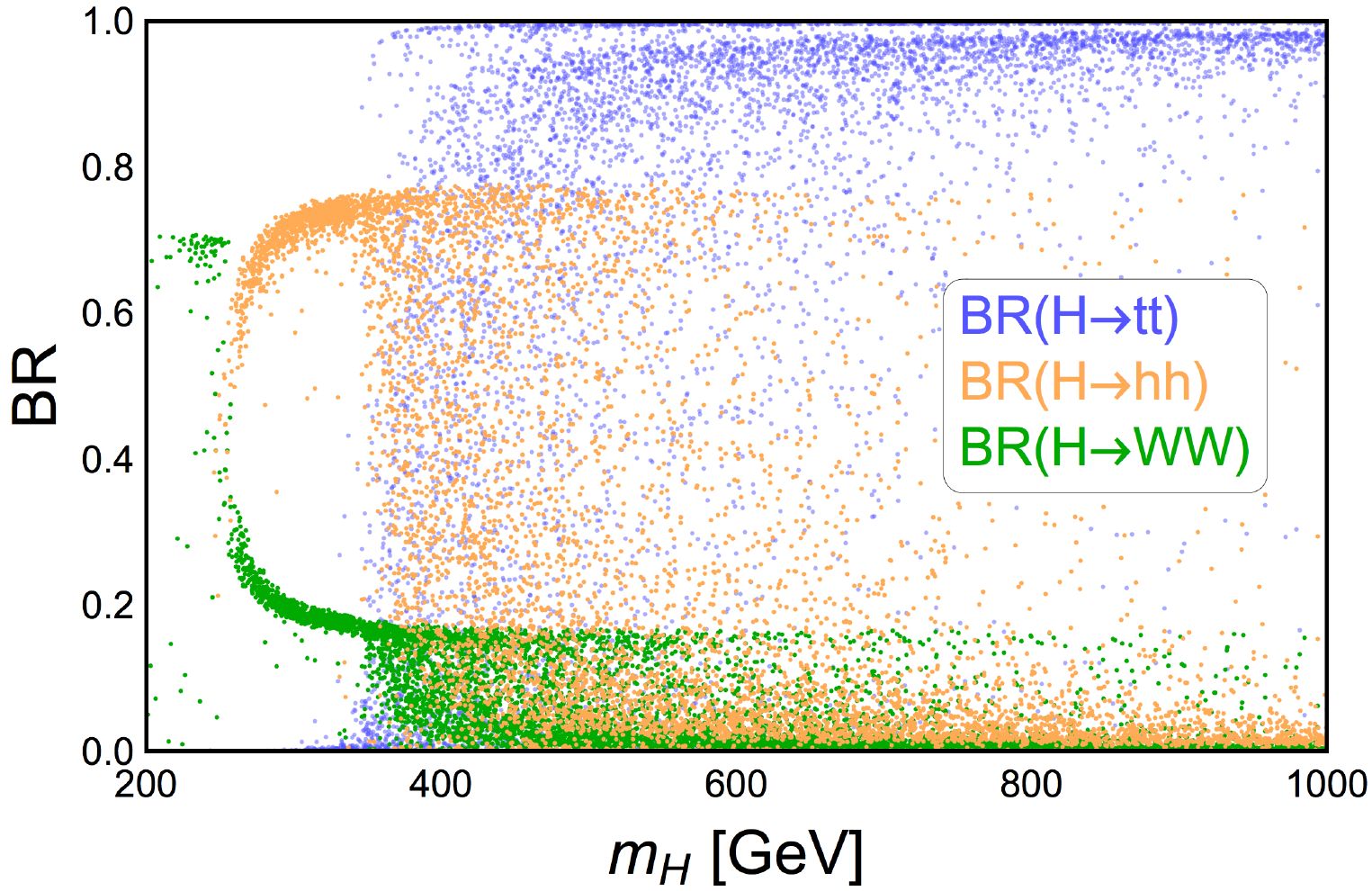}
\includegraphics[scale=0.5]{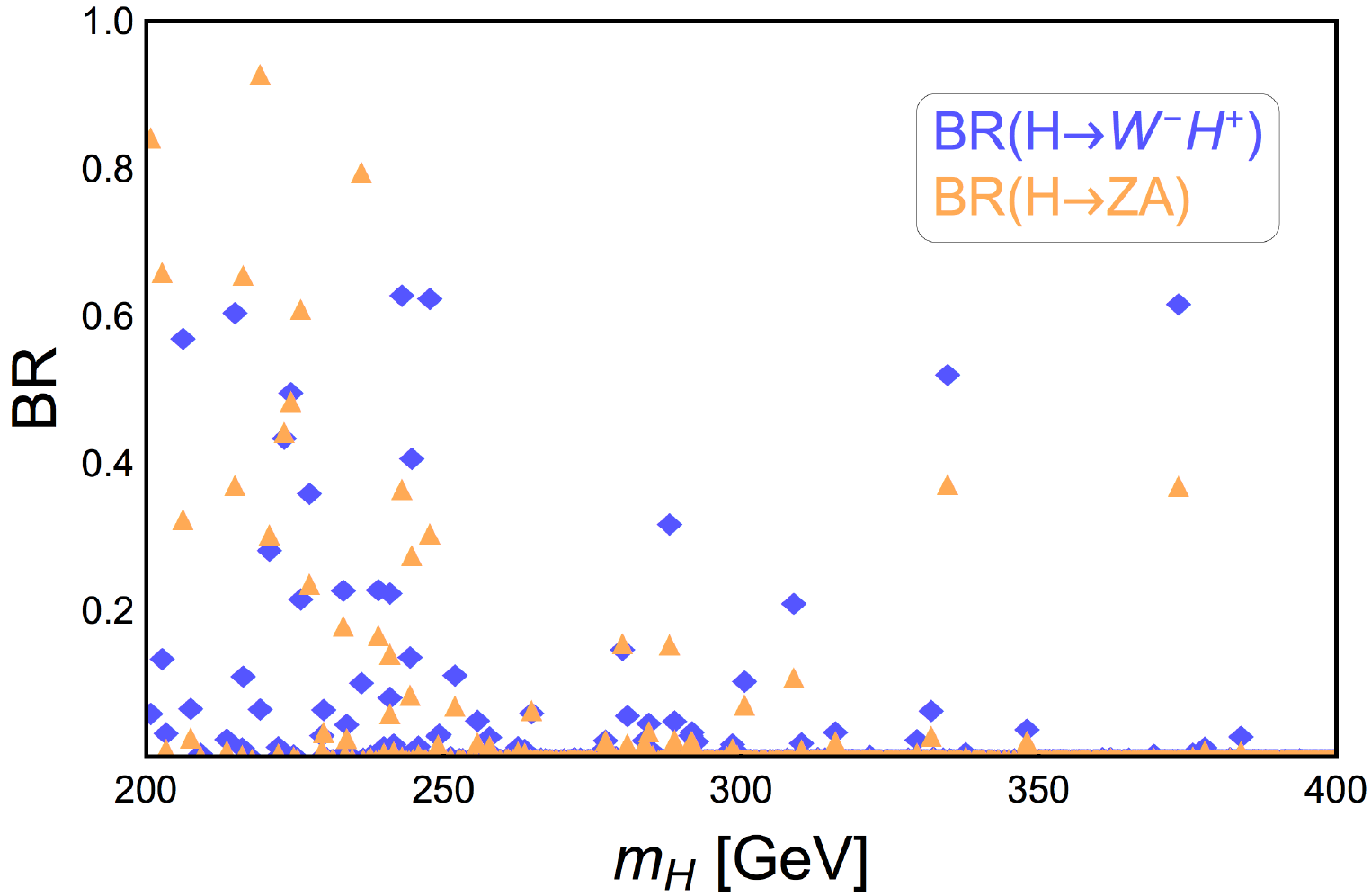}
\caption{(Left) The BRs of the $H$ state in the $t \bar t$ (blue), $hh$ (orange) and $W^+W^-$ (green) channels. (Right) The BRs of $H$ into $W^{- *}H^+$ (blue) and $Z^* A$ (orange). For readability, we refrain from presenting here  BR$(H\to ZZ)$: in virtue of Eq. (\ref{H-BR}) this is about half that of BR$(H\to W^+W^-)$. 
\label{fig:BRs-H}}
\end{figure}

\begin{itemize}
\item Scalar $H$. Our numerical analysis allows us to fully compute the relevant observables for the $H$ state and, amongst these,  it is instructive to study first  the interplay between SM-like decays to di-boson final states and those with third generation fermions.  While it is true that  $H$ couplings to fermions have several contributions (see Eq.~\eqref{eq:fermion-coupling-0}), for $f\gg v_\textrm{SM}$, the leading contribution to the $H\to t\bar{t}$ decay rate is given by $\zeta_t$, with 
\be 
\Gamma(H\to t\bar t) \approx \frac{3y_t^2}{16\pi}|\zeta_t|^2 m_H,
\end{equation}
which represents the main decay mode above the $t\bar t$ threshold. The other important decay channel is $H\to hh$ that, when the $t \bar t$ mode is kinematically closed, can reach a BR of $80\%$, with the remaining decay space saturated by  $H\to Z Z, W^+ W^-$. The corresponding BRs are shown in Fig.~\ref{fig:BRs-H} (left).
In the $H \to hh$ case, there are potentially many contributions due to quartic couplings in the scalar potential and higher dimensional operators from the  strong sector Lagrangian. 
The leading contributions are rather simple, though, since, for $f$ larger than $v_\textrm{SM}$, $\lambda_{Hhh} \simeq 3/2 v_\textrm{SM} \, \Lambda_6$, where $\Lambda_6$ is the scalar quartic coupling previously defined in the Higgs basis. In this regime we find
\begin{align}\label{H-BR}
\Gamma(H \to hh) &\approx \frac{9}{32\pi m_H}(v_\textrm{SM}^2 \Lambda_6^2) \,, \nonumber \\
\Gamma(H \to W^+W^-) &\approx 2\Gamma(H \to ZZ)\approx \frac{1}{16\pi m_H} \sin^2 \theta \frac{m_H^4}{v_\textrm{SM}^2} \,,
\end{align}
hence, the study of these three decay topologies would enable one direct access to three key parameters of the C2HDM, i.e., $\zeta_t$, $\theta$ and $\lambda_{Hhh}$, as well as to attest their correlations which are shown in Fig.~\ref{fig:couplings-correl}. 

Furthermore, according to the predicted mass splittings shown in  
Fig.~\ref{figure:splittings}, also non-SM-like, i.e., 
$H\to AZ^*$ and $H\to H^+W^{-*}$ (off-shell) decays are possible but limited to $m_H \lesssim 400$ GeV, in which regime the mass splitting between $H$ and $A$ or $H^+$ is larger. The corresponding BRs are shown in Fig.~\ref{fig:BRs-H} (right).
These two decay modes are controlled by $\cos^2 \theta$ which delineates the region of the parameter space where the two decay modes can be sizeable, namely, small $\theta$, that also closes the other SM-like decay modes. 

The $H$ production cross section is simply dominated by gluon fusion, where $H$ is produced via  its coupling to the top. So it is simply obtained from  SM gluon fusion Higgs production (calculated at $m_H$) rescaled by $\zeta_t^2$.

We conclude this part by showing some prospects for  $H$ phenomenology at the forthcoming runs of the LHC. Here we focus on the $H \to hh$ channel and on its $bb\gamma \gamma$ final state which has been recently addressed in \cite{CMS:2017ihs} by the CMS collaboration using data from the LHC at the collider energy $\sqrt{s} = 13$ TeV and (integrated) luminosity $L = 35.9$ fb$^{-1}$.
In particular, we illustrate in Fig.~\ref{fig:bbgaga} the interplay between direct and indirect searches and the ability of the High-Luminosity LHC
(HL-LHC) and High-Energy LHC (HE-LHC) upgrades to investigate the $gg\to H\to hh\to b\bar b\gamma\gamma$ signal over regions of the C2HDM parameter space projected onto the $(m_H,\zeta_t)$ plane, even when no deviations are visible in the coupling strength modifiers $\kappa_i$ of the SM-like Higgs state $h$ (red points) at the end of Run 3 at $L=300$ fb$^{-1}$ and at the end of the HL-LHC at $L=3000$ fb$^{-1}$. The compliance with the coupling modifiers is achieved by asking that $|1 - \kappa_i|$ is less than the uncertainty quoted in Ref.~\cite{CMS:2013xfa}, where $i=VV,\gamma\gamma$ and $gg$, while the 95\% CL exclusion limits are extracted from the $gg\to H\to hh\to b\bar b\gamma\gamma$ search by adopting the sensitivity projections of \cite{CMS:2017ihs} (orange points).
The HE-LHC, assuming $\sqrt s=27$ TeV and $ L=15$ ab$^{-1}$, will improve the reach in the $H$ high mass region up to 1.3 TeV by studying the process $gg\to H\to hh\to b\bar b\gamma\gamma$ (yellow points). 

The dashed line in Fig.~\ref{fig:bbgaga} delimits the excluded region from the measurement of the $B_s \to X_s \gamma$ transition under the assumption $\zeta_b = \zeta_t$ which has been employed to compute the Higgs branching ratios discussed in this section. We stress again that other scenarios with $\zeta_b < \zeta_t$ would significantly relax the flavour bound from $B_s \to X_s \gamma$ measurements while will not change, at the same time, the distribution of points allowed by current direct and indirect searches. 
Fig.~\ref{fig:bbgaga} shows the interplay between the HE-LHC reach and the impact of present flavour constraints which can be independently exploited to explore same regions of the parameter space. 
For example, if the $gg\to H\to hh\to b\bar b\gamma\gamma$ signal will be established at the future upgraded phases of the LHC in a region of parameter space with large and negative $\zeta_t$ (namely, below the dashed line in Fig.~\ref{fig:bbgaga}), this would clearly point to a scenario with $\zeta_b < \zeta_t$. 
\begin{figure}
\centering
{\includegraphics[scale=0.5]{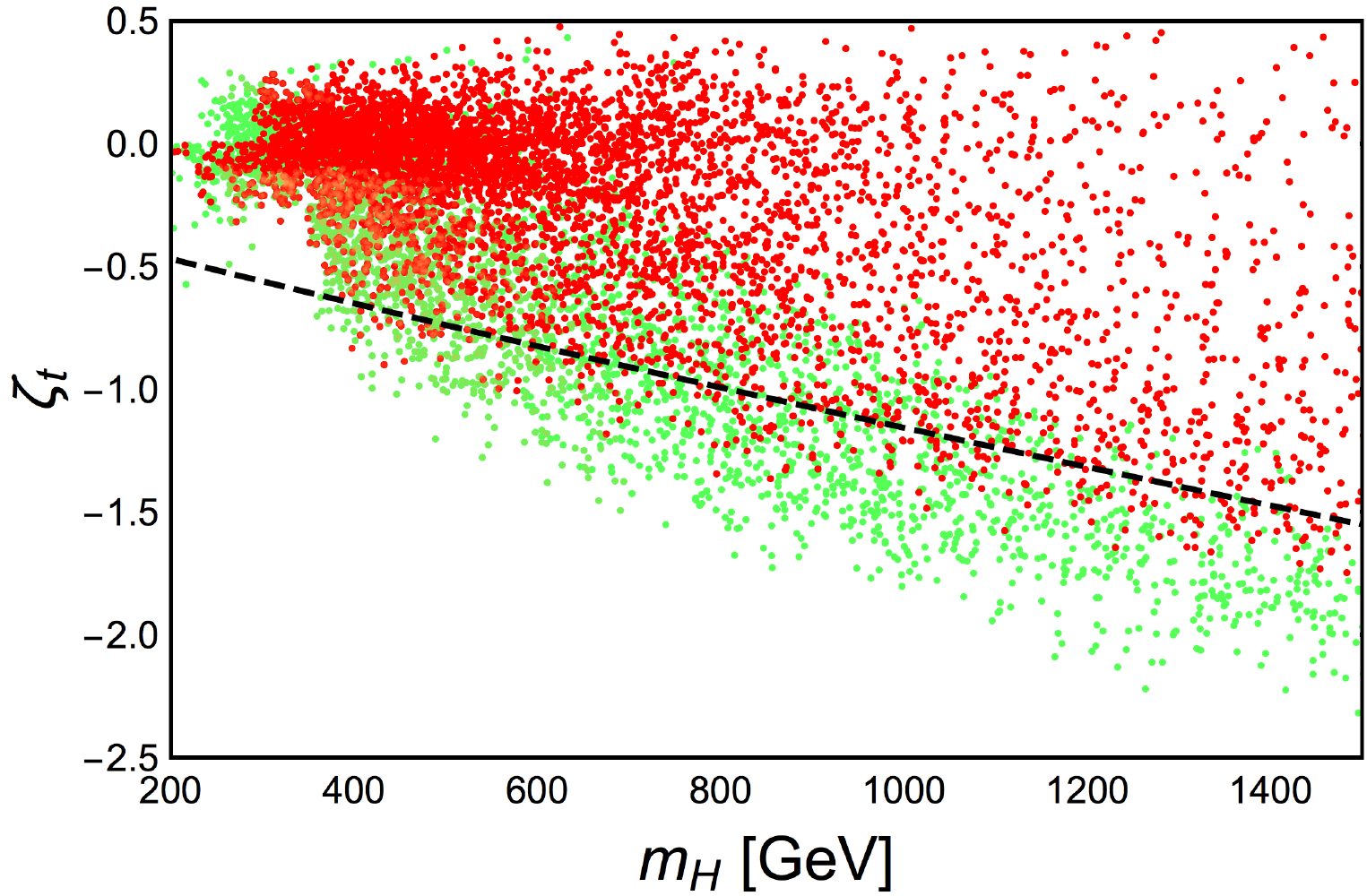}} \qquad
{\includegraphics[scale=0.5]{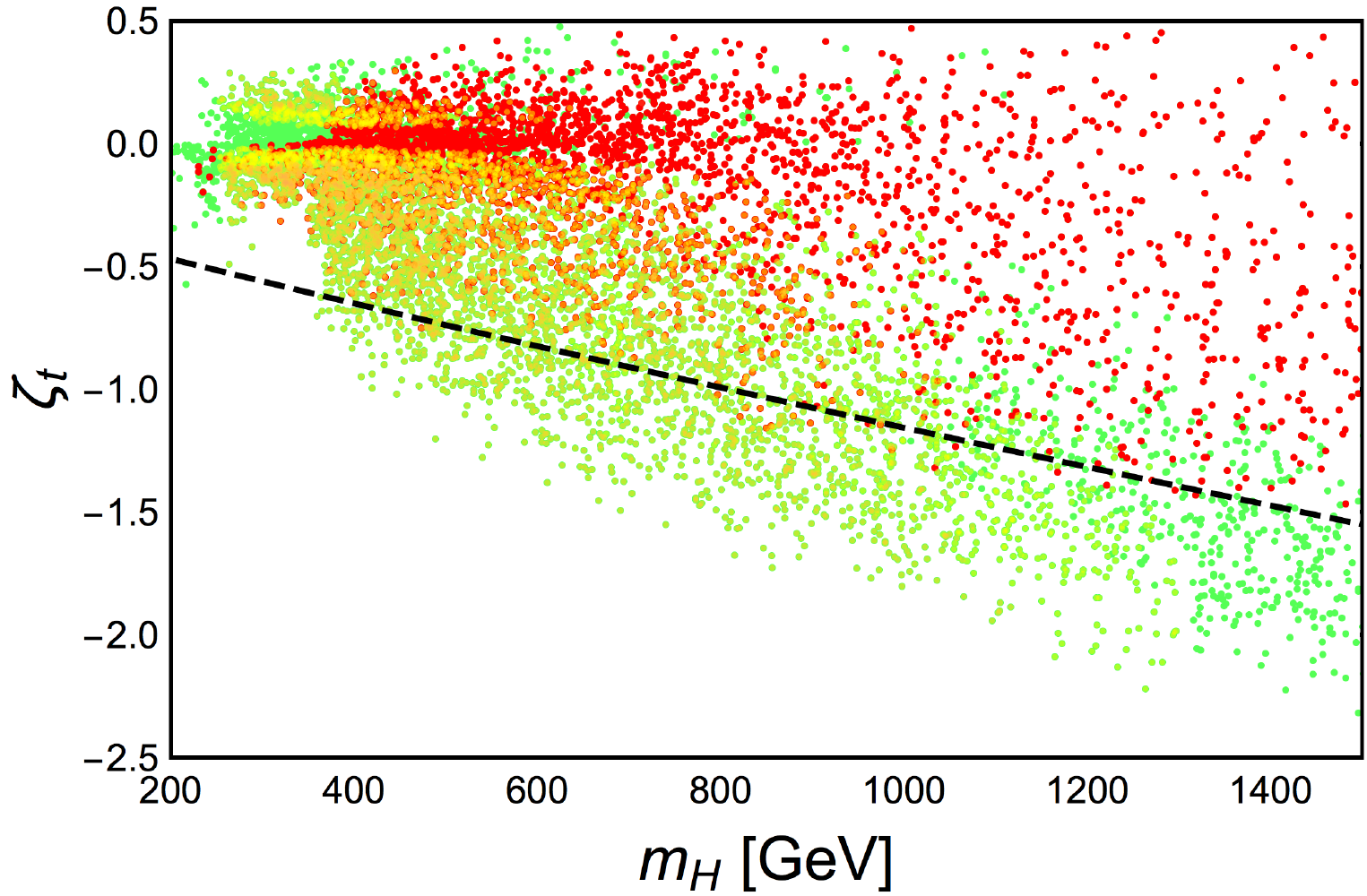}}
\caption{Interplay between direct and indirect C2HDM searches projected onto the $(m_H,\zeta_t)$ plane. Colour coding is as follows.
Green: all points that pass present constraints at 13 TeV. 
Red: points that, in addition to the above, have $\kappa_{VV}$, $\kappa_{\gamma\gamma}$ and $\kappa_{gg}$ within the 95\% CL projected uncertainty at $L=300$ fb$^{-1}$ (left) and $L=3000$ fb$^{-1}$ (right).
Orange: points that, in addition to the above,  are 95\% CL excluded  by the direct search $gg\to H\to hh\to b\bar b\gamma\gamma$, at $L=300$ fb$^{-1}$ (left) and $L=3000$ fb$^{-1}$ (right). In the right plot the yellow points are 95\% CL excluded by the same search  at the HE-LHC with $L=15$ ab$^{-1}$. Points below the dashed line are excluded by the measurement of $B_s \to X_s \gamma$ under the assumption $\zeta_b = \zeta_t$.
\label{fig:bbgaga}}
\end{figure}

 \begin{figure}
\centering
\includegraphics[scale=0.5]{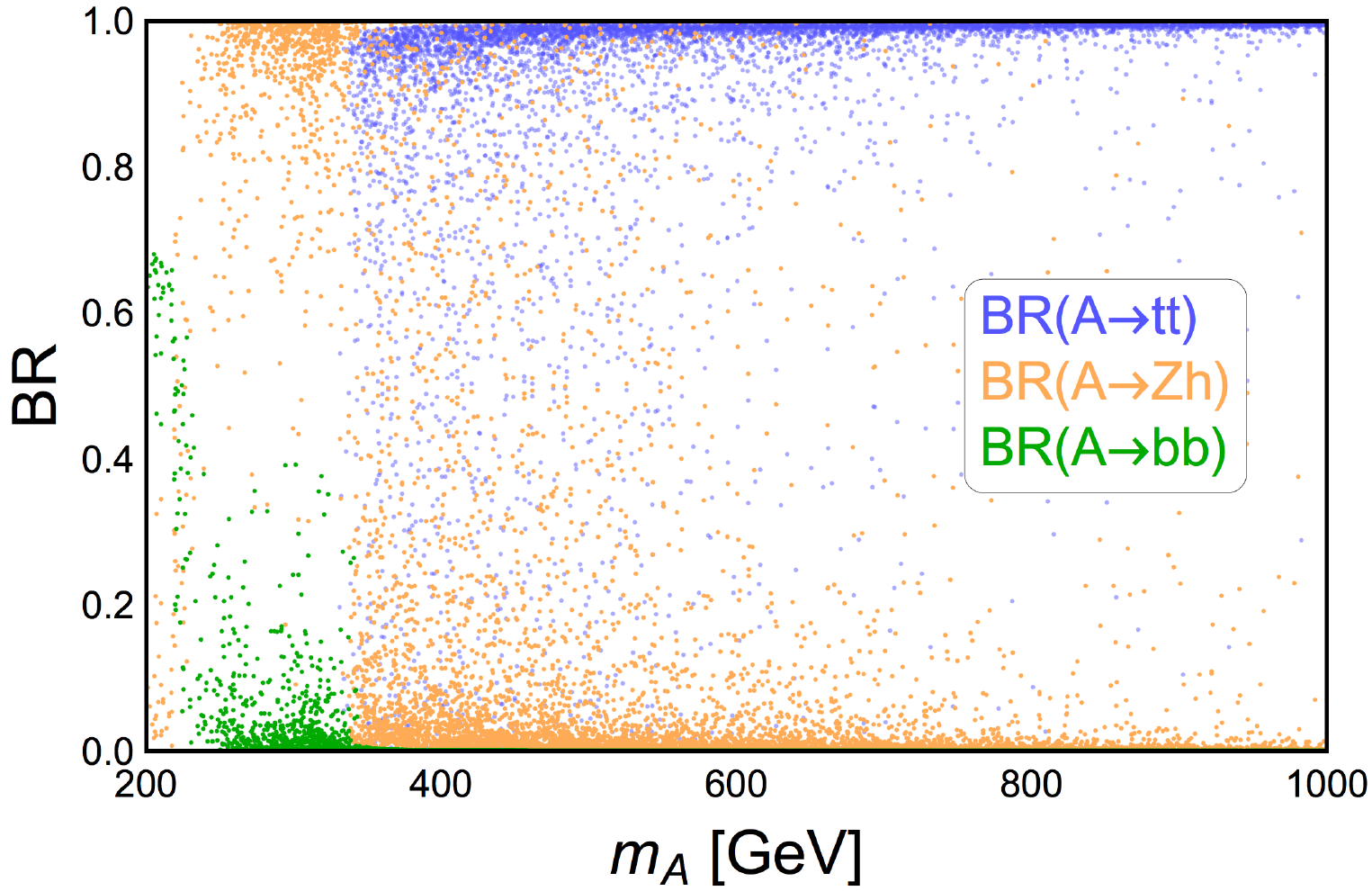}
\includegraphics[scale=0.5]{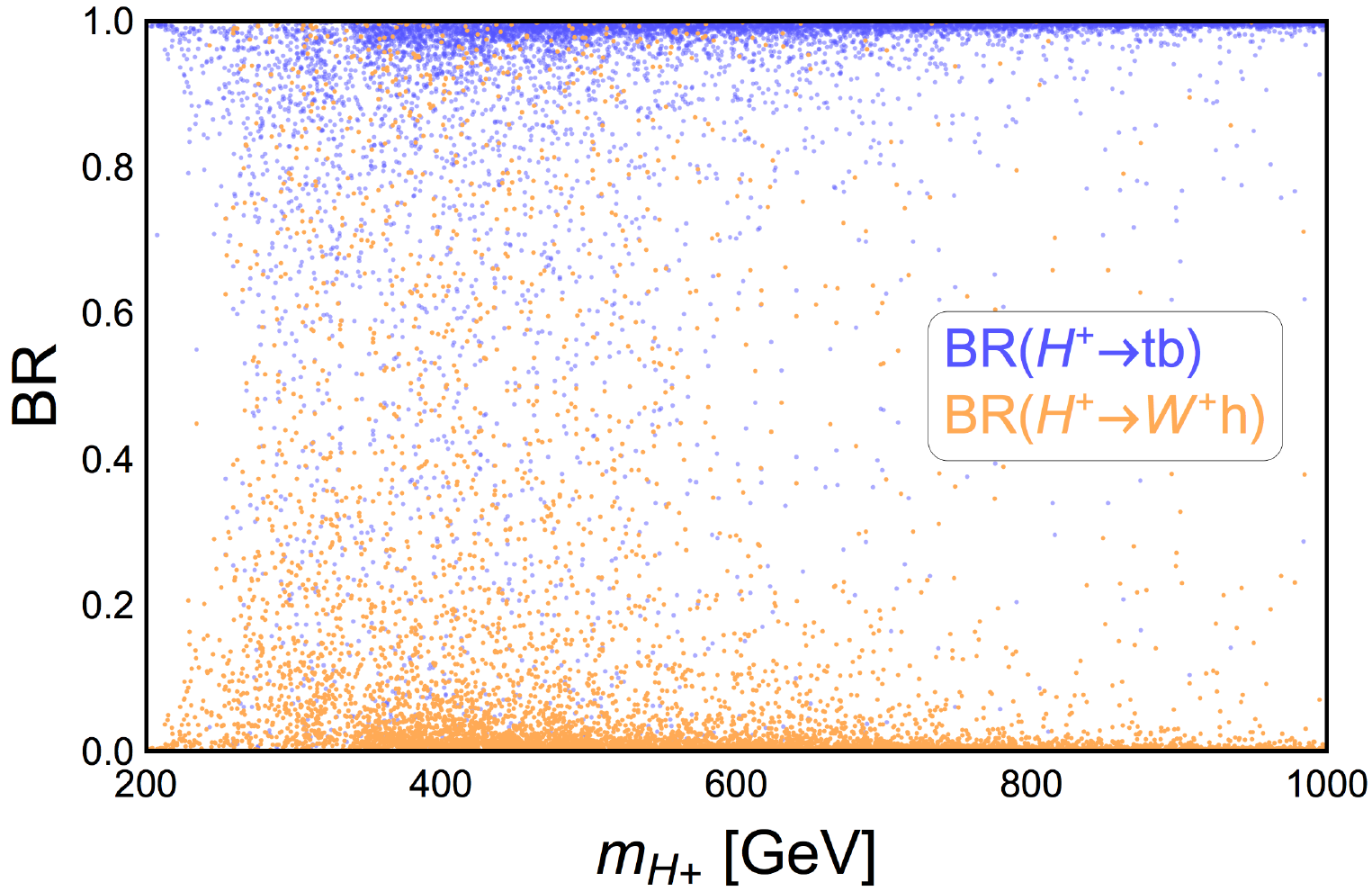}
\caption{(Left) The BRs of the $A$ state in the $t \bar t$ (blue), $Zh$ (orange) and $bb$ (green) channels. (Right) The BRs of the charged Higgs in the $t b$ (blue) and $W h$ (orange) channels.
\label{fig:BRs-A}}
\end{figure}

\item Pseudoscalar $A$. Since we consider a CP-symmetric composite sector, the phenomenology of the CP-odd scalar is, as far as decays to SM states are concerned, very constrained, since it can basically only decay to SM fermions and $Zh$. Indeed, according to Fig.~\ref{figure:splittings}, the exotic off-shell decays $A\to H Z^{*}$ and $A\to H^\pm W^{\mp*}$ are strongly suppressed by the very tight phase space available. 
In Fig.~\ref{fig:BRs-A} (left) we show the BRs for the leading decay modes of $A$ as function of its mass. The $A \to t \bar t$ channel clearly dominates when kinematically open while, below the $t \bar t$ threshold, the main channel is represented by $A \to Z h$, with $A \to b \bar b$ reaching $70\%$ at most for $m_A \sim 200$ GeV. The latter depends on the Higgs coupling to $b$ quarks, namely,  on the parameter $\zeta_b$ which has been fixed to $\zeta_t$ in our scan under the reasonable assumption that $C_2$ is broken with the same strength in all the strong fermionic sectors.
In the limit $m_A\gg m_{t}$, the BRs into SM fermions are particularly simple:
\be
\br(A\to t\bar t)\approx 1,\quad \br(A\to b\bar b)\approx 8 \, \times 10^{-4} \large(\frac{\zeta_b^2}{\zeta_t^2}\large), \quad \br(A\to \tau^+\tau^-)\approx 4 \,\times 10^{-5} \large(\frac{\zeta_\tau^2}{\zeta_t^2}\large).
\ee
Searches for $A\to t\bar{t}$ can then be used to constrain the couplings of $A$ to top quarks (as the production cross section is again $\propto \zeta_t^2$) and then one can access $\zeta_b$ and $\zeta_\tau$. 
The decay channel $A \to Z h$ is, instead, controlled by the square of the coupling $(1 - \xi/2)\sin \theta$. 
Suitable search strategies are, for instance, performed by reconstructing the $Z$ boson from its leptonic decays and the SM-like Higgs from $h \to bb$ or $h \to \tau \tau$.

\item Charged $H^\pm$. As far as decays into SM objects are concerned, here, the  phenomenology is dictated by $H^+ \to t \bar b$ and $H^+ \to W^+ h$, while $H^+\to \tau^+\nu_\tau$ is found to be negligible with $\br(H^+\to \tau^+\nu_\tau) \approx 4 \times 10^{-5} \zeta_\tau^2/\zeta_t^2$ in the large $m_{H^+}$ limit. The non-SM-like modes $H^+ \to W^{+*} A$ and $H^+ \to W^{+*} H$ are also suppressed due to the small mass splittings between the heavy scalars, see Fig.~\ref{figure:splittings}. The BRs of the two dominant channels are shown in Fig.~\ref{fig:BRs-A} (right) with $H^+ \to t \bar b$ being the leading one as $m_{H^\pm} > m_t$. The partial decay width of $H^+ \to t \bar b$ is determined by $\zeta_t^2$ (the contribution of $\zeta_b$ is suppressed by the ratio $m_b/m_t$) while $H^+ \to W^+ h$ is driven by the square of $(1 - \xi/2)\sin \theta$.

As for the $H^\pm$ production cross section, in the relevant mass range (i.e., $m_{H^\pm} > m_t$), this is governed by the $bg\to tH^-$ + c.c. channel, which is proportional to the same 
coupling entering $H^+\to t\bar b$ decays.

\end{itemize}

In short,  if deviations will be established at the LHC in the couplings of the discovered Higgs state to  
either SM  gauge bosons or matter fermions, then, not only a thorough investigation of the 2HDM hypothesis is called for (as one of the simplest non-minimal version of EWSB induced by the Higgs mechanism via {\it doublet} states, like the one already
discovered) but also a dedicated scrutiny of the production and  decay patterns of all potentially accessible heavy Higgs states, including data from the HL-LHC and HE-LHC options of the CERN machine, could enable one
to separate the E2HDM from the C2HDM. In this endeavour,      key roles will be played by interactions amongst the Higgs bosons themselves (with or without gauge bosons intervening) and with top quarks \cite{DeCurtis:2016tsm}.

\section{Conclusions}
\label{sec:conclusions}

In summary, in this paper, we have used compositeness as a possible  remedy to the hierarchy problem of the SM,  in particular,  assuming a pNGB nature of the discovered Higgs state.  In this respect, an intriguing setting   is the C2HDM, as it builds  upon the experimentally established existence of a doublet structure with a SM-like Higgs state $h$ triggering EWSB, and the need for BSM physics.
This scenario in fact surpasses the SM by providing one more composite doublet of Higgs states that can be searched for at the LHC, i.e.,  the familiar $H$, $A$ and $H^\pm$ states of a 2HDM, alongside additional composite gauge bosons and fermions. 
In fact, in order to obtain an acceptable fine-tuning at the EW scale, the compositeness scale $f$, which drives the masses of these (heavy)  non-SM Higgs states, must be in the TeV region.    The C2HDM framework advocated here is  thus a BSM setting which is both natural and minimal, offering as byproducts Higgs mass and coupling spectra within the LHC reach.

In fact, the entire  physical content  of the C2HDM, unlike the case of an elementary 2HDM, is actually predicted by a new confining strong dynamics. Our set up is  based on the spontaneous symmetry breaking of the global symmetry of a new strong sector, ${\rm SO}(6)\to {\rm SO}(4)\times {\rm SO}(2)$, with the residual symmetry in turn explicitly broken by the linear mixing between the (elementary) SM and the (composite) strong sector  fields via the so-called partial compositeness mechanism.

In this construct, the scalar potential  emerges at one-loop level and governs the dynamics of the aforementioned Higgs states, all realised as pNGBs, i.e., it predicts their properties. 
We have therefore calculated the mass and coupling spectra of the Higgs sector of this C2HDM explicitly and for the first time in literature. In particular,  we have  truncated the potential to the quartic order in the (pseudo)scalar fields, further assuming that the dynamics discussed here is CP symmetric yet $C_2$ broken, so that, in order to avoid Higgs-mediated FCNCs, we had to enforce an alignment in the Yukawa couplings. In order to do so, we had to also account for the corresponding gauge and fermionic spectra. 
While the structure of the former is dictated by the gauge symmetry  and the breaking pattern, there is arbitrariness in the choice of the latter. Here, we have placed the   fermions in the fundamental representation of SO(6) and we have further required a LR structure, as it guarantees UV finiteness and reduces the number of free parameters in the new strong sector. 
Ultimately, this implies that the scalar potential, and thus the couplings amongst (pseudo)scalars and fermions, depend on the composite fermion mass spectrum.

We have highlighted here the presence of correlations among the Higgs sector parameters and the strong sector ones, in particular, the dependence of the extra-Higgs mass scale on the extra-fermion masses.  In fact, within this framework, we have obtained that the mass  $m_{H_2}$ of the second Higgs doublet is mainly proportional to the $m_3$ parameter in the scalar potential, which arises from the breaking of $C_2$ in the composite sector and correlates with the masses of the two lightest top-partners. Then, as $m_3$  is driven by $f$, we have highlighted the fact that all heavy Higgs masses decouple for large $f$  from the  SM-light Higgs one, which is maintained light as it emerges from the SM VEV. Hence, as $f\to\infty$, the SM is recovered.
 
 Another interesting limit that we discussed is obtained when the above Yukawa alignment is  dismissed and one of the two relevant couplings is set to zero (i.e.,  $Y_1 \to 0$) to prevent tree level FCNCs. This generates a $C_2$ symmetric case, wherein one VEV is zero (i.e.,  
$\tan \beta = 0$) and no mixing exists between the first and second doublet (i.e., $\sin \theta = 0$).
This  generates a 2HDM structure with one inert doublet,  thus offering a DM candidate (in the form of the lightest between the additional neutral Higgs states, $H$ and $A$).  
The proposed scenario thus provides a concrete realisation of models with two doublets (one playing the role of the SM Higgs and the other being active or inert) originating from a strong confining dynamics.  

We have then proceded to discuss the phenomenological implications of this C2HDM at the LHC, in relation to both measurements of the discovered SM-like Higgs state and the (potential) discovery of its heavy companions. In this respect, we have assessed that the mass and coupling patterns that emerge from the strong dynamics embedded in the C2HDM are rather prescrictive, so that simultaneous $h$ measurements and detection of heavy Higgs decays into themselves (e.g., $H\to AZ^*$ and $H\to H^\pm W^{\mp *}$ or $A\to HZ^*$ and $H^\pm\to HW^{\pm *}$, depending on the actual mass hierarchy), with absence of any decays involving $A$ and $H^\pm$, would be a hallmark signature of the C2HDM. With this in mind, we have finally tested the scope of the standard LHC, HL-LHC and HE-LHC in accessing the C2HDM in both the above self-interacting Higgs channels and others involving (primarily) production and decay modes with top (anti)quarks involved. 

As we have produced all these phenomelogical results in presence of up-to-date  experimental  (notably including limits from both void searches for heavy Higgs bosons and measurements of the 125 GeV discovered Higgs state as well as flavour data) constraints, we are confident to have set the stage for pursuing dedicated analyses aimed at separating the C2HDM from the E2HDM hypothesis, in turn potentially enabling one to distinguish the composite from the fundamental nature of any Higgs boson accessible at the LHC by the end of all its already scheduled and currently discussed future stages.      

\section{Outlook}

In this work we have studied a scenario based on the coset $\rm SO(6)/SO(4) \times SO(2)$ with fermions embedded into the fundamental $\rm SO(6)$ representations. This represents the simplest scenario with maximal subgroup $H$ providing exactly two Higgs doublets as pNGBs, which was the Higgs sector content conceived here as natural next step up from the discovery of a `doublet' Higgs field. However, despite its simplicity, some of the phenomenological features described by this setup (as, in particular, the scenario with a composite inert Higgs) are also typical of other realisations.
More involved patterns can be constructed by allowing for subsequent breakings as in the case of $\rm SO(6) \to SO(4)$ in which the two scalar doublets are accompanied by an extra SM singlet state.
Focusing on scenarios with only two Higgs doublets, another interesting model is realised by the coset $\rm Sp(6)/SU(2) \times Sp(4)$ in which large corrections to the $T$ parameter are automatically avoided by the extended custodial symmetry of the corresponding subgroup. In this setup the left-handed $q_L$ and right-handed $t_R$ components of the top quark can be embedded in the $\bm 1_{2/3}$ and $\bm{14}_{2/3}$, respectively. 
Since two inequivalent embeddings are present for $q_L$, even though only one invariant can be build among the two representations and the Higgs doublets, the absence of dangerous tree-level FCNCs in the scalar sector must be ensured by enforcing the alignment in flavour space as in our case.
A scenario that shares even more similarities with the one addressed in this paper is instead described by the same coset $\rm SO(6)/SO(4) \times SO(2)$ but with fermions in the $\bm 1_{2/3}$ for the $t_R$ and in the $\bm{20}'_{2/3}$, containing two different embeddings, for the $q_L$ \cite{Mrazek:2011iu}. Either requiring $CP$ or $C_2$ unavoidably selects one of the two $q_L$ embeddings. This also avoids the presence of Higgs-mediated FCNCs without resorting to the flavour alignment. 
If $C_2$ is preserved, the composite inert 2HDM, described in section \ref{sec:c2symm}, is obtained.
In contrast, if only $CP$ is required, $C_2$ still arises at leading order as an accidental symmetry in the scalar potential with an almost-inert Higgs doublet. As such its phenomenology is expected to be vey similar to the one discussed in section \ref{sec:c2symm}.
The construction discussed in this paper has the advantage to encompass several scenarios with very different phenomenological features like,
for instance, the $C_2$ symmetric case that provides an inert Higgs doublet which, as discussed above, is also common to other realisations of CHMs,
but also much more peculiar scenarios like the $CP$-invariant and $C_2$-broken case that has been extensively studied in this work.

\section*{Acknowledgements}
SM is funded in part through the NExT Institute, the STFC CG ST/L000296/1 and the H2020-MSCA-RISE-2014 grant no.  645722
 (NonMinimalHiggs). We thank Andrea Tesi for very fruitful discussions and for reading the manuscript.
We are grateful to the Galileo Galilei Institute for Theoretical Physics for the hospitality during the program 
"Beyond Standard Model: where do we go from here".

\newpage
\appendix

\section{The C2HDM with the CCWZ formalism}
\label{sec:ccwz}

Let us review some of the main topics discussed in the present work by using the CCWZ formalism and focusing only on the fermion sector. 
This methodology is based on an effective Lagrangian approach and, as such, does not require the specification of an UV completion and only relies on the features of the symmetry breaking pattern. 
Therefore, it is possible to draw general conclusions without specifying the details of the model in the strong sector, in contrast to what we have done in the sections above, where we provided an explicit realisation of the C2HDM. Needless to say, the two approaches are completely equivalent at low energy. 

In order to keep the discussion as general as possible, we introduce two families of resonances, $\psi_4$ and $\psi_2$, the first one transforming in the fundamental of ${\rm SO}(4)$ and the other one in the fundamental of ${\rm SO}(2)$. 
The Lagrangian for the elementary and composite fermions is
\begin{eqnarray}
\label{eq:lagr_2}
\mathcal L_{\rm CCWZ}^{\rm fermion} &=& i \bar q_L^i \gamma^\mu \mathcal D_\mu q_L^i    +   i \bar u_R^i \gamma^\mu \mathcal D_\mu u_R^i  \nonumber \\
&+& i \bar \psi_4^i  \gamma^\mu \mathcal D_\mu \psi_4^i - m_{4,i} \bar \psi_4^i \psi_4^i 
+ i \bar \psi_2^i  \gamma^\mu \mathcal D_\mu \psi_2^i - m_{2,i} \bar \psi_2^i \psi_2^i  \nonumber \\
&+& y_{L4}^{ik} \, f \left( \bar q_L^{\bold 6 \, i} U\right)_4 \psi_4^k + y_{L2}^{ik} \, f \left( \bar q_L^{\bold 6 \, i} U\right)_2 \psi_2^k + y_{L\hat 2}^{ik} \, f \left( \bar q_L^{\bold 6 \, i} U\right)_{\hat 2} \psi_{ 2}^k \nonumber \\
&+& y_{R4}^{ik} \, f \left( \bar u_R^{\bold 6 \, i} U\right)_4 \psi_4^k + y_{R2}^{ik} \, f \left( \bar u_R^{\bold 6 \, i} U\right)_2 \psi_2^k + y_{R\hat 2}^{ik} \, f \left( \bar u_R^{\bold 6 \, i} U\right)_{\hat 2} \psi_{ 2}^k + \textrm{h.c.}
\end{eqnarray}
In order to simplify the discussion, the down-quark $d_R$ has not been explicitly written in the previous equation but it will be reintroduced back when needed.
The index $i$ runs over the three SM families while $k$ runs over the fermionic resonances for which the multiplicity can be, in principle, different for each of the two families. 
The notation $q_L^{\bold 6}$ and $u_R^{\bold 6}$ denotes, as usual, the embedding of the SM fields in the fundamental of ${\rm SO}(6)$, 
while the subscripts on the pNGB matrix $U$ represent its projections onto the different ${\rm SO}(4) \times {\rm SO}(2)$ representations. 
In particular, the subscripts $2$ and $\hat 2$ denote how the two invariants $2 \cdot 2$ and $2 \wedge 2$ are built from the fundamental representation of $\rm SO(2)$, specifically, by exploiting the contraction with $\delta^{\alpha\beta}$ and $\epsilon^{\alpha\beta}$, respectively. 
Restricting to the top-quark sector and considering only two families of resonances, Eq.~(\ref{eq:lagr_2}) can be easily mapped onto the Lagrangian given in Eq.~(\ref{eq:lag_ferm0}). 
After integrating out the heavy resonances, the momentum space Lagrangian for the quark fields reads as
\begin{eqnarray}
\mathcal L_\textrm{eff} &=& \bar q_L^i  \,\slashed{p} \, \Pi_L^{ij} q_L^j + \bar u_R^i \, \slashed{p} \, \Pi_R^{ ij}  u_R^j  - \left( \bar q_L^i  \Pi_{LR}^{ij} u_R^j     + \textrm{h.c.}   \right),
\end{eqnarray}
where the gauge interactions arising from the fermionic covariant derivates have been neglected as they play no role in the present discussion.
The form factors are
\begin{eqnarray}
 \Pi_L^{ij} &=& \delta^{ij} - f^2 \sum_k \Upsilon_{L}^\dag  \bigg\{ \frac{ y_{L4}^{ik} \, y_{L4}^{jk *} }{p^2 - m_{4,k}^2} [U_4 \cdot {}_4U^\dag ]    +     \frac{ y_{L2}^{ik} \, y_{L2}^{jk *} + y_{L \hat 2}^{ik} \, y_{L \hat 2}^{jk *} }{p^2 - m_{2,k}^2} [U_2 \cdot {}_2U^\dag]    \nonumber \\
 &+&   \frac{ y_{L\hat{2}}^{ik} \, y_{L2}^{jk *}  -  y_{L2}^{ik} \, y_{L\hat 2}^{jk *}  }{p^2 - m_{2,k}^2} [U_{2} \cdot {}_{\hat{2}} U^\dag]   \bigg\} \Upsilon_{L},  \nonumber \\
\Pi_R^{ij}  &=& \delta^{ij} - f^2 \sum_k \Upsilon_{R}^{i \dag} \bigg\{ \frac{ y_{R4}^{ik} \, y_{R4}^{jk *} }{p^2 - m_{4,k}^2} [U_4 \cdot {}_4U^\dag]    +     \frac{ y_{R2}^{ik} \, y_{R2}^{jk *}  +  y_{R \hat 2}^{ik} \, y_{R \hat 2}^{jk *} }{p^2 - m_{2,k}^2} [U_2 \cdot {}_2U^\dag]    \nonumber \\
&+&   \frac{ y_{R\hat{2}}^{ik} \, y_{R2}^{jk *}  -  y_{R2}^{ik} \, y_{R\hat 2}^{jk *}   }{p^2 - m_{2,k}^2} [U_{2} \cdot {}_{\hat{2}} U^\dag]   \bigg\}  \Upsilon_{R}^j, \nonumber 
\end{eqnarray}
\begin{eqnarray}
\Pi_{LR}^{ij} &=& f^2 \sum_k \Upsilon_{L}^\dag \bigg\{ \frac{ y_{L4}^{ik} \, m_{4,k} \, y_{R4}^{jk *} }{p^2 - m_{4,k}^2} [U_4 \cdot {}_4U^\dag]    +     \frac{ y_{L2}^{ik} \, m_{2,k} \, y_{R2}^{jk *}  +  y_{L\hat{2}}^{ik} \, m_{2,k} \, y_{R\hat{2}}^{jk *} }{p^2 - m_{2,k}^2} [U_2 \cdot {}_2U^\dag]    \nonumber \\
&+&   \frac{ y_{L\hat{2}}^{ik} \, m_{2,k} \, y_{R2}^{jk *} -  y_{L2}^{ik} \, m_{2,k} \, y_{R \hat 2}^{jk *}  }{p^2 - m_{2,k}^2} [U_{2} \cdot {}_{\hat{2}} U^\dag]   \bigg\} \Upsilon_{R}^j \,, 
\label{eq:app:ff}
\end{eqnarray}
where $\Upsilon_{L}$ and $\Upsilon_{R}^i$ are the spurions with their VEV defined as in Eq.~(\ref{spurion61}). Notice that, differently from its L-handed counterpart, the $\Upsilon_{R}^i$ spurion carries a flavour index
because each of the three R-handed fields $u_R^i$ can be embedded into a $\bm 6$ of $\rm SO(6)$ in two different inequivalent ways. This freedom is parameterised by an angle $\theta_i$.
The contractions of the pNGB matrices in the form factors above encode all the dependence on the Higgs fields and are explicitly defined as
\begin{eqnarray}
\label{eq:app:inv}
 [  U_{4} \cdot {}_{4} U^\dag ]  = U_{i} \delta_{ij} U^\dag_j, \qquad   [  U_{2} \cdot {}_{2} U^\dag ]  = U_{\alpha} \delta_{\alpha \beta} U^\dag_\beta,  \qquad   [  U_{2} \cdot {}_{\hat 2} U^\dag ]  = U_{\alpha} \epsilon_{\alpha \beta} U^\dag_\beta,
\end{eqnarray}
where $i,j$ and $\alpha,\beta$ run over the fundamental representations of ${\rm SO}(4)$ and ${\rm SO}(2)$, respectively. 

In order to make a closer contact with the results presented in the previous sections, where the model in the strong sector has been explicitly specified, we notice that
the contractions of the pNGB matrix are related to the $\Sigma$ matrix by
\begin{eqnarray}
 [U_4 \cdot {}_4 U^\dag] = \bold 1 + \Sigma^2, \qquad
 [U_2 \cdot {}_2 U^\dag] = - \Sigma^2,  \qquad
 [U_{ 2} \cdot {}_{\hat 2} U^\dag] =  \Sigma \,.
\end{eqnarray}
Considering only the top-quark and two families of heavy resonances, the effective Lagrangian after the integration of the heavy resonances can be schematically written in the same form as Eq.~(\ref{fermion6}), namely,
\begin{eqnarray}
\mathcal L_\textrm{eff} 
&=&  \bar q_L^\bold 6 \slashed{p} \left[ \tilde \Pi^{q_t}_0 + \tilde \Pi^{q_t}_1 \, \Sigma + \tilde \Pi^{q_t}_2 \, \Sigma^2 \right] q_L^\bold 6 
+  \bar t_R^\bold 6 \slashed{p} \left[ \tilde \Pi^t_0 + \tilde \Pi^t_1 \,  \Sigma + \tilde \Pi^t_2 \, \Sigma^2 \right] t_R^\bold 6 \nonumber \\
&+&  \bar q_L^\bold 6  \left[ \tilde M^t_0 + \tilde M^t_1 \,  \Sigma + \tilde M^t_2 \, \Sigma^2 \right] t_R^\bold 6 + \textrm{h.c.}, 
\end{eqnarray}
where the individual form factors are extracted from Eq.~(\ref{eq:app:ff}) as follows:
\begin{eqnarray}
\tilde \Pi^{q_t}_0 &=& 1 - f^2 \sum_k \frac{|y_{L4}^k|^2 }{p^2 - m_{4,k}^2} \,, \qquad
\tilde \Pi^{q_t}_1 = - 2 i f^2  \sum_k \frac{ \textrm{Im} (y_{L\hat{2}}^{k} \, y_{L2}^{k *} )   }{p^2 - m_{2,k}^2} \,, \nonumber \\
\tilde \Pi^{q_t}_2 &=&  - f^2 \sum_k \left[ \frac{|y_{L4}^k|^2 }{p^2 - m_{4,k}^2} +  \frac{|y_{L 2}^k|^2 +  |y_{L \hat 2}^k|^2 }{p^2 - m_{ 2,k}^2} \right] \,, \nonumber \\
\tilde \Pi^t_0 &=& 1 - f^2 \sum_k \frac{|y_{R4}^k|^2 }{p^2 - m_{4,k}^2} \,, \qquad
\tilde \Pi^t_1 = -2 i f^2 \sum_k \frac{   \textrm{Im} (y_{R\hat{2}}^{k} \, y_{R2}^{k *} )   }{p^2 - m_{ 2,k}^2} \,, \nonumber \\
\tilde \Pi^t_2 &=&  - f^2 \sum_k \left[ \frac{|y_{R4}^k|^2 }{p^2 - m_{4,k}^2} +  \frac{|y_{R 2}^k|^2 + |y_{R \hat 2}^k|^2 }{p^2 - m_{ 2,k}^2} \right] \,, \nonumber \\
\tilde M^t_0 &=& - f^2 \sum_k \frac{y_{L4}^k \, m_{4,k}  \, y_{R4}^{k*}}{p^2 - m_{4,k}^2} \,, \qquad
\tilde M^t_1 = - f^2 \sum_k \frac{  m_{2,k} ( y_{L\hat{2}}^{k}  \, y_{R2}^{k *} -  y_{L2}^{k} \, y_{R \hat 2}^{k *} )  }{p^2 - m_{ 2,k}^2}  \,, \qquad \nonumber \\
\tilde M^t_2 &=&  - f^2 \sum_k \left[ \frac{y_{L4}^k  \, m_{4,k}  \, y_{R4}^{k*}}{p^2 - m_{4,k}^2} +  \frac{y_{L 2}^k  \, m_{2,k}  \,  y_{R 2}^{k*}  +   y_{L \hat 2}^k  \, m_{2,k}  \,  y_{R \hat 2}^{k*} }{p^2 - m_{ 2,k}^2} \right]  \,,
\label{eq:ff_ccwz}
\end{eqnarray}
with $k$ running only on two families of heavy resonances. 
After integrating out the heavy degrees of freedom,  as naively expected, the CCWZ approach leads to an effective Lagrangian with the same structure of the one derived from an explicit model. The only difference appears in the parameterisation of the form factors in terms of masses and couplings. For instance, we immediately notice that the dependence on the proto-Yukawa couplings is the same, 
that $\tilde \Pi_0$ and $\tilde \Pi_2$ are real while $\tilde \Pi_1$ is imaginary and that $\tilde \Pi_1 = 0$ in a CP invariant scenario. 
The calculation of the effective potential in terms of the form factors remains completely unchanged, so are Eqs.~(\ref{eq:parameters_g}) and (\ref{eq:parameters_f}). 
As we anticipated above, the form factors obtained in the 2-site model can be easily mapped onto the ones in Eq.~(\ref{eq:ff_ccwz}) once the Lagrangian of Eq.~(\ref{eq:lag_ferm0}) is recasted into the basis in which, for  $\Delta_{L,R} \to 0$, the resonance fields $\Psi_t$ are mass eigenstates. Notice also that, despite the equivalence of the form factors obtained after the integration of the heavy resonances, the effective Lagrangian in Eq.~(\ref{eq:lagr_2}), which is used as the starting point of the CCWZ construction, does not capture all the information encoded in Eq.~(\ref{eq:lag_ferm0}) such as, for instance, the interaction among the resonances and the NGBs. 
\subsection*{An issue with flavour changing neutral currents}

The most general effective Lagrangian in the fermion sector can be easily constructed using the CCWZ formalism and reads as
\begin{equation}
\label{eq:app:yuk}
- \mathcal L_\textrm{yuk} = a_{ij}^A \bar \psi_L^i U P_A U^\dag \psi_R^j + \textrm{h.c.},
\end{equation}
where $\psi_{L,R}$ denote SM quark fields embedded into incomplete $\mathcal G$ multiplets while the index $A$ spans over all the possible invariants in the subgroup $\mathcal H$ with $P_A$ being the corresponding projector operator.
As described above, we embed the L- and R-handed components of the elementary SM quarks into the fundamental representation of ${\rm SO}(6)$ using the spurion fields.
At this level, all the information on the interactions with the composite resonances, which have been integrated out, is encoded in the coefficients $a_{ij}^A$.
With the fermions in the $\bold 6$ of ${\rm SO}(6)$, where $\bold 6 = \bold 4 \oplus \bold 2$ under ${\rm SO}(4) \times {\rm SO}(2)$, three different invariants of $\mathcal H$ can be constructed with
$\delta_{ij}, \delta_{\alpha \beta}$ and $\epsilon_{\alpha \beta}$, respectively, where latin (greek) indices belong to ${\rm SO}(4)$ (${\rm SO}(2)$).
The sum of the first two invariants is trivial such that only two of them are independent. These can be chosen to be, for instance, $\delta_{\alpha \beta}$ and $\epsilon_{\alpha \beta}$.  
The possibility to build several invariants naturally introduces dangerous FCNCs in the Higgs sector, 
unless one imposes particular conditions on the coefficients $a_{ij}^A$ or discrete symmetries to single out only one invariant.
One of the two invariants can be removed by imposing a $C_2$ symmetry in the strong sector. 
But, even in that case, since the R-handed quarks can be embedded in the $\bold 6$ of ${\rm SO}(6)$ into two independent ways (as already mentioned, this freedom is described by the $\theta_i$ angle), an extra symmetry must be advocated in order to select one of the two embeddings. For instance, the CP symmetry uniquely picks up the $\theta_i = 0$ direction. Therefore, Higgs mediated FCNCs are avoided if one considers a $C_2$ and CP invariant scenario.
If the strong sector does not enjoy such symmetries that prevent the appearance of multiple invariants, 
FCNCs can only be avoided by restricting the structure of the $a_{ij}^A$ matrices as in the flavour alignment limit. 
The absence of leading contributions to FCNCs is thus ensured if the three matrices (in flavour space) appearing in the form factor $\Pi_{LR}^{ij}$, namely the ones proportional to the three independent invariants 
in Eq.~(\ref{eq:app:inv}), are proportional to each other in the small momentum regime. 
This condition can be satisfied by requiring, e.g., the following: (a) $a_1 y_{L4}^{ik} = a_2 y_{L2}^{ik} = a_3 y_{L\hat 2}^{ik} \equiv y_L^{ik}$ and 
$b_1 y_{R4}^{ik} = b_2 y_{R2}^{ik} = b_3 y_{R\hat 2}^{ik} \equiv y_{R}^{ik}$, so that $ y_{L}^{ik}  \, y_{R}^{jk *}$ can be factorised, (b) the alignment of the R-handed 
spurion fields, $\theta^i \equiv \theta$ and (c) the proportionality of the resonance masses $c_1 m_{4, k} = c_2 m_{2, k} \equiv m_{k}$ for all $k = 1 \ldots N_\psi$, where the number of resonances $N_\psi$ has been chosen to be the same for the two families.

Under the flavour alignment assumption, the Yukawa Lagrangian can be finally recast in the following form:
\begin{eqnarray}
\label{eq:yukawalagr}
- \mathcal L_\textrm{yuk} = q_L^i    Y_u^{ij}  \, g_u(H_1, H_2) u_R^j   +     q_L^i   Y_d^{ij}  \, g_d(H_1, H_2) d_R^j, 
\end{eqnarray}
where we have reintroduced back the $d_R$ quark field. The $g_{u,d}(H_1, H_2)$ are functions of the Higgs fields and the SM Yukawa couplings $Y_{u,d}^{ij}$ are defined by
\begin{eqnarray}
Y_u^{ij} = \sum_{k=1}^{N_\psi} \frac{y_{L}^{ik} \, y_{u \, R}^{jk *}}{m_{k}} \,, \qquad   
Y_d^{ij} = \sum_{k=1}^{N_\psi} \frac{y_{L}^{ik} \, y_{d \, R}^{jk *}}{m_{k}} \,.
\end{eqnarray}
The Yukawa structure in Eq.~(\ref{eq:yukawalagr}) clearly ensures the absence of Higgs mediated FCNCs.

\section{Form factors}
\label{sec:formfactors}

The form factors characterising the gauge part of the effective Lagrangian in Eq.~(\ref{eff6}) are
\begin{align}
\tilde{\Pi}_0    =   - \frac{m_\rho^2}{g_\rho^2(q^2-m_\rho^2)}, \quad
\tilde{\Pi}_1   =  -\frac{2m_\rho^4(m_{\hat{\rho}}^2 - m_\rho^2 )}{f^2g_\rho^2(q^2-m_{\rho}^2)(q^2-m_{\hat{\rho}}^2)}, \quad
\tilde{\Pi}_2   =  - \tilde{\Pi}_{1}, \quad
\tilde{\Pi}_X   =  - \frac{m_{\rho_X}^2}{g_{\rho_X}^2(q^2-m_{\rho_X}^2)}
\end{align}
and, by choosing for simplicity $g_{\rho_X} = g_\rho$, which implies $m_{\rho_X} = m_\rho$ as shown in Eq.~(\ref{eq:rhomass}), we get the normalised form factors
 \begin{align}
 \Pi_W &=  \frac{\tilde{\Pi}_1}{\tilde{\Pi}_W} = -\frac{2r_W^{}m_\rho^4(m_{\hat{\rho}}^2-m_\rho^2)}{f^2(q^2-m_{\hat{\rho}}^2)[q^2-(1+r_W^{})m_{\rho}^2]}, \nonumber \\
 \Pi_B &=  \frac{\tilde{\Pi}_1}{\tilde{\Pi}_B} = -\frac{2r_B^{}m_\rho^4(m_{\hat{\rho}}^2-m_\rho^2)}{f^2(q^2-m_{\hat{\rho}}^2)[q^2-(1+2r_B^{})m_{\rho}^2]}\Bigg|_{g_{\rho_X} = g_{\rho} },
 \end{align}
 with
 \begin{align}
 \tilde \Pi_W & = \tilde \Pi_0 + \frac{1}{g_W^2}, \nonumber \\
 \tilde \Pi_B & = \tilde \Pi_0 + \tilde \Pi_X + \frac{1}{g_B^2},
 \end{align}
 where $r_{W,B}^{} = g_{W,B}^2/g_\rho^2$. 
The couplings $g_{W,B}$ are implicitly defined by
\begin{equation}
\frac{1}{g^2} = \frac{1}{g_W^2} + \frac{1}{g_\rho^2} \,, \qquad \frac{1}{g^{'2}} =  \frac{1}{ g_B^2} + \frac{1}{g_\rho^2} +  \frac{1}{g_{\rho_X}^2},
\end{equation}
where $g$ and $g'$ are the SM ${\rm SU}(2)_L$ and ${\rm U}(1)_Y$ coupling constants. 
The form factors appearing in Eq.~(\ref{eq:parameters_g}) are normalised as $\bar \Pi_{W,B} = f^2/q^2 \, \Pi_{W,B}$. Finally, 
the form factors extracted after the integration of the heavy spin-1/2 resonances coupled to the top quark are explicitly given by
\begin{align}
&\tilde{\Pi}_0^{q_t} =1 -\frac{q^2\Delta_L^t\Delta_L^{t \, T} + (\Delta_L^t\sigma_2) M_{\Psi_t}^T M_{\Psi_t} (\Delta_L^t\sigma_2)^T}{q^4-q^2\text{tr}M_{\Psi_t} M_{\Psi_t}^T + \text{det}M_{\Psi_t}^2},   \qquad
\tilde{\Pi}_0^{t} = 1 -\frac{q^2\Delta_R^t\Delta_R^{t \, T} + (\Delta_R^t\sigma_2) M_{\Psi_t}   M_{\Psi_t}^T (\Delta_R^t\sigma_2)^T }{q^4-q^2\text{tr}M_{\Psi_t} M_{\Psi_t}^T + \text{det}M_{\Psi_t}^2},   \nonumber \\
&\tilde{\Pi}_1^{q_t} = \tilde{\Pi}_1^t = 0, \nonumber \\
&\tilde{\Pi}_2^{q_t} = \frac{q^2\Delta_L^t \Delta_L^{t \,T} + (\Delta_L^t\sigma_2)(Y_1^{t \,T} Y_1^t + \bar Y_2^{t \, T} \bar Y_2^t) (\Delta_L^t \sigma_2)^T}
{q^4-(\text{tr}Y_1^t Y_1^{t \,T} + \text{tr}\bar Y_2^t \bar Y_2^{t \,T})q^2 + f(Y_1^t, \bar Y_2^t) }
-\frac{q^2\Delta_L^t \Delta_L^{t \, T} + (\Delta_L^t\sigma_2) M_{\Psi_t}^T M_{\Psi_t} (\Delta_L^t \sigma_2)^T }{q^4-q^2\text{tr}M_{\Psi_t}^T M_{\Psi_t} + \text{det}M_{\Psi_t}^2},   \nonumber \\
&\tilde{\Pi}_2^t = \frac{q^2\Delta_R^t \Delta_R^{t \,T} + (\Delta_R^t\sigma_2)(Y_1^tY_1^{t \,T} + \bar Y_2^t \bar Y_2^{t \,T}) (\Delta_R^t \sigma_2)^T}
{q^4-(\text{tr}Y_1^t Y_1^{t \, T} + \text{tr}\bar Y_2^t \bar Y_2^{t \,T})q^2 + f(Y_1^t, \bar Y_2^t) }
-\frac{q^2\Delta_R^t\Delta_R^{t \, T} + (\Delta_R^t \sigma_2) M_{\Psi_t} M_{\Psi_t}^T (\Delta_R^t\sigma_2)^T}{q^4-q^2\text{tr}M_{\Psi_t} M_{\Psi_t}^T + \text{det}M_{\Psi_t}^2},    \nonumber  \\
&\tilde{M}_0^t = -\frac{q^2 \Delta_L^t M_{\Psi_t}^T \Delta_R^{t \, T} + (\Delta_L^t\sigma_2)( M_{\Psi_t} ^T\sigma_2 M_{\Psi_t} \sigma_2 M_{\Psi_t}^T)  (\Delta_R^t \sigma_2)^T  }
{q^4-q^2\text{tr}M_{\Psi_t} M_{\Psi_t}^T + \text{det}M_{\Psi_t}^2} ,  \nonumber   \\
&\tilde{M}_1^t = \frac{q^2\Delta_L^t Y_1^{t \,T} \Delta_R^{t \, T} + \Delta_L^t \sigma_2 (Y_1^{t \, T}\sigma_2 Y_1^t \sigma_2 Y_1^{t \, T} + \bar Y_2^{t \,T} \sigma_2Y_1^t \sigma_2 \bar Y_2^{t \, T})  (\Delta_R^t\sigma_2)^T  }
{q^4-(\text{tr}Y_1^t Y_1^{t \,T} + \text{tr}\bar Y_2^t \bar Y_2^{t \, T})q^2 + f(Y_1^t,\bar Y_2^t) } ,  \nonumber  \\
&\tilde{M}_2^t = \frac{q^2\Delta_L^t \bar Y_2^{t \, T} \Delta_R^{t \, T} + \Delta_L^t \sigma_2 (\bar Y_2^{t \,T} \sigma_2 \bar Y_2^t \sigma_2 \bar Y_2^{t \,T} + Y_1^{t \,T} \sigma_2 \bar Y_2^t \sigma_2 Y_1^{t \,T}) (\Delta_R^t\sigma_2)^T }
{q^4-(\text{tr}Y_1^t Y_1^{t \,T} + \text{tr}\bar Y_2^t \bar Y_2^{t \, T})q^2 + f(Y_1^t, \bar Y_2^t) }   +   \tilde{M}_0^t , 
\end{align}
where $\bar Y_2^t = M_{\Psi_t}- Y^t_2$ and $f(Y_1^t,\bar Y_2^t) = [\det (Y_1^t) - \det (\bar Y_2^t)]^2 + \text{tr}(\sigma_2 Y_1^t \sigma_2 \bar Y_2^{t \,T})^2$. 
Similar results are obtained for the bottom quark and the tau lepton. The previous equations are valid in the scenario with two sixplet fermions and, as such, $M_{\Psi_t}$, $Y_1^t$, $Y_2^t$ are understood as $2\times 2$ matrices, while $\Delta_{L,R}^t$ are two-components vectors. From those we can define the canonically normalised form factors as
\begin{equation}
\Pi_{1,2}^{q_t} = \frac{\tilde{\Pi}_{1,2}^{q_t}}{\tilde{\Pi}_0^{q_t}} \,, \qquad
\Pi_{1,2}^{t} = \frac{\tilde{\Pi}_{1,2}^{t}}{\tilde{\Pi}_0^{t} -\tilde\Pi_2^t} \,, \qquad
M_{1,2}^{t} = \frac{\tilde{M}_{1,2}^{t}}{\sqrt{ \tilde{\Pi}_0^{q_t} \tilde{\Pi}_0^{t} }} \,,
\end{equation}
which are then given by, under the LR assumption,
\begin{align}
& \Pi_2^{q_t} = m_{\Psi_1}^2 \Delta_L^2 (Y^2-m_{\Psi_{12}}^2)(q^2-m_{\Psi_2}^2) 
  H^{-1}[q^2;Y^2,0]\,  H^{-1}[q^2;m_{\Psi_{12}}^2 +\Delta_L^2,(m_{\Psi_2}^2 +m_{\Psi_{12}}^2)\Delta_L^2],  \nonumber \\
& \Pi_2^t = m_{\Psi_2}^2 \Delta_R^2 (Y^2-m_{\Psi_{12}}^2)(q^2-m_{\Psi_1}^2)   H^{-1}[q^2;m_{\Psi_{12}}^2,0]\,  H^{-1}[q^2;Y^2 + \Delta_R^2,(m_{\Psi_1}^2 + Y^2)\Delta_R^2], \nonumber \\
& M_1^t = Y_1 \frac{H^{1/2}[q^2; m^2_{\Psi_{12}}, 0]}{H^{1/2}[q^2, Y^2,0]} M^t \,, \nonumber \\
& M_2^t =  \left[ - \bar Y_2 \frac{H^{1/2}[q^2; m^2_{\Psi_{12}}, 0]}{H^{1/2}[q^2, Y^2,0]}  + m_{\Psi_{12}}   \frac{H^{1/2}[q^2; Y^2, 0]}{H^{1/2}[q^2, m^2_{\Psi_{12}} ,0]}  \right] M^t , \nonumber \\
& M_1^t M_2^t = Y_1 \left[ - \bar Y_2 \frac{H[q^2; m^2_{\Psi_{12}}, 0]}{H[q^2, Y^2,0]}  + m_{\Psi_{12}}  \right] (M^t)^2
\label{eq:ff_appendix}
\end{align}
with
\begin{align}
M^t =  \frac{m_{\Psi_1} m_{\Psi_2} \Delta_L \Delta_R }{ H^{1/2}[q^2;m_{\Psi_{12}}^2 +\Delta_L^2,(m_{\Psi_2}^2 +m_{\Psi_{12}}^2)\Delta_L^2]  H^{1/2}[q^2;Y^2 +\Delta_R^2,(m_{\Psi_1}^2 + Y^2)\Delta_R^2]}
\end{align}
and 
\begin{align}
H[q^2 ; a, b] =  q^4 - q^2 (m_{\Psi_1}^2 + m_{\Psi_2}^2 + a) + m_{\Psi_1}^2 m_{\Psi_2}^2 + b \,. 
\end{align}
In the previous equations we used the shorthand notation
\begin{align}
& Y_1 = (Y_1^t)_{12} \,, ~ \bar Y_2 =  (\bar Y_2^t)_{12} = (M_{\Psi_t})_{12} - ( Y_2^t)_{12} \,,  ~ Y^2 = Y_1^2 +\bar Y_2^2 \,, \nonumber \\
 & \Delta_L = (\Delta_L^t)_1 \,, ~ \Delta_R = (\Delta_R^t)_2 ~ m_{\Psi_1} = (M_{\Psi_t})_{11} \,,  ~ m_{\Psi_2} = (M_{\Psi_t})_{22} \,, ~ m_{\Psi_{12}} = (M_{\Psi_t})_{12}
\end{align}
and we required
\begin{align}
(\Delta_L^t)_2 = (\Delta_R^t)_1 = (M_{\Psi_t})_{21} = (Y_1^t)_{11} = (Y_1^t)_{22} = (Y_2^t)_{11} = (Y_2^t)_{22}  = (Y_1^t)_{21} = (Y_2^t)_{21} = 0
\end{align}
to enforce the LR symmetry. 

Notice that, when $m_{\Psi_{12}}^2 \simeq Y_1^2+ \bar Y_2^2$, which is realised in our scan with a good approximation as a result of the minimisation of the scalar potential, the ratio of $H$ functions appearing in the square brackets of $M_1^t M_2^t$ in Eq.~(\ref{eq:ff_appendix}) reduces to 1 and therefore
\begin{align}
\label{eq:M1M2ff}
M_1^t M_2^t = Y_1 Y_2 (M^t)^2 \,.
\end{align}

\section{The scalar potential in the Higgs basis}
\label{sec:Higgsbasis}
In this Appendix we provide the relations between the parameters of the scalar potential in the general and Higgs bases. We use capital letters for the latter case.
In both cases, the normalisation of the parameters follows from Eq.(\ref{2HDM-potential}). In the CP invariant scenario we obtain
\begin{eqnarray}
M_{11}^2 &=& m_{1}^2 c_\beta^2 + m_{2}^2 s_\beta^2  - m_{3}^2 s_{2\beta} \,, \nonumber \\
M_{12}^2 &=&  \frac{1}{2} (m_1^2 - m_2^2) +  m_{3}^2 c_{2\beta}  \,, \nonumber \\
M_{22}^2 &=&  m_{1}^2 s_\beta^2 + m_{2}^2 c_\beta^2  + m_{3}^2 s_{2\beta}  \,, \nonumber \\
\Lambda_1 &=& \lambda_1 c_\beta^4 + \lambda_2 s_\beta^4 + \frac{1}{2} \lambda_{345} s_{2\beta}^2 + 2 s_{2 \beta} (\lambda_6 c_\beta^2 + \lambda_7 s_\beta^2) \,, \nonumber \\
\Lambda_2 &=& \lambda_1 s_\beta^4 + \lambda_2 c_\beta^4 + \frac{1}{2} \lambda_{345} s_{2\beta}^2 - 2 s_{2 \beta} (\lambda_6 s_\beta^2 + \lambda_7 c_\beta^2) \,, \nonumber \\
\Lambda_3 &=& \frac{1}{4} s_{2 \beta}^2 (\lambda_1 + \lambda_2 - 2 \lambda_{345}) + \lambda_3 - s_{2 \beta} c_{2 \beta} (\lambda_6 - \lambda_7) \,, \nonumber \\
\Lambda_4 &=& \frac{1}{4} s_{2 \beta}^2 (\lambda_1 + \lambda_2 - 2 \lambda_{345}) + \lambda_4 - s_{2 \beta} c_{2 \beta} (\lambda_6 - \lambda_7)  \,, \nonumber \\
\Lambda_5 &=& \frac{1}{4} s_{2 \beta}^2 (\lambda_1 + \lambda_2 - 2 \lambda_{345}) + \lambda_5 - s_{2 \beta} c_{2 \beta} (\lambda_6 - \lambda_7) \,, \nonumber \\
\Lambda_6 &=& -\frac{1}{2} s_{2\beta} (\lambda_1 c_\beta^2 - \lambda_2 s_\beta^2 - \lambda_{345} c_{2\beta}) + \lambda_6 c_\beta c_{3 \beta} + \lambda_7 s_\beta s_{3 \beta}  \,, \nonumber \\
\Lambda_7 &=& -\frac{1}{2} s_{2\beta} (\lambda_1 s_\beta^2 - \lambda_2 c_\beta^2 + \lambda_{345} c_{2\beta}) + \lambda_6 s_\beta s_{3 \beta} + \lambda_7 c_\beta c_{3 \beta}   \,, 
\end{eqnarray}
with $\lambda_{345} = \lambda_3 + \lambda_4 + \lambda_5$.

\providecommand{\href}[2]{#2}\begingroup\raggedright\endgroup

\end{document}